\documentclass[fleqn,usenatbib]{mnras}
    
\usepackage{newtxtext,newtxmath}
\usepackage[T1]{fontenc}
\usepackage{ae,aecompl}
\usepackage{placeins}

\usepackage{graphicx}	 
\usepackage{amsmath}	
\usepackage{bbding}	
\usepackage{color}
\usepackage{gensymb}
\usepackage{comment}
\usepackage{empheq}

\usepackage{scalerel,tikz}
\usetikzlibrary{svg.path}
\definecolor{orcidlogocol}{HTML}{A6CE39}
\tikzset{orcidlogo/.pic={
 \fill[orcidlogocol] svg{M256,128c0,70.7-57.3,128-128,128C57.3,256,0,198.7,0,128C0,57.3,57.3,0,128,0C198.7,0,256,57.3,256,128z};
 \fill[white] svg{M86.3,186.2H70.9V79.1h15.4v48.4V186.2z}
 svg{M108.9,79.1h41.6c39.6,0,57,28.3,57,53.6c0,27.5-21.5,53.6-56.8,53.6h-41.8V79.1z M124.3,172.4h24.5c34.9,0,42.9-26.5,42.9-39.7c0-21.5-13.7-39.7-43.7-39.7h-23.7V172.4z}
 svg{M88.7,56.8c0,5.5-4.5,10.1-10.1,10.1c-5.6,0-10.1-4.6-10.1-10.1c0-5.6,4.5-10.1,10.1-10.1C84.2,46.7,88.7,51.3,88.7,56.8z};
}}
\newcommand\orcidicon[1]{\href{https://orcid.org/#1}{\mbox{\scalerel*{
\begin{tikzpicture}[yscale=-1,transform shape]
\pic{orcidlogo};
\end{tikzpicture}
}{|}}}}

\newcommand{\aref}[1]{\hyperref[#1]{Appendix~\ref{#1}}}

\defcitealias{2018MNRAS.477.2716K}{KBFC18}

\usepackage[normalem]{ulem}

\definecolor{darkgreen}{rgb}{0.13, 0.55, 0.13}
\definecolor{brown}{rgb}{0.65, 0.16, 0.16}

\title[Gas-phase metal distribution in galaxies]{The interplay between feedback, accretion, transport and winds in setting gas-phase metal distribution in galaxies}

\author[P. Sharda et al.]{Piyush Sharda$^{\orcidicon{0000-0003-3347-7094}\,1}$\thanks{sharda@strw.leidenuniv.nl (PS)},
Omri Ginzburg$^{\orcidicon{0000-0002-0311-2206}\,2}$\thanks{omry.ginzburg@mail.huji.ac.il (OG)},
Mark R. Krumholz$^{\orcidicon{0000-0003-3893-854X}\,3,4}$\thanks{mark.krumholz@anu.edu.au (MRK)},
John C. Forbes$^{\orcidicon{0000-0002-1975-4449}\,5}$,
\newauthor
Emily Wisnioski$^{\orcidicon{0000-0003-1657-7878}\,3,4}$,
Matilde Mingozzi$^{\orcidicon{0000-0003-2589-762X}\,6}$,
Henry R. M. Zovaro$^{\orcidicon{0000-0003-4334-9811}\,3,4}$, and
\newauthor
Avishai Dekel$^{\orcidicon{0000-0003-4174-0374}\,2,10}$
\\
$^{1}$Leiden Observatory, Leiden University, P.O. Box 9513, 2300 RA Leiden, The Netherlands\\
$^{2}$Racah Institute of Physics, The Hebrew University of Jerusalem, Jerusalem 91904, Israel\\
$^{3}$Research School of Astronomy and Astrophysics, Australian National University, Canberra, ACT 2611, Australia\\
$^{4}$Australian Research Council Centre of Excellence for All Sky Astrophysics in 3 Dimensions (ASTRO 3D), Australia\\
$^{5}$Center for Computational Astrophysics, Flatiron Institute, New York, NY 10010, USA\\
$^{6}$Space Telescope Science Institute, Baltimore, MD 21218, USA\\
$^{7}$Santa Cruz Institute for Particle Physics, University of California, Santa Cruz, CA 95064, USA\\
}

\date{Accepted 2024 January 08. Received 2023 December 04; in original form 2023 March 21}
    
\pubyear{2023}

\begin{document}
\label{firstpage}
\pagerange{\pageref{firstpage}--\pageref{lastpage}}
\maketitle

\begin{abstract}
The recent decade has seen an exponential growth in spatially-resolved metallicity measurements in the interstellar medium (ISM) of galaxies. To first order, these measurements are characterised by the slope of the radial metallicity profile, known as the metallicity gradient. In this work, we model the relative role of star formation feedback, gas transport, cosmic gas accretion, and galactic winds in driving radial metallicity profiles and setting the mass-metallicity gradient relation (MZGR). We include a comprehensive treatment of these processes by including them as sources that supply mass, metals, and energy to marginally unstable galactic discs in pressure and energy balance. We show that both feedback and accretion that can drive turbulence and enhance metal-mixing via diffusion are crucial to reproduce the observed MZGR in local galaxies. Metal transport also contributes to setting metallicity profiles, but it is sensitive to the strength of radial gas flows in galaxies. While the mass loading of galactic winds is important to reproduce the mass metallicity relation (MZR), we find that metal mass loading is more important to reproducing the MZGR. Specifically, our model predicts preferential metal enrichment of galactic winds in low-mass galaxies. This conclusion is robust against our adopted scaling of the wind mass-loading factor, uncertainties in measured wind metallicities, and systematics due to metallicity calibrations. Overall, we find that at $z \sim 0$, galactic winds and metal transport are more important in setting metallicity gradients in low-mass galaxies whereas star formation feedback and gas accretion dominate setting metallicity gradients in massive galaxies.
\end{abstract}


\section{Introduction}
\label{s:intro}
Metals act as natural tracers of galaxy evolution, and provide some of the most important constraints on the physical properties of baryons in galaxies \citep{2020ApJ...900..179K}. This has resulted in the development of several galactic chemical evolution models over the last three decades to reproduce, fit, or explain trends in the metal content of galaxies. Most of the literature thus far has focused on understanding the physics of galaxy-integrated (or global) metallicities, and fundamental trends such as the mass-metallicity relation  \citep[MZR,][]{2004ApJ...613..898T,2014ApJ...791..130Z,2020arXiv200907292S}. However, thanks to integral field unit (IFU) spectroscopy, the last decade has seen immense advancements in mapping the spatially-resolved metal content in galaxies. It is now clear that, at least at $z = 0$, galaxies typically exhibit a gradient in metallicity whereby their inner parts are more metal-rich than their outskirts. Such a negative metallicity gradient is a natural consequence of inside-out galaxy formation \citep[e.g.,][]{1983MNRAS.204...53S}. 

This simplified picture is complicated by various factors internal and external to galaxies, as is evident from the vast diversity of metallicity gradients discovered in numerous galaxies across cosmic time \citep[see reviews by][]{2019ARA&A..57..511K,2019A&ARv..27....3M,2020ARA&A..5812120S}. High resolution data from IFU spectroscopy has also revealed the existence of resolved versions of the MZR: the mass-metallicity gradient relation (MZGR), and the resolved mass-metallicity relation (rMZR). The former describes the evolution of metallicity gradients with stellar mass \citep[e.g.,][]{2014A&A...563A..49S,2017MNRAS.469..151B,2018A&A...609A.119S,2018MNRAS.479.5235P,2020A&A...636A..42M,2020aMNRAS.xxx..xxxS,2021ApJ...923...28F} whereas the latter describes the evolution of local metallicity with the local stellar surface density \citep[e.g.,][]{2012ApJ...756L..31R,2013A&A...554A..58S,2016MNRAS.463.2513B,2019MNRAS.485.5715T}.

IFU observations have started to reveal the detailed 2D structure of metal distribution in galaxies, finding evidence not just for radial, but also azimuthal variations \citep[e.g.,][]{2017ApJ...846...39H,2018A&A...618A..64H,2019ApJ...887...80K,2020arXiv200902342K,2021MNRAS.504.5496L,2023MNRAS.518..286L}. However, progress in theoretical work on modeling these $1^{\rm{st}}$ and $2^{\rm{nd}}$ order metallicity variations for a wide range of galaxies has remained rather limited \citep{2015MNRAS.450..342K,2015MNRAS.448.2030H,2018MNRAS.476.3883L,2018MNRAS.475.2236K,2019MNRAS.487..456B,2021MNRAS.502.5935S,2021MNRAS.508..489M}. A key challenge in modeling resolved metal distributions is that they are impacted by both local and global processes in galaxies (\citealt{2022arXiv221003755B}; see, however, \citealt{2022MNRAS.514.2298B}). Additionally, while the global metallicity is only sensitive to the total metal content of galaxies, the spatially-resolved metallicity structure is also affected by transport processes within the ISM, which redistribute metals from the time they are produced. Thus, metal dynamics are as important as metal production and ejection in setting spatially-resolved metal distributions. We sub-divide and introduce these two categories below (see also, \autoref{fig:lilly}).

\subsection{Gas and metal flows in/out of galaxies}
\label{s:intro_metalcontent}
To understand the physics of spatially-resolved metal distribution and metallicity gradients, it is helpful to first explore the life cycle of baryons in galaxies. Gas regulator models paint a simple yet powerful picture of this life cycle: gas accretes onto the galaxy from the cosmic web in the form of cold or hot streams, or through mergers and interactions \citep[e.g.,][]{2001ApJ...562..605F,2006MNRAS.368....2D,2009ApJ...694L.158B,2009Natur.457..451D,2010ApJ...710L.156R}. The accreted gas is then transported across the galactic disc via angular momentum exchange due to viscous torques or encounters with giant molecular clouds (GMC) and bars or spiral arms \citep[e.g.,][]{1990ApJ...363..391P,2002A&A...394L..35B,2005MNRAS.364L..18B,2009ApJ...703..785D,2010ApJ...724..895K,2011MNRAS.415.1027H}. The accreted gas also acts as a fuel for star formation, and dictates subsequent galactic life cycles \citep[e.g.,][]{2005MNRAS.363....2K,2015ARA&A..53...51S}. Star formation leads to metal production dictated by the mass distribution of massive stars following a stellar initial mass function (IMF). Some of these metals are locked in low mass stars that live for a long time, others are returned to the ISM on relatively shorter timescales \citep{1980FCPh....5..287T}, or are depleted onto dust \citep[e.g.,][]{2022A&A...666A..12K}. Star formation and active galactic nuclei (AGN) feedback drives winds that eject mass and metals out of the galaxy \citep[e.g.,][]{2005ARA&A..43..769V,2015MNRAS.454.2691M,2018MNRAS.473.4077P,2020MNRAS.494.3971M,2020A&ARv..28....2V,2021MNRAS.508.2979P}. A fraction of the ejected mass makes its way back onto the disc in the form of fountains \citep[e.g.,][]{1976ApJ...205..762S,2008MNRAS.386..935F,2019MNRAS.490.4786G,2020ApJ...898..148L}, whereas the rest enriches the circumgalactic medium (CGM) with metals \citep{2017ARA&A..55..389T}. Sufficiently strong winds can also deposit metals beyond the virial radii of galaxies into the intergalactic medium (IGM; \citealt{2004ApJ...606..862O,2008MNRAS.386.1464R,2011Sci...334..948T}). Thus, the life cycle of metals is quite dynamic, and it is now known that all these different physical processes alter the metal content of galaxies \citep[e.g.,][]{2007ApJ...658..941D,2008MNRAS.385.2181F,2013ApJ...772..119L,2019MNRAS.490.4786G,2021MNRAS.502.5935S,2022ApJ...929...95W,2022arXiv220809008G}.

\subsection{Metal dynamics within galaxies}
\label{s:intro_metaldynamics}
We define metal dynamics as processes that alter the spatial distribution of metals within galaxies without altering the total metal mass. Broadly speaking, we can further classify metal dynamics into categories: (1.) transport of metals with the bulk flow of the gas, and (2.) redistribution of metals. The former results in advection of metals, and is primarily driven by radial gas flows generated due to gravitational instabilities \citep{2009ApJ...703..785D,2010ApJ...724..895K,2012ApJ...754...48F}, magnetic stresses \citep{2022ApJ...927..217W}, stellar bars \citep{2013MNRAS.436.1479K,2015A&A...580A.127K}, or mismatch in the specific angular momentum of accreting gas and the disc \citep{1981A&A....98....1M,2016MNRAS.455.2308P}. The latter is caused by turbulence in galactic discs that mix metals produced in different parts of the galaxy \citep[e.g., due to diffusion or thermal instabilities,][]{2012ApJ...758...48Y,Petit15a,2020MNRAS.496.1891B}. Except for a handful of galactic chemical evolution models, most models to date still lack a treatment for metal advection. The few that do include this process find it to be important for setting resolved metallicity trends in galaxies \citep{2021MNRAS.502.5935S,2022arXiv220411413C}. However, the treatment of radial gas flows is often decoupled from star formation and gas accretion, which creates inconsistency in the treatment of metals and the gas, or allows for tunable parameters that are then matched to an observable to constrain metal advection. This is further complicated by the fact that direct evidence for radial gas flows is scarce (\citealt{2016MNRAS.457.2642S,2021ApJ...923..220D}; Sharda et al. in prep.), even though it has been long suspected to play an important role in galaxy evolution \citep{1980FCPh....5..287T,1985ApJ...290..154L}.

The diffusion of metals due to turbulence warrants a detailed discussion since it is a relatively new component of galactic chemical evolution models \citep{2021MNRAS.502.5935S}, and is usually implemented as sub-grid physics in hydrodynamical simulations \citep{2018MNRAS.474.2194E}. Galactic discs typically exhibit supersonic turbulence, as is now known from measurements of the multi-phase gas velocity dispersion across cosmic time \citep{2012AJ....144...96I,2015MNRAS.449.3568M,2015ApJ...799..209W,2016PASA...33....6V,2020MNRAS.495.2265V,2017ApJ...843...46S,2019ApJ...874...59S,2019ApJ...886..124W,2018ApJS..238...21F,2019ApJ...880...48U,2021ApJ...909...12G}. Simulations have shown that, in the absence of strong anisotropies \citep{2011ApJ...738...88H,2020MNRAS.492..668B} or strong magnetic fields oriented perpendicular to the disc \citep{2013ApJ...778...88K}, supersonic turbulence decays very rapidly compared to the dynamical time of the disc \citep{1998PhRvL..80.2754M,1998ApJ...508L..99S,2009ApJ...691.1092L,2009ApJ...701.1258D,2011ApJ...730...12D,2015ApJ...805..118B,2015ApJ...815...67K}. Therefore, turbulence needs to be continuously driven in galactic discs in order to explain the observed high gas velocity dispersions that cannot be a result of thermal gas motions.  cite \cite{2023arXiv230801015S}

The two most common sources of turbulent energy injection into the interstellar medium (ISM) in galactic discs are star formation feedback due to supernovae \citep{2011ApJ...731...41O,2013MNRAS.433.1970F} and radial gas flows. \cite{2018MNRAS.477.2716K} (hereafter, \citetalias{2018MNRAS.477.2716K}) present a unified model of galaxies that includes both these processes, and show that star formation feedback alone can maintain a floor velocity dispersion of $6-10\,\rm{km\,s^{-1}}$, explaining why the gas velocity dispersion tends to settle down to this value. \citetalias{2018MNRAS.477.2716K} also find that the high gas velocity dispersions seen in starburst galaxies are primarily driven by gravity due to radial mass transport. This model has been used to reproduce the observed star formation rate (SFR) - gas velocity dispersion trend in several surveys (\citealt{Johnson18a,2019MNRAS.486.4463Y,2020MNRAS.495.2265V,2021MNRAS.505.5075Y,2021ApJ...909...12G}; see also, \citealt{2021arXiv211109322E}). A number of other authors have also published models to explain the role of turbulence in galaxy evolution but without the dual role of feedback and transport while simultaneously considering energy and momentum balance \citepalias[][section 1.2 and references therein]{2018MNRAS.477.2716K}.

However, \citetalias{2018MNRAS.477.2716K} only consider gas accretion from outside the galaxy as a source that supplies mass to the galactic disc, but did not consider the possibility that it might also drive turbulence in the disc \citep{2022MNRAS.TMP.1282G}. Such driving is possible because the accreting gas can convert some of its kinetic energy into turbulence, as has been demonstrated in several simulations \citep{2010A&A...520A..17K,2010ApJ...712..294E,2012MNRAS.425..788G,2014MNRAS.437L..56G,2020MNRAS.495..758H,2022arXiv220405344F}. Some works \citep[e.g.,][]{2013MNRAS.432.2639H} argue that this effect is negligible; however, they do not consider all regimes (both in terms of halo mass and redshift) where accretion can be dominant. Some simulations are inconclusive about the importance of accretion in driving turbulence in discs \citep{2020MNRAS.496.1620O}, while others find it to be quite important, at least for certain halo masses and redshifts \citep{2013MNRAS.434..606G,2022arXiv220405344F,2022arXiv221009673J,2023MNRAS.521.2949G}. Accretion is less efficient at driving turbulence if it is more spherical rather than in the form of thin filaments, or if the accreting streams are coherent and co-rotating with the disc, which might be the case for thin-disc star-forming galaxies \citep{2015MNRAS.449.2087D,2022MNRAS.514.5056H}, but this hypothesis is complicated by the multi-phase nature of accreting streams \citep{2018A&A...610A..75C}. Regardless of the predictions of these simulations, it is clearly the case that if gas accretion drives turbulence in the disc, several assumptions about the origin of gas velocity dispersions (and other quantities that are influenced by the gas velocity dispersion, like metallicity) need to be revisited \citep{2013MNRAS.432.2639H}.

\subsection{This work and its motivation}
\label{s:intro_thiswork}
Only a handful of theoretical works have focused on the spatially-resolved metal distribution and MZGR \citep{2015MNRAS.448.2030H,2015MNRAS.450..342K,2018MNRAS.476.3883L,2019MNRAS.487..456B,2021MNRAS.503.4474Y}, and none of these investigate metal dynamics due to both radial gas flows and turbulent diffusion. \citet{2015MNRAS.448.2030H}, \citet{2015MNRAS.450..342K}, and \citet{2019MNRAS.487..456B} further assume that galactic winds are not metal enriched as compared to the ISM, which has been well known to be a severe limitation of chemical evolution models \citep[e.g.,][]{2007ApJ...658..941D,2011MNRAS.417.2962P,2018MNRAS.481.1690C,2021ApJ...918L..16C}. Several studies have discussed the role of winds in the context of chemical evolution models and shown that they are crucial to explaining trends in the MZR \citep[e.g.,][]{2008MNRAS.385.2181F,2015MNRAS.449.3274F,2018MNRAS.481.1690C,2020arXiv200910184C,2022A&A...657A..19T}, but the corresponding role of winds in setting the spatially-resolved metal distribution remains largely unexplored.

Cosmological simulations have also investigated the MZGR, with varying degrees of success in reproducing the observed trends \citep{2016MNRAS.456.2982T,2019MNRAS.482.2208T,2017MNRAS.466.4780M,2020MNRAS.495.2827C,2020arXiv200710993H,2022MNRAS.515.3555P}. While these simulations provide an excellent overview of the impacts of large-scale physical processes (including the role of CGM/IGM) and environment on metallicity profiles, they rely on subgrid recipes for star formation and metal diffusion. This is due to limitations imposed by finite resolution and the adopted temperature floor that does not capture the cold ISM. It is also difficult to control for important variables that influence metallicity profiles in these simulations, which leads to uncertainty around the relative contribution of different physical processes in setting the MZGR.

In a series of previous works, \cite{2021MNRAS.502.5935S} introduced a first principles model for the physics of gas-phase metallicity and metallicity gradients in galaxies. The authors built upon the unified disc model of \citetalias{2018MNRAS.477.2716K} by including metal dynamics to show how various galaxy properties govern the gas-phase metallicity profiles in galaxies \citep{2021MNRAS.502.5935S,2020aMNRAS.xxx..xxxS,2020bMNRAS.xxx..xxxS}. However, \citeauthor{2021MNRAS.502.5935S} only incorporated gas accretion as a process that adds mass to the disc, and did not self-consistently model gas turbulence and its relationship to accretion. In this work, following the updates to the \cite{2018MNRAS.477.2716K} model by \cite{2022MNRAS.TMP.1282G}, we self-consistently model the role of gas accretion by also including it as a source of turbulence, properly capturing the relation between metal dilution by accretion of metal-poor gas and driving of turbulence by the kinetic energy of incoming material. Such a comprehensive treatment of gas accretion has not been included in existing chemical and metallicity evolution models, which renders our understanding of the importance of gas accretion in setting ISM metallicities incomplete. As we will show in this work, gas velocity dispersion plays a central role in setting spatially-resolved gas-phase metallicities, and metal diffusion due to turbulence becomes dominant over other processes in some galaxies.

We arrange the rest of the paper as follows: \autoref{s:model} presents the model for the evolution of gas-phase metallicity, \autoref{s:results} presents resulting metallicity gradients, \autoref{s:mzgr} presents the local MZGR we obtain from the fiducial model, and \autoref{s:galactic_winds} discusses the role of galactic winds in setting the MZGR. Finally, we present our conclusions in \autoref{s:conclusions}. For the purpose of this paper, we use $Z_{\odot} = 0.0134$, corresponding to $12 + \log_{10}\rm{O/H} = 8.69$ \citep{2021A&A...653A.141A}, Hubble time $t_{\rm{H(0)}} = 13.8\,\rm{Gyr}$ \citep{2018arXiv180706209P}, and follow the flat $\Lambda$CDM cosmology: $\Omega_{\rm{m}} = 0.27$, $\Omega_{\rm{\Lambda}} = 0.73$, $h=0.71$, and $\sigma_8 = 0.81$.

\section{Model for spatially-resolved gas-phase metallicity}
\label{s:model}

In this work, we piece together models for galactic discs and gas-phase metallicities from \cite{2014MNRAS.438.1552F}, \citetalias{2018MNRAS.477.2716K}, \cite{2021MNRAS.502.5935S}, and \cite{2022MNRAS.TMP.1282G} to present a semi-analytic model for gas-phase metallicity gradients that self-consistently includes the role of accretion, feedback, and transport. We list all the mathematical symbols representing different physical parameters for the galactic disc in \autoref{tab:tab1}. Similarly, we provide parameters for the metallicity model in \autoref{tab:tab2}. For the remainder of this work, we classify galaxies according to their stellar mass, $M_{\star}$, as follows: low-mass galaxies -- $\log_{10} M_{\star}/\rm{M_{\odot}} \leq 9$, and massive galaxies -- $\log_{10} M_{\star}/\rm{M_{\odot}} \geq 10.5$.

Any chemical evolution model should also be able to reproduce (or, be consistent with) gas-phase galaxy scaling relations since the prescription of metals in these models is inherently tied to that of the gas. The galactic disc model that we incorporate from \citetalias{2018MNRAS.477.2716K} and \citet{2022MNRAS.TMP.1282G} has been verified against numerous scaling relations, including the (resolved and unresolved) Kennicutt-Schmidt relation, and the trend between star formation rate and galaxy kinematics (\citealt{2019MNRAS.486.4463Y,2020MNRAS.495.2265V,2021MNRAS.505.5075Y,2021ApJ...909...12G,2023A&A...673A.153P}; Rowland et al. in, prep.). Similarly, modeling metallicity gradients has an additional constraint as compared to modeling integrated/unresolved metallicities -- metallicity gradient models should also be able to simultaneously reproduce the scaling relations of integrated metallicity with galaxy properties. We have shown in previous works how our model also reproduces the MZR \citep{2020aMNRAS.xxx..xxxS}. Since the updates we present in this work provide a better description of metal dynamics within galaxies but does not change the global metal content in the disc, we only discuss and apply the new model to study metallicity gradients.

We base our model on the premise that galactic discs remain marginally stable due to gas transport driven by gravitational instabilities \citep{2010ApJ...724..895K,2012ApJ...754...48F,2014MNRAS.438.1552F}. In other words, we assume that galactic discs are able to self regulate to a particular value of the Toomre $Q$ parameter \citep{1964ApJ...139.1217T}. Following \citet[equation 7]{2018MNRAS.477.2716K}, we set the Toomre $Q$ parameter of the gas
\begin{equation}
Q_{\rm{g}} = \frac{\kappa \sigma_{\rm{g}}}{\pi G \Sigma_{\rm{g}}}
\label{eq:Qg}
\end{equation}
where $\kappa$ is the epicyclic frequency given by $\kappa = \sqrt{2(1+\beta)}\Omega$; here, $\beta = d \,\mathrm{ln} v_{\phi}/d\,\mathrm{ln}\,r$ is the rotation curve index of the galaxy, $v_{\phi}$ is the rotational velocity at the outer edge of the disc, and $\Omega$ is the angular velocity. $\sigma_{\rm{g}}$ is the gas velocity dispersion, and $\Sigma_{\rm{g}}$ is the gas surface density. We assume a flat rotation curve for massive galaxies ($\beta = 0$) and a rising rotation curve for low-mass galaxies ($\beta = 0.5$). Following \cite{2020aMNRAS.xxx..xxxS}, we create a linear ramp to interpolate between these two for setting $\beta$ for intermediate-mass galaxies ($9 \leq \log_{10} M_{\star}/\rm{M_{\odot}} \leq 10.5$).

Following \cite{2013MNRAS.433.1389R}, we also express the total Toomre $Q$ parameter of the gas and the stars as
\begin{equation}
Q = f_{\rm{g,Q}}\times Q_{\rm{g}}\,,
\label{eq:Q}
\end{equation}
where $f_{\rm{g,Q}}$ is the effective gas fraction in the disc \citepalias[equation 9]{2018MNRAS.477.2716K}. For discs to be marginally stable, we require that $Q \geq Q_{\rm{min}}$, where $Q_{\rm{min}}$ is the minimum Toomre $Q$ parameter needed for stability (see, however, \citealt{2023arXiv230207823F} for a more sophisticated approach). As a fiducial value, we set $Q_{\rm{min}}=1$ as appropriate for relatively quiescent discs.\footnote{Note that, unlike this work, both \cite{2021MNRAS.502.5935S} and \cite{2022MNRAS.TMP.1282G} set $Q=Q_{\rm{min}}$ for simplicity. } We will discuss the role of $Q_{\rm{min}}$ in detail in \autoref{s:model_gasmass}.

\begin{figure*}
\includegraphics[width=\textwidth]{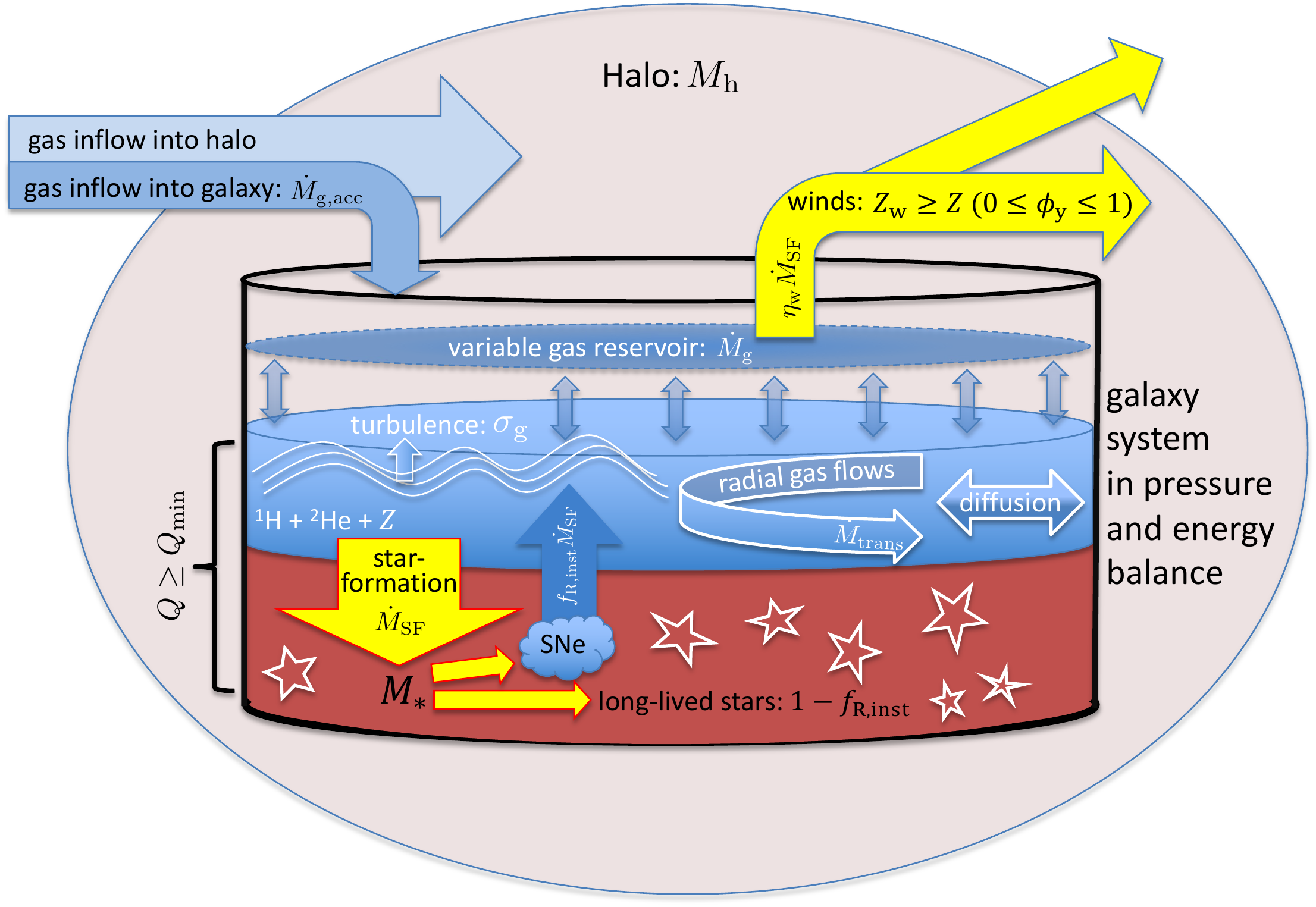}
\caption{Bathtub model of chemical evolution, adapted from \protect\citet[figure 2]{2013ApJ...772..119L}, and modified to reflect the spatially-resolved metallicity model presented in this work. The galaxy is modeled as a two component reservoir: stellar (red) and gas-phase (blue). The gas reservoir is made up of H, He and metals (denoted by the ISM metallicity $Z$). We consider marginally unstable galactic discs (quantified by the Toomre $Q$ parameter) in pressure and energy balance. The evolution of gas mass, $\dot M_{\rm{g}}$, is described by \autoref{eq:dotMg}. It depends on the gas accretion rate ($\dot M_{\rm{g,acc}}$), star formation rate ($\dot M_{\rm{SF}}$), radial gas flows (or, gas transport, $\dot M_{\rm{trans}}$), and galactic winds ($\eta_{\rm{w}}\dot M_{\rm{SF}}$, where $\eta_{\rm{w}}$ is the mass-loading factor). Gas turbulence (quantified by the gas velocity dispersion, $\sigma_{\rm{g}}$) is driven by energy injected into the ISM via accretion, star formation feedback, and gas transport. Metals are produced as star formation leads to supernovae (SNe). Some fraction of the metals produced is returned almost instantaneously to the ISM ($f_{\rm{R,inst}}$) whereas the rest is locked in long-lived stars. Metals are also ejected via galactic winds with metallicity $Z_{\rm{w}}$, which can be greater than or equal to $Z$. $\phi_{\rm{y}}$ denotes the preferential metal enrichment of winds. Once produced, metals are also advected with radial gas flows, and diffused into the ISM due to turbulence and thermal instabilities. © AAS. Reproduced with permission.}
\label{fig:lilly}
\end{figure*}

We begin by describing the evolution of the halo mass and gas mass (\autoref{s:model_gasmass}), from which we infer the gas velocity dispersion (\autoref{s:model_gasvelocitydispersion}) based on the Toomre criterion. We then incorporate these parameters into the metallicity model, solve for the metallicity at each radius, and check if the radial metallicity distribution is in equilibrium. If it is, we find radial metallicity profiles (\autoref{s:model_gasphasemetal}) and calculate the metallicity gradient. This way, we can self consistently find the gas velocity dispersion and metallicity at each epoch without violating the total mass, energy and metal budget of the disc. We provide a schematic that captures the most important ingredients of our model in \autoref{fig:lilly}.

\begin{table*}
\centering
\caption{List of parameters in the galaxy evolution model.}
\begin{tabular}{|l|l|c|c}
\hline
Parameter & Description & Reference equation & Fiducial value  \\
\hline
$M_{\rm{h}}$ & Halo mass & ... & ...\\
$M_{\star}$ & Stellar mass & ... & ...\\
$M_{\rm{g}}$ & Gas mass & ... & ...\\
$\Sigma_{\rm{g}}$ & Gas surface density & \autoref{eq:Qg} & ...\\
$\kappa$ & Epicyclic frequency & \autoref{eq:Qg} & ...\\
$\beta$ & Rotation curve index & \autoref{eq:Qg} & $-0.5 - 0.5$\\
$\sigma_{\rm{g}}$ & Gas velocity dispersion & \autoref{eq:Qg} & ...\\
$Q$ & Toomre $Q$ parameter of stars + gas & \autoref{eq:Q} & $\geq Q_{\rm{min}}$\\
$\dot M_{\rm{h}}$ & Dark matter accretion rate & \autoref{eq:dotMh} & ...\\
$v_{\phi}$ & Rotational velocity & \autoref{eq:vphi} & ...\\
$c$ & Halo concentration parameter & \autoref{eq:vphi} & $6-17$ \\
$\dot M_{\rm{g}}$ & Rate of change of gas mass & \autoref{eq:dotMg} & ...\\
$\eta_{\rm{w}}$ & Wind mass loading factor & \autoref{eq:dotMg} & $0$\\
$\dot M_{\rm{g,acc}}$ & Gas accretion rate & \autoref{eq:dotMgacc} & ...\\
$f_{\mathrm{B}}$ & Universal baryonic fraction & \autoref{eq:dotMgacc} & $0.17$\\
$\epsilon_{\rm{in}}$ & Baryonic accretion efficiency & \autoref{eq:dotMgacc} & $0.1 - 1$\\
$\dot M_{\rm{SF}}$ & Star formation rate & \autoref{eq:dotMsf} & ...\\
$\phi_{\rm{a}}$ & Ratio of unresolved and resolved star formation rate & \autoref{eq:dotMsf} & $2$\\
$t_{\rm{SF,max}}$ & Maximum gas depletion timescale & \autoref{eq:dotMsf} & $2\,\rm{Gyr}$\\
$\epsilon_{\mathrm{ff}}$ & Star formation efficiency per free-fall time & \autoref{eq:dotMsf} & $0.015$\\
$\phi_{\mathrm{mp}}$ & Ratio of the total to the turbulent pressure at the disc midplane & \autoref{eq:dotMsf} & $1.4$\\
$f_{\rm{g,Q}}$ & Effective gas fraction in the disc & \autoref{eq:dotMsf} & $0 -1$\\
$f_{\mathrm{sf}}$ & Fraction of star-forming gas & \autoref{eq:dotMsf} & $0-1$\\
$f_{\rm{g,P}}$ & Effective mid-plane pressure due to self-gravity of gas & \autoref{eq:dotMsf} & $0-1$\\
$\dot M_{\rm{trans}}$ & Gas transport due to radial flows & \autoref{eq:dotMtrans} & ...\\
$\eta$ & Scaling factor for the rate of turbulent dissipation & \autoref{eq:dotMtrans} & $1.5$ \\
$\phi_{\rm{Q}}$ & 1 $+$ ratio of gas to stellar Toomre $Q$ parameter & \autoref{eq:phiQ} & $2$\\
$\phi_{\rm{nt}}$ & Fraction of gas velocity dispersion due to non-thermal motions & \autoref{eq:phint} & $0-1$\\
$\sigma_{\rm{SF}}$ & Turbulence driven by star formation feedback & \autoref{eq:sigmasf} & ...\\
$\langle p/m \rangle_{\star}$ & Average supernova momentum per stellar mass formed & \autoref{eq:sigmasf} & $3000\,\rm{km\,s^{-1}}$ \\
$\sigma_{\rm{acc}}$ & Turbulence driven by accretion & \autoref{eq:sigmaacc} & ... \\
$\xi_{\rm{a}}$ & Accretion-induced turbulence injection efficiency & \autoref{eq:sigmaacc} & $0.2$\\
\hline
\end{tabular}
\label{tab:tab1}
\end{table*}

\subsection{Evolution of the halo mass and gas mass}
\label{s:model_gasmass}
To describe the evolution of the halo mass, $M_{\rm{h}}$, we first estimate the halo accretion rate, $\dot M_{\rm{h}}$, following \cite{2008MNRAS.383..615N} and \cite{2010ApJ...718.1001B}
\begin{equation}
    \frac{\dot M_{\rm{h}}}{M_{\rm{h}}} = -a \left(\frac{M_{\rm{h}}}{10^{12}\,\rm{M_{\odot}}}\right)^b \dot w\,,
\label{eq:dotMh}
\end{equation}
where $a=0.628$, $b=0.14$ and the self-similar time variable $w$ is such that

\begin{equation}
\dot w = -0.0476 \times \left(1 + z + 0.093(1+z)^{-1.22}\right)^{2.5}\, \rm{Gyr^{-1}}\,.
\label{eq:dotw}
\end{equation}
We integrate \autoref{eq:dotMh} to find the halo mass, $M_{\rm{h}}$ as a function of redshift. Note that we use the stellar mass -- halo mass relation from \citet[table 1]{2013MNRAS.428.3121M} to estimate $M_{\star}$. Once we determine $M_{\rm{h}}$, we follow \cite{2013MNRAS.435..999D} to find the rotational velocity at the outer edge of the disc, $v_{\phi}$, assuming a \cite{1997ApJ...490..493N} profile for the dark matter density as \citep[equations 69$-$71]{2018MNRAS.477.2716K}
\begin{equation}
  v_{\phi} = 76.17\sqrt{\frac{c}{\ln(1+c) - c/(1+c)}}\left(\frac{M_{\rm{h}}}{10^{12}\,\rm{M_{\odot}}}\right)^{1/3}\left(1+z\right)^{1/2}
\label{eq:vphi}
\end{equation}
where $c$ is the halo concentration parameter that encodes the merger history of dark matter haloes. We follow \citet[fig. 16]{2009ApJ...707..354Z} to vary $c$ as a function of $M_{\rm{h}}$, ignoring the $\sim 0.1$ dex scatter in $c$ at fixed $M_{\rm{h}}$ \citep[e.g.,][]{2000ApJ...535...30J,2002ApJ...568...52W,2019ApJ...871..168D}, since changes by an order of magnitude in $c$ only impact $v_{\phi}$ by $\sim 50$ per cent. We also evolve the size of the galactic disc following \citep{2022MNRAS.TMP.1282G}
\begin{equation}
R = 2\times1.89  \left(\frac{M_{\rm{h}}}{10^{12}\,\rm{M_{\odot}}}\right)^{1/3}\left(\frac{1+z}{3}\right)^{-1}\,,
\label{eq:R}
\end{equation}
where we have added the leading factor of two to ensure we cover most of the star-forming part of the disc.

The rate of change of gas mass in the disc is controlled by processes that supply mass to the disc (e.g., accretion) and that remove mass from the disc (e.g., star formation and outflows). Thus
\begin{equation}
\dot M_{\rm{g}} = \dot M_{\rm{g,acc}} - \dot M_{\rm{trans}} - \dot M_{\rm{SF}} - \eta_{\rm{w}}\dot M_{\rm{SF}}
\label{eq:dotMg}
\end{equation}
where the terms on the right hand side represent, respectively, the accretion rate of gas onto the galactic disc ($\dot M_{\rm{g,acc}}$), radial transport of gas through the disc ($\dot M_{\rm{trans}}$, also referred to as radial gas flows), star formation rate ($\dot M_{\rm{SF}}$), and mass outflow rate given by the product of the mass-loading factor $\eta_{\rm{w}}$ and $\dot M_{\rm{SF}}$. In writing \autoref{eq:dotMg}, we have assumed that the gas is transported down the potential well of the galaxy and ends up in a bulge that forms a distinct component of the galaxy. We, however, do not track the evolution of the bulge in the model; the bulge is only intended to act as a boundary condition for the galactic disc rather than a separate reservoir. We have neglected the recycling of metals in the form of galactic fountains \citep{2013A&A...551A.123S,2016ApJ...824...57C,2019MNRAS.485.2511T,2019MNRAS.490.4786G,2020A&ARv..28....2V}. Assuming it is the gas expelled by star formation feedback that gets recycled through a fountain, including galactic fountains would add a source term of the form $\alpha_{\rm{GF}}\dot M_{\rm{SF}}$ in \autoref{eq:dotMg}, where $\alpha_{\rm{GF}}$ describes mass-loading of material returned to the galaxy \citep[e.g.,][]{2020ApJ...897...81L}. However, it is unclear if this is sufficient to describe galactic fountains because the mass and dynamics of fountains are not just set by star formation-driven winds, but also by material already present in the CGM \citep[e.g.,][]{2022arXiv221109755P}. Further, there is no consensus on the timescales on which fountains re-supply mass and metals to the galaxy \citep[e.g.,][]{2009A&A...504...87S,2017MNRAS.470.4698A,2019MNRAS.490.4786G,2020MNRAS.497.4495M}.

\subsubsection{Gas accretion}
To define the gas accretion rate, we re-write it as
\begin{equation}
    \dot M_{\rm{g,acc}} = \dot M_{\mathrm{h}}f_{\mathrm{B}}\epsilon_{\mathrm{in}}
\label{eq:dotMgacc}
\end{equation}
where $f_{\rm{B}}=0.17$ is the universal baryonic fraction \citep{1995MNRAS.273...72W,2010ApJ...725.2324B,2016A&A...594A..13P}, and $\epsilon_{\rm{in}}$ is the accretion efficiency of baryons that we adopt from \citet[equation 22]{2014MNRAS.438.1552F}, which is based on fits to cosmological simulations of \cite{2011MNRAS.417.2982F}. As \citetalias{2018MNRAS.477.2716K} and \cite{2022MNRAS.TMP.1282G} mention, these fits likely overestimate the accretion rate in the most massive halos at $z = 0$.

\subsubsection{Star formation}
Following \cite{2012ApJ...745...69K} and \cite{2014MNRAS.438.1552F}, we assume that the star formation rate, $\dot M_{\rm{SF}}$, is different in the Toomre-regime (where it is set by global disc instabilities -- \citealt{2008ApJ...687...59G,2020arXiv200306245T,2020RSOS....700556H}) and in the GMC-regime (where it is set by local instabilities, as observed in the Milky Way and nearby spiral galaxies -- \citealt{2008AJ....136.2846B,2008AJ....136.2782L,2020MNRAS.498..385J}). In other words, star formation occurs in a continuous, volume-filling ISM in the Toomre regime whereas it occurs in individual molecular clouds in the GMC regime. Note that both \cite{2021MNRAS.502.5935S} and \cite{2022MNRAS.TMP.1282G} only consider the Toomre regime for calculating $\dot M_{\rm{SF}}$. In our present model, we take the SFR to be \citep[equation 60]{2018MNRAS.477.2716K}
\begin{align}
\nonumber \dot M_{\rm{SF}} = \sqrt{\frac{2}{1+\beta}} \frac{\phi_{\rm{a}}f_{\rm{sf}}f_{\rm{g,Q}}v^2_{\phi}\sigma_{\rm{g}}}{\pi G Q}\times \\
\mathrm{max} \left[\sqrt{\frac{2(1+\beta)}{3f_{\rm{g,P}}\phi_{\rm{mp}}}}\frac{8\epsilon_{\rm{ff}}f_{\rm{g,Q}}}{Q}, \frac{t_{\rm{orb,out}}}{t_{\rm{SF,max}}}\right]
\label{eq:dotMsf}
\end{align}
where $f_{\rm{sf}}$ is the fraction of gas in the star-forming molecular phase, $\phi_{\rm{a}}$ is a parameter of order unity that accounts for the integral of the star formation rate per unit area for galaxies with varying $\beta$ \citep[section 3.1.2]{2018MNRAS.477.2716K}, $\epsilon_{\rm{ff}}$ is the star formation efficiency per freefall time, $f_{\rm{g,P}}$ denotes the ratio of the mid-plane pressure due to self gravity of the gas to the total mid-plane pressure, $\phi_{\rm{mp}}$ denotes the ratio of the total to the turbulent pressure at the disc mid-plane, $t_{\rm{orb,out}}$ is the orbital timescale at the outer edge of the disc, and $t_{\rm{SF,max}} \approx 2\,\rm{Gyr}$ is the maximum gas depletion timescale. Thus, terms in parentheses represent star formation feedback in the Toomre and the GMC regimes, respectively.

In practice, we set $f_{\rm{g,Q}}=f_{\rm{g,P}}$, $\phi_{\rm{mp}}=1.4$, and $\phi_{\rm{a}} = 2$. Following numerous results \citep{2012ApJ...745...69K,2012ApJ...761..156F,2015ApJ...806L..36S,2018MNRAS.477.4380S,2019MNRAS.487.4305S,2022MNRAS.511.1431H}, we keep $\epsilon_{\rm{ff}}$ = 0.015, ignoring possible intrinsic scatter \citep{2021MNRAS.502.5997H,2022MNRAS.511.1431H}, variations with radius \citep{2018ApJ...869..126L,2022ApJ...928..169F} or redshift \citep{2020arXiv200306245T}. We express $f_{\rm{sf}}$ as the ratio of the molecular to the molecular and atomic gas mass, and find it following a compilation of observations by \citet[table 3]{2020arXiv200306245T} and \citet[figure 4]{2022arXiv220200690S}. We also consider a theoretical model that can specify radial variations in $f_{\rm{sf}}$ \citep{2013MNRAS.436.2747K} in \aref{s:app_fsf}.

\subsubsection{Gas transport}
As gravitational instabilities and non-axisymmetric torques transport angular momentum outwards from the galaxy centre, mass is transported inwards to conserve angular momentum (a process similar to that observed in accretion discs around protostars and black holes -- \citealt{1973A&A....24..337S,1998ApJ...495..385H}). The effective transport rate can be estimated either from considerations of energy conservation \citep{2010ApJ...724..895K} or, in the case of gas-rich discs that produce massive clumps, from analysis of clump kinematics \citep{2009ApJ...703..785D}. Since the two approaches differ only by a small factor in their predictions (and not at all for galaxies with flat rotation curves), we adopt the transport rate by \citetalias{2018MNRAS.477.2716K}, who adopt the energy conservation approach. Their formulation is based on the premise that star formation feedback can also drive turbulence in the disc in addition to gas transport. \cite{2022MNRAS.TMP.1282G} extend the \citetalias{2018MNRAS.477.2716K} formulation to include gas accretion as another source of turbulence in the disc. We combine the two approaches set in (\citetalias{2018MNRAS.477.2716K}, equation 49) and \citet[equation 39]{2022MNRAS.TMP.1282G} to define the gas transport rate as
\begin{equation}
\dot M_{\rm{trans}} = \frac{4\eta\phi_{\rm{Q}}\phi^{3/2}_{\rm{nt}}f^2_{\rm{g,Q}}}{Q^2_{\rm{min}}}\left(\frac{1+\beta}{1-\beta}\right)\left(\frac{\sigma^3_{\rm{g}}}{G}\right)F(\sigma_{\rm{g}})\,,
\label{eq:dotMtrans}
\end{equation}
where $\dot M_{\rm{trans}} > 0$ corresponds to inward mass flow, $\eta = 1.5$ defines the dissipation of turbulence across the scale height of the disc in one crossing time, $\phi_{\rm{Q}}$ is given by
\begin{equation}
\phi_{\rm{Q}} = 1 + \frac{Q_{\rm{g}}}{Q_{\star}}\,,
\label{eq:phiQ}
\end{equation}
and $\phi_{\rm{nt}}$ parameterises the contribution of thermal motions to the total gas velocity dispersion,
\begin{equation}
\phi_{\rm{nt}} = 1 - \frac{\sigma^2_{\rm{th}}}{\sigma^2_{\rm{g}}}\,,
\label{eq:phint}
\end{equation}
where $\sigma_{\rm{th}}$ is the thermal gas velocity dispersion. Following \citet[section 2.4.3]{2018MNRAS.477.2716K}, we estimate $\sigma_{\rm{th}} = f_{\rm{sf}}\sigma_{\rm{th,mol}} + (1-f_{\rm{sf}}\sigma_{\rm{th,WNM}})$. Here, $\sigma_{\rm{th,mol}} \approx 0.2\,\rm{km\,s^{-1}}$ is the thermal velocity dispersion of molecular gas assuming a gas temperature $\approx 10\,\rm{K}$, and $\sigma_{\rm{th,WNM}} \approx 5.4\,\rm{km\,s^{-1}}$ is the thermal velocity dispersion of the warm neutral medium \citep[WNM,][]{2003ApJ...587..278W}.

The last term in \autoref{eq:dotMtrans}, $F (\sigma_{\rm{g}})$, warrants a detailed discussion. It represents the fractional contribution of star formation feedback and gas accretion as compared to transport in driving turbulence in the disc. Below, we analyze different cases where we only include star formation feedback or accretion as drivers of turbulence in the disc, to study their impact on the gas-phase metallicity gradients.

\begin{enumerate}
    \item {\textit{Turbulence driven by feedback and transport.}}  
    First, we only consider the case where star formation feedback and gas transport due to radial flows drive $\sigma_{\rm{g}}$, and accretion only adds mass to the disc but does not drive turbulence in the disc. This is equivalent to setting 
    
    \begin{equation}
    F(\sigma_{\rm{g}}) = 1 - \frac{\sigma_{\rm{SF}}}{\sigma_{\rm{g}}}\,.
    \end{equation}
    where $\sigma_{\rm{SF}}$ is the gas velocity dispersion that can be maintained by star formation feedback. We define $\sigma_{\rm{SF}}$ as \citep[equation 39]{2018MNRAS.477.2716K}
    \begin{align}
    \nonumber \sigma_{\rm{SF}} = \frac{4f_{\rm{sf}\epsilon_{\rm{ff}}}}{\sqrt{3f_{\rm{g,P}}}\pi\eta\phi_{\rm{mp}}\phi_{\rm{Q}}\phi^{3/2}_{\rm{nt}}} \left\langle \frac{p}{m} \right\rangle_{\star}\,\times\\
    \mathrm{max}\left[1, \sqrt{\frac{3f_{\rm{g,P}}}{8(1+\beta)}}\frac{Q_{\rm{min}}\phi_{\rm{mp}}}{4f_{\rm{g,Q}}\epsilon_{\rm{ff}}}\frac{t_{\rm{orb}}}{t_{\rm{SF,max}}}\right]\,,
    \label{eq:sigmasf}
    \end{align}
    where we maximise over terms that represent star formation feedback in the Toomre and the GMC regimes, respectively. $\langle p/m \rangle_{\star} \approx 3000\,\rm{km\,s^{-1}}$ represents the average momentum per unit stellar mass formed that is injected into the ISM by non-clustered core-collapse supernovae \citep{1988ApJ...334..252C,1998ApJ...500...95T,2010ApJ...721..975O,2015MNRAS.451.2757W,2015ApJ...802...99K,2017MNRAS.465.1682H}. Although simulations of clustered supernovae that lead to the formation of superbubbles \citep[e.g.,][]{2014MNRAS.442.3013K} suggest a value higher by a factor of few \citep{2015ApJ...802...99K,2017MNRAS.465.2471G,2019MNRAS.483.3647G,2017ApJ...834...25K}, discrepancies up to an order of magnitude exist between different simulations \citep[see the discussion in][]{2022arXiv220810528H}. These discrepancies stem from a mixture of numerical (treatment of shocks and contact discontinuities in Lagrangian versus Eulerian codes) and physical (impact of magnetic fields) issues. Supernova clustering is also sensitive to the local ISM metallicity due to metallicity-dependent cooling and associated expansion of superbubbles \citep{2020ApJ...896...66K}. It is also not yet clear what fraction of this momentum drives turbulence in the ISM as compared to driving winds \citep[e.g.,][]{2022ApJ...932...88O}. Given these caveats, we continue to use momentum injection from non-clustered supernovae in our model, but emphasize that further exploration of supernova clustering (especially at non-Solar metallicities) is highly desirable to more accurately model the role of star formation feedback for ISM metal distribution.

    \item{\textit{Turbulence driven by accretion and transport.}} 
    Next, we consider the case where only accretion and transport drive $\sigma_{\rm{g}}$. Here, 
    \begin{equation}
    F(\sigma_{\rm{g}}) = 1 - \left(\frac{\sigma_{\rm{acc}}}{\sigma_{\rm{g}}}\right)^3\,,
    \end{equation}
    where $\sigma_{\rm{acc}}$ is the gas velocity dispersion due to turbulence induced by gas accretion
    \begin{equation}
    \sigma_{\rm{acc}} = \left(\frac{(2+\beta) \xi_{\rm{a}}G\dot M_{\rm{g,acc}}}{8(1+\beta)\eta\phi_{\rm{Q}}\phi^{3/2}_{\rm{nt}}}\frac{Q^2_{\rm{min}}}{f^2_{\rm{g,Q}}}\right)^{1/3}
    \label{eq:sigmaacc}
    \end{equation}
    where $\xi_{\rm{a}}$ is the fraction of kinetic energy of the accreted gas that drives turbulent motions in the disc.\footnote{In writing \autoref{eq:sigmaacc}, we have equated mass transport rates due to turbulent viscosity from \citet[equation 49]{2018MNRAS.477.2716K} and \citet[equation 39]{2022MNRAS.TMP.1282G} to re-define the parameter $\gamma_{\rm{diss}}$ in \cite{2022MNRAS.TMP.1282G} as $\gamma_{\rm{diss}} = 3/\left(2\eta\phi_Q \phi^{3/2}_{\rm{nt}}\right)$. Our expression differs from \citet[equation 37]{2022MNRAS.TMP.1282G} by a factor $(2+\beta)/2$ because we also study galaxies where $v_{\phi}$ is not constant across the disc (i.e., galaxies with $\beta \neq 0$). For $\beta = 0.5$, this makes a 7 per cent difference for $\sigma_{\rm{acc}}$.} \cite{2010A&A...520A..17K} argue that $\xi_{\rm{a}}$ is set by the density contrast between the accreting material and the material in the disc. The density of the accreting material depends on the clumpiness of the accreting streams \citep{2020MNRAS.498.2415M}. Taking inspiration from models that find the clumpiness of accreting streams to strongly vary with redshift \citep{2018ApJ...861..148M}, \cite{2022MNRAS.TMP.1282G} propose $\xi_{\rm{a}} = 0.2(1+z)$. Further, \cite{2022arXiv220405344F} also find $\xi_{\rm{a}} \approx 0.1-0.2$ on average across the disc at $z \approx 0$ for a $M_{\star} = 5\times10^{9}\,\rm{M_{\odot}}$ galaxy they study in the Illustris TNG50 suite of simulations \citep{2019ComAC...6....2N}. However, there are no direct observational measurements of $\xi_{\rm{a}}$ to date. Keeping these facts in mind, we fix $\xi_{\rm{a}} =0.2(1+z)$, and show how our results remain unchanged for a higher $\xi_{\rm{a}}$ in \aref{s:app_xia}.

    \item{\textit{Turbulence driven by feedback, accretion and transport.}}
    In this case, star formation feedback, gas accretion and gas transport all drive turbulence in the galaxy. The equivalent expression for $F(\sigma_{\rm{g}})$ then becomes \citep{2022MNRAS.TMP.1282G}
    
    \begin{equation}
        F(\sigma_{\rm{g}}) = 1 - \frac{\sigma_{\rm{SF}}}{\sigma_{\rm{g}}} - \left(\frac{\sigma_{\rm{acc}}}{\sigma_{\rm{g}}}\right)^3\,.
    \label{eq:Fsigmag}
    \end{equation}
    It is worth noting that $\sigma_{g}$ equilibrates to a higher value when both accretion and feedback drive turbulence as compared to the cases above. In the analysis that follows, we use different versions of $F(\sigma_{\rm{g}})$ to understand the impacts of various drivers of turbulence on gas-phase metallicity gradients.

\end{enumerate}

\subsubsection{Winds}
\label{s:model_gasphase_winds}
The role of galactic winds for the evolution of $M_{\rm{g}}$ is described by $\eta_{\rm{w}}$ (see \autoref{eq:dotMg}). Several works have focused on developing models for $\eta_{\rm{w}}$ \citep{2013MNRAS.429.1922C,2015MNRAS.446.2125C,2014MNRAS.443..168F,2017MNRAS.465.1682H,2019MNRAS.484.5587T,2020MNRAS.497..698T}, as well as measuring it in simulations \citep{2015MNRAS.454.2691M,2016ApJ...824...57C,2018MNRAS.475..648P,2020ApJ...900...61K,2021MNRAS.508.2979P}. These works find $\eta_{\rm{w}}$ that spans more than three orders of magnitude \citep[fig. 13]{2020MNRAS.494.3971M}, which is a substantially larger range than that estimated from direct observations of galactic winds. This scatter is a complex combination of varying details of (star formation and/or AGN) feedback models, accurately resolving the multi-phase nature of galactic winds, the location at which $\eta_{\rm{w}}$ is measured (in isolated versus cosmological simulations), treatment of the CGM, and burstiness of star formation. Observations estimate $\eta_{\rm{w}} \approx 0-30$ \citep{2012MNRAS.426..801B,2012ApJ...761...43N,2014ApJ...792L..12K,2015ApJ...804...83S,2019MNRAS.490.4368S,2017MNRAS.469.4831C,2019ApJ...873..122D,2019ApJ...875...21F,2019ApJ...886...74M}, but are plagued by systematics originating from the location as well as the thermal phase in which it is measured.

In our fiducial model, we leave $\eta_{\rm{w}}$ undefined because no models exist that self-consistently treat the role of outflows in driving turbulence in galactic discs together with the other sources we describe above.\footnote{Leaving $\eta_{\rm{w}}$ undefined is not the same as setting it to 0 since the latter would constrain $\phi_{\rm{y}}$ to unity as per \autoref{eq:phiy}, not permitting variations in $\phi_{\rm{y}}$. Since our aim in this section is to explore the full range of $\phi_{\rm{y}}$, we leave $\eta_{\rm{w}}$ undefined.} In practice, this means that we do not consider the wind term in \autoref{eq:dotMg} in the fiducial model, but we explore the full possible range of preferential metal ejection via winds below in \autoref{s:model_gasphasemetal}. To rectify this inconsistency, we will later consider three different scalings of $\eta_{\rm{w}}$ derived from theoretical models and simulations to discuss the effects of a non-zero $\eta_{\rm{w}}$ in \autoref{s:galactic_winds}. Note that we do not consider AGN feedback in our model. In massive haloes, AGN feedback can boost $\eta_{\rm{w}}$ \citep{2020MNRAS.494.3971M}, so it is likely that some of the scalings we consider based on studies excluding AGN feedback underestimate $\eta_{\rm{w}}$ at the high mass end.

\subsection{Gas velocity dispersion}
\label{s:model_gasvelocitydispersion}
With all the terms in \autoref{eq:dotMg} defined, we can now integrate the equation to obtain the evolution of the gas mass as a function of redshift. The only input parameter that we need is the halo mass at some high redshift, which we use to evolve the galaxy down to $z=0$.\footnote{We use \texttt{SCIPY} solve\_ivp to numerically integrate \autoref{eq:dotMg} from high-$z$ to $z = 0$. Given the stiffness in the differential equation that arises from the different timescales of accretion and star formation, we employ the backwards differentiation formula \citep[BDF,][]{bdf_ref} method to search for stable solutions at each $z$.} The exact choice of redshift is not important since the solution quickly converges to its steady-state value within a few orbital times \citep[fig. B1]{2022MNRAS.TMP.1282G}. 

In cases where $F(\sigma_{\rm{g}}) \geq 0$, we use the Toomre $Q$ criterion (\autoref{eq:Qg}) to obtain $\sigma_{\rm{g}}(z)$. However, we also encounter cases where $F(\sigma_{\rm{g}}) < 0$; physically, this occurs when mass transport is not needed to drive the required level of gas velocity dispersion, and the disc self-regulates the Toomre $Q$ parameter. In such cases, $Q \geq Q_{\rm{min}}$, and we find $\sigma_{\rm{g}}$ by solving for $F(\sigma_{\rm{g}}) = 0$.\footnote{The equation we solve is a sixth degree polynomial in $\sigma_{\rm{g}}$, as compared to the cubic equation in \cite{2022MNRAS.TMP.1282G} because $\sigma_{\rm{SF}} \propto \phi^{-3/2}_{\rm{nt}}$ (cf. \autoref{eq:sigmasf}) and $\sigma_{\rm{acc}} \propto \phi^{-3/2}_{\rm{nt}}$ (cf. \autoref{eq:sigmaacc}). We retain the dependence of $\phi_{\rm{nt}}$ on $\sigma_{\rm{g}}$ whereas \cite{2022MNRAS.TMP.1282G} assumed $\phi_{\rm{nt}} = 1$.} This is a major improvement over \cite{2021MNRAS.502.5935S} because we self-consistently derive $\sigma_{\rm{g}}$ based on the overall mass and energy budget in galactic discs instead of setting it to \textit{ad-hoc} values. Although only a handful of theoretical studies have focused on possible correlations between $\sigma_{\rm{g}}$ and gas-phase metallicity \citep{2017MNRAS.466.4780M,2018MNRAS.475.2236K,2020bMNRAS.xxx..xxxS,2020arXiv200710993H}, it is expected that the ISM metal distribution is sensitive to $\sigma_{\rm{g}}$ \citep{2012A&A...539A..93Q,2021MNRAS.500.4229G}, so it is crucial to self-consistently find $\sigma_{\rm{g}}$. Solving for $\sigma_{\rm{g}}$ in this manner means that we do not take any radial variations in $\sigma_{\rm{g}}$ into account; however, these are expected to be minor, since both modeling \citepalias{2018MNRAS.477.2716K} and observations \citep[e.g.,][]{2011MNRAS.410.1409W,2016AJ....151...15M} show that $\sigma_\mathrm{g}$ varies with radius by at most a factor of two in local galaxies. We do not make a distinction between $\sigma_{\rm{g}}$ in the star-forming molecular phase versus the ionized phase, though we caution that recent observations and simulations have shown the former can be systematically lower \citep{2021ApJ...909...12G,2021arXiv211109322E,2022arXiv221115419R,2023arXiv230200030L}.

\begin{figure}
\includegraphics[width=\columnwidth]{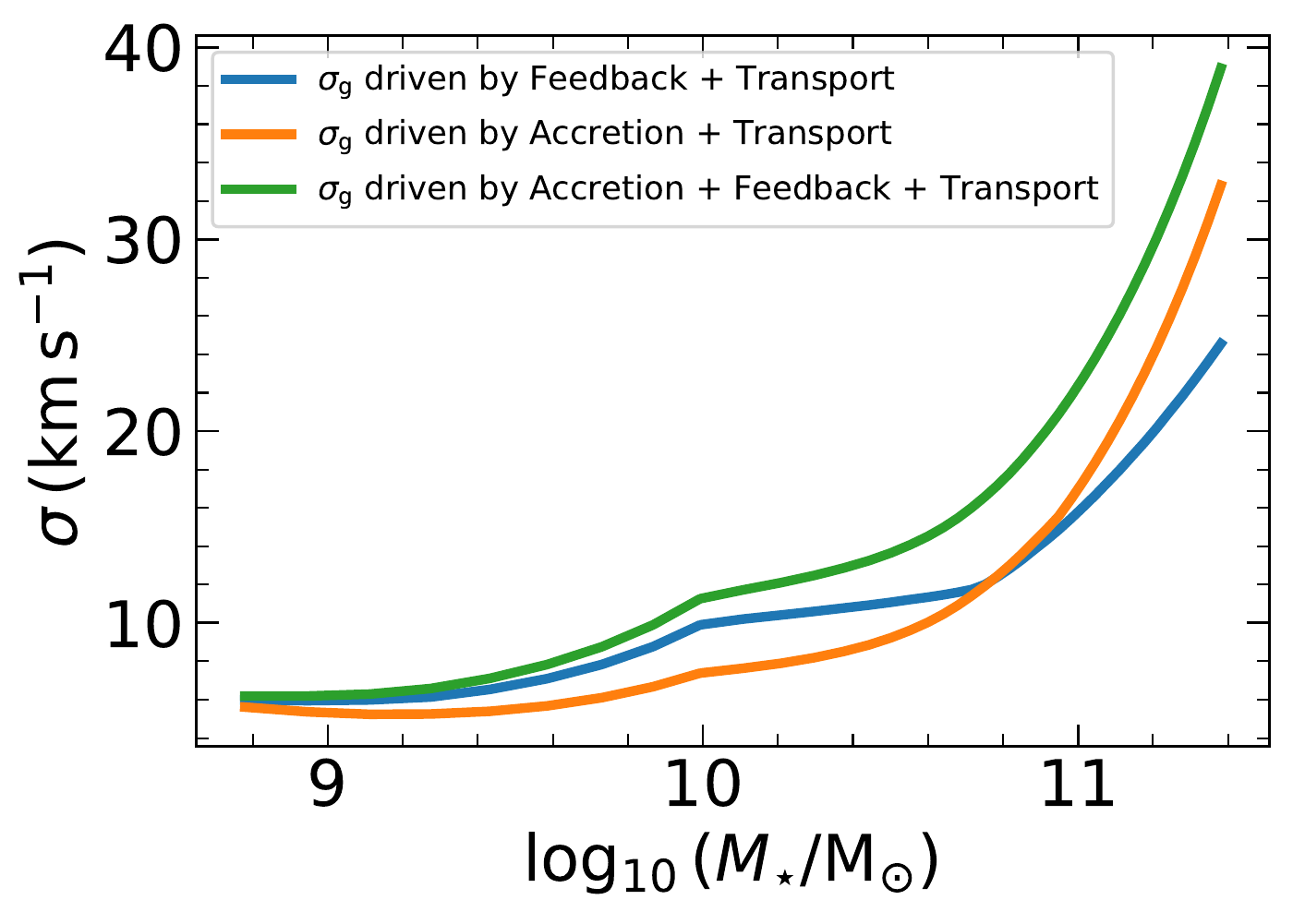}
\caption{Gas velocity dispersion, $\sigma_{\rm{g}}$, as a function of stellar mass, at $z = 0$. The three curves corresponds to models where turbulence is driven by either feedback due to star formation, gas accretion, gas transport, or a combination of the three.}
\label{fig:sigmag}
\end{figure}

\autoref{fig:sigmag} shows the resulting $\sigma_{\rm{g}}$ for the three cases where gas turbulence is driven by feedback + transport (blue), accretion + transport (orange), and feedback + accretion + transport (green). We find that, in energy balance, accretion-driven turbulence (with efficiency $\xi_{\rm{a}}$ set to 0.2) plays a minor role in setting $\sigma_{\rm{g}}$ in low-mass galaxies, but it is the dominant contributor to turbulence in massive galaxies. The high $\sigma_{\rm{acc}}$ at high $M_{\star}$ is partially caused by the overestimation of $\dot M_{\rm{g,acc}}$ in massive haloes as we mention in \autoref{s:model_gasmass}. Nevertheless, the value of $\sigma_{\rm{g}}$ at the high-mass end is in good agreement with that observed in local galaxies \citep{2008MNRAS.390..466E,2015MNRAS.449.3568M,2020MNRAS.495.2265V}. 

Conversely, feedback plays a major role in driving turbulence in low-mass galaxies but becomes sub-dominant compared to accretion in massive galaxies. This is seemingly contrary to the findings of both \citetalias{2018MNRAS.477.2716K} and \cite{2022MNRAS.TMP.1282G} who find that mass transport is the dominant driver of turbulence in massive local galaxies. However, \citetalias{2018MNRAS.477.2716K} did not consider accretion as a possible source of turbulence that would reduce the amount of transport needed to maintain $\sigma_{\rm{g}}$, and \cite{2022MNRAS.TMP.1282G} did not consider the GMC regime of star formation, which compared to their assumption that all galaxies are in the Toomre regime yields a higher $\dot M_{\rm{sf}}$ and hence reduces the transport rate $\dot M_{\rm{trans}}$ required to maintain energy equilibrium (cf. \autoref{eq:dotMg}). Mass transport is thus very sensitive to model parameters that govern $\dot M_{\rm{sf}}$ and $\eta_{\rm{w}}$ at $z = 0$. As we will discuss later, this has important consequences for the metallicity distribution and gradient in local galaxies.

\subsubsection{Other potential drivers of turbulence}
We pause here to briefly mention other potential sources of turbulence in the disc that we have not taken into account in this work.

Feedback from supermassive black holes is another potential source of turbulent energy injection in galaxies \citep[e.g.,][]{2004MNRAS.347...29C,2014Natur.515...85Z,2020ApJ...889L...1L}. For instance, both AGN-driven winds \citep{2013ApJ...763L..18W} and radio jets \citep{2018MNRAS.479.5544M,2019MNRAS.484.3393Z,2021A&A...654A...8N} can enhance turbulence throughout the galactic disc on kpc scales. We also ignore turbulence injected by magnetorotational instabilities (MRI) which could be important in the outskirts of galactic discs \citep{1999ApJ...511..660S,2005ApJ...629..849P}, since the current evidence remains rather inconclusive \citep{2009AJ....137.4424T,2019ApJ...871...17U,2020A&A...641A..70B}.

Another class of potential turbulence drivers are stellar components of the galaxy that can transfer energy to the gas, for example, spiral arms and bars. While these can drive turbulence in the gas \citep{2007ApJ...660.1232K,2016A&A...589A..66H,2018A&A...609A..60K,2022arXiv220914159O}, simulations spanning a variety of redshifts and halo masses find that they remain sub-dominant compared to feedback, transport, and accretion \citep{2007ApJ...670..237B,2009ApJ...694L.158B,2010MNRAS.404.2151C,2015ApJ...814..131G,2016ApJ...827...28G}. Nevertheless, the impact of these drivers of turbulence on the gas-phase metallicity remains largely unexplored \citep[e.g.,][]{2020MNRAS.tmp.2303Z,2023MNRAS.519.4801C}. 

\begin{table}
\centering
\caption{List of parameters in the ISM metallicity model.}
\begin{tabular}{|l|l|c}
\hline
Parameter & Description & Reference\\
\hline
$s_{\rm{g}}$ & Radial profile of gas surface density & \autoref{eq:ode_Z}\\
$k$ & Radial profile of metal diffusion & \autoref{eq:ode_Z}\\
$\dot s_{\star}$ & Radial profile of star formation rate & \autoref{eq:ode_Z}\\
& surface density & \\
$\dot c_{\star}$ & Radial profile of cosmic accretion & \autoref{eq:ode_Z}\\
& surface density & \\
$x_{\rm{b}}$ & Radius where star formation in the Toomre & \autoref{eq:xb}\\
& regime equals that in the GMC regime &  \\
$\mathcal{T}$ & Orbital to metal diffusion timescale & \autoref{eq:T} \\
$\mathcal{P}$ & Ratio of metal advection to metal diffusion & \autoref{eq:P} \\
$\mathcal{A}$ & Ratio of gas accretion to metal diffusion & \autoref{eq:P} \\
$\mathcal{S}$ & Ratio of star formation to metal diffusion & \autoref{eq:S} \\
$\phi_{\rm{y}}$ & Preferential enrichment of galactic winds & \autoref{eq:phiy}\\
$\mathcal{Z}_{\rm{w}}$ & Metallicity of galactic winds & \autoref{eq:phiy}\\
$\xi_{\rm{w}}$ & Fraction of newly produced metals ejected  & \autoref{eq:Zw}\\
& with the wind &\\
$t_{\rm{eqbm}}$ & Metallicity equilibration timescale & \autoref{eq:teqbm}\\
$t_{\rm{fluc}}$ & Metallicity fluctuation timescale & \autoref{s:model_gasphasemetal_teqbm}\\
$\mathcal{Z}$ & ISM metallicity normalized to solar & \autoref{eq:Z_Toomre}\\
$\mathcal{Z}_{r_0}$ & Central ISM metallicity normalized to solar & \autoref{eq:Z_Toomre} \\
$c_1$ & Constant of integration in the & \autoref{eq:Z_Toomre}\\
& metallicity equation & \\
\hline
\end{tabular}
\label{tab:tab2}
\end{table}

\subsection{Evolution of gas-phase metallicity}
\label{s:model_gasphasemetal}
The metallicity model of \cite{2021MNRAS.502.5935S} is a standalone model that uses a galaxy evolution model as an input to predict the metallicity profile in a wide variety of galaxies. The model includes several key physical processes -- gas accretion, radial gas flows, metal advection and diffusion, star formation, metal consumption in long-lived stars, and outflows. In its primitive form, the evolution of spatially-resolved gas-phase metallicity in this model is given by \citep[equation 14]{2021MNRAS.502.5935S}
\begin{equation}
\underbrace{\mathcal{T} s_{\rm{g}} \frac{\partial \mathcal{Z}}{\partial \tau}}_{\substack{\text{Equilibration} \\ \text{time}}} - \underbrace{\frac{\mathcal{P}}{x} \frac{\partial \mathcal{Z}}{\partial x}}_{\text{Advection}} - \underbrace{\frac{1}{x}\frac{\partial}{\partial x}\left(x k s_{\rm{g}} \frac{\partial \mathcal{Z}}{\partial x}\right)}_{\text{Diffusion}}\,\,
= \underbrace{\mathcal{S} \dot{s}_\star}_{\substack{\text{Star formation} \\ + \\ \text{Outflows}}} - \underbrace{\mathcal{Z}\mathcal{A}\dot c_{\star}}_{\text{Accretion}}.
\label{eq:ode_Z}
\end{equation}
where $x$ and $\tau$ are the dimensionless galactocentric distance and time variables, respectively. $x = r/r_0$, where $r_0 = 1\,\rm{kpc}$ is the reference radius that we assume to be the inner edge of the disc. Similarly, $\tau = t \,\Omega_0$, where $\Omega_0$ is the angular velocity at $r_0$. $\mathcal{Z}$ is the gas-phase metallicity $Z$ normalised to $\rm{Z_{\odot}}$. The parameters $s_{\rm{g}} (x), k(x), \dot s_{\star}(x)$, and $\dot c_{\star}(x)$ respectively represent the functional forms of spatial variations in the gas surface density $\Sigma_{\rm{g}}$, epicyclic frequency $\kappa$, star formation rate surface density $\dot \Sigma_{\star}$, and gas accretion rate surface density $\dot \Sigma_{\rm{g,acc}}$.

We pause here to mention a key difference between our approach and that of \cite{2021MNRAS.502.5935S}. These authors find $s_{\rm{g}}(x) = 1/x$, $k(x) = x$ by assuming that $v_{\phi}$ is the same throughout the galactic disc. However, this is only valid as long as $\beta \sim 0$. We further generalize their results by re-writing $v_{\phi} = v_{\phi,\rm{r}} (r/R)^{-\beta}$ where $v_{\phi,\rm{r}}$ is the rotational velocity at a distance $r$ from the galaxy center, and $R$ is the disc size we define in \autoref{eq:R}. The advantage of re-writing the rotational velocity in this way is that we retain the meaning of $v_{\phi}$ as that defined at the outer edge of the disc. With this generalization, we obtain
\begin{subequations}
\begin{equation}
s_{\rm{g}}(x) = \frac{x^{\beta}}{x}\,,
\end{equation}
\begin{equation}
k (x) = \frac{x}{x^{\beta}}\,.
\end{equation}
\label{eq:littlex}
\end{subequations}
It is expected that $\dot c_{\star}$ declines with $x$ \citep{2001ApJ...554.1044C,2009ApJ...696..668F,2010AIPC.1240..131C,2014MNRAS.443..168F,2016MNRAS.455.2308P,2016MNRAS.462.1329M,2017MNRAS.467.2066S}. We retain $\dot c_{\star}(x)=1/x^2$ from \cite{2021MNRAS.502.5935S}. While the motivation for such a radial variation in $\dot \Sigma_{\rm{acc}}$ is that it can reproduce the present-day stellar surface density of the Milky Way \citep{2008A&A...483..401C}, there is no reason why all galaxies should follow this profile. Nevertheless, \cite[appendix A]{2021MNRAS.502.5935S} show that a different functional form does not qualitatively change the resulting metallicity gradients in local spirals, and does not matter at all for local dwarfs, and we therefore adopt a single functional form for simplicity. 

Since we include both the Toomre and the GMC regime of star formation, we update the definition of $\dot s_{\star}$ such that it is a continuous function of $x$
\begin{subequations}
\begin{empheq}[left={\dot s_{\star}(x)=}\empheqlbrace]{align}
 x^{2\left(\beta-1\right)};\,\, x \leq x_{\rm{b}} \\
  \left(x_{\rm{b}}x\right)^{\beta-1};\,\, x > x_{\rm{b}}\,,
\end{empheq}
\label{eq:dotsstar}
\end{subequations}
where $x_{\rm{b}}$ is the critical location in the disc where $\dot \Sigma_{\star}$ in the GMC regime equals that in the Toomre regime \citep[equation 32]{2018MNRAS.477.2716K}
\begin{equation}
    x_{\rm{b}} = \left[4\frac{\sqrt{2(1+\beta)} f_{\rm{g,Q}} \epsilon_{\rm{ff}} v_{\phi} t_{\rm{SF,max}}}{\pi Q \sqrt{3 f_{\rm{g,P}} \phi_{\rm{mp}}}} \left(\frac{r_0}{R}\right)^{\beta}\frac{1}{r_0}\right]^{\frac{1}{1-\beta}}\,.
\label{eq:xb}
\end{equation}
Note that the star formation rate profile is less steep in the GMC regime, which will lead to more metal production in the outer regions of the galaxy as compared to the Toomre regime. As we will discuss later in \autoref{s:mzgr}, this feature is partially responsible for giving rise to steady-state inverted metallicity gradients in the model.

The four dimensionless ratios that connect the metallicity model to the evolution of gas and stars are: (1). $\mathcal{T} -$ the ratio of the orbital and metal diffusion timescales, (2). $\mathcal{P} -$ the ratio of metal advection to metal diffusion, well known as the Péclet Number in fluid dynamics \citep{patankar1980numerical}, (3). $\mathcal{S} -$ the ratio of metal production (star formation) to metal diffusion, and (4). $\mathcal{A} -$ the ratio of gas accretion to metal diffusion. As in \autoref{s:model_gasmass}, we have not taken into account the effect of galactic fountains in returning metals to the disc. As in \autoref{eq:littlex} and \autoref{eq:dotsstar}, we also generalize the expressions for $\mathcal{T}, \mathcal{P}, \mathcal{S}$ and $\mathcal{A}$ from \cite{2021MNRAS.502.5935S}
\begin{equation}
\mathcal{T} = \frac{3\sqrt{2\left(\beta+1\right)}\phi_{\rm{Q}}f_{\rm{g,Q}}}{Q}\left[\left(\frac{v_{\phi}}{\sigma_{\rm{g}}}\right)\left(\frac{r_0}{R}\right)^{\beta}\right]^2\,,
\label{eq:T}
\end{equation}

\begin{equation}
\mathcal{P} = \frac{6\eta\phi^2_{\rm{Q}}\phi^{3/2}_{\rm{nt}} f^2_{\rm{g,Q}}}{Q^2_{\rm{min}}}\left(\frac{1+\beta}{1-\beta}\right) F(\sigma_{\rm{g}})\,,
\label{eq:P}
\end{equation}

\begin{equation}
\mathcal{S} = \frac{24 \phi_Q  f^2_{\rm{g,Q}} \epsilon_{\rm{ff}} f_{\rm{sf}}}{\pi Q^2 \sqrt{3f_{\rm{g,P}} \phi_{\rm{mp}}}}\left(\frac{\phi_{\rm{y}} y}{Z_{\odot}}\right)\left(1+\beta\right)\left[\left(\frac{v_{\phi}}{\sigma_{\rm{g}}}\right)\left(\frac{r_0}{R}\right)^{\beta}\right]^2\,,
\label{eq:S}
\end{equation}

\begin{equation}
\mathcal{A} = \frac{3G\phi_{\rm{Q}} \dot M_{\rm{g,acc}}}{2\sigma^3_{\rm{g}} \ln x_{\rm{max}}}\,,
\label{eq:A}
\end{equation}
where $x_{\rm{max}} = R/r_0$ is the disc size normalized by $r_0$, and $\phi_{\rm{y}}$ is a fractional quantity that describes the preferential metal enrichment of galactic winds.\footnote{Note that \cite{2021MNRAS.502.5935S} missed a factor $1/Q$ in their definition of $\mathcal{S}$ (equation 39 in their paper). This error has no impact on their results since they assumed $Q = Q_{\rm{min}} = 1$.} Most galactic chemical evolution models to date assume that the metallicity of galactic winds is the same as that of the ISM. However, this is not necessarily the case for all galaxies. $\phi_{\rm{y}}$ accounts for the fact that the wind metallicity can be higher than the ISM metallicity in a galaxy if the ejecta from supernovae are imperfectly mixed with the ISM \citep{1986ApJ...303...39D,1999ApJ...513..142M,2008A&A...489..555R,2011MNRAS.417.2962P,2013MNRAS.434.1531F,2018ApJ...869...94E,2019MNRAS.482.1304E,2020MNRAS.496.4433T,2021MNRAS.502.5935S,2021MNRAS.503.4474Y,2022arXiv220906218A}, as has now been observed in galactic winds of low mass galaxies \citep{2018MNRAS.481.1690C,2020ApJ...904..152L,2021ApJ...918L..16C,2022arXiv220409181X,2022A&A...657A..19T}. Following \citet[equation A1]{2020aMNRAS.xxx..xxxS}, we write
\begin{equation}
\phi_{\rm{y}} = 1 - \eta_{\rm{w}}\mathcal{Z}\left(\frac{\rm{Z_{\odot}}}{y}\right)\left(\frac{\mathcal{Z}_{\rm{w}}}{\mathcal{Z}}-1\right)\,,
\label{eq:phiy}
\end{equation}
where $y = 0.028$ is the yield of newly formed Type II metals \citep{2019MNRAS.487.3581F} following the instantaneous recycling approximation \citep{1980FCPh....5..287T} and $\mathcal{Z}_{\rm{w}}$ is the metallicity of galactic winds normalized to $\rm{Z_{\odot}}$. It is given by \citep[equation 41]{2019MNRAS.487.3581F}
\begin{equation}
    \mathcal{Z}_{\rm{w}} = \mathcal{Z} + \left(\frac{y}{\rm{Z_{\odot}}}\right)\frac{\xi_{\rm{w}}}{\rm{max}\,\left(\eta_{\rm{w}}, 1 - f_{\rm{R,inst}}\right)}\,,
\label{eq:Zw}
\end{equation}
where $f_{\rm{R,inst}} = 0.77$ is the fraction of metals locked in low mass stars \citep{1980FCPh....5..287T}, and $\xi_{\rm{w}}$ is a fractional parameter bounded between 0 and 1 that describes the preferential enrichment of galactic winds if some of the newly produced metals are ejected with winds before they mix with the ISM. Thus, \autoref{eq:Zw} implies that common assumption of wind metallicity being equal to the ISM metallicity is likely only a lower bound in reality. Further, since a fraction $f_{\rm{R,inst}}$ of metals are locked in stars, the minimum metal mass ejected is $1 - f_{\rm{R,inst}}$.

We see from \autoref{eq:phiy} that $\phi_{\rm{y}} \sim 0$ corresponds to all newly produced metals being entrained in the winds, whereas if $\phi_{\rm{y}} \sim 1$, newly produced metals are perfectly mixed with the ISM before they are ejected such that the wind metallicity equals the ISM metallicity. In the absence of any scaling relations of $\phi_{\rm{y}}$ with galaxy properties, we explore $\phi_{\rm{y}}$ in the range $0.1-1$, and show how the resulting metallicity gradients are quite sensitive to the choice of $\phi_{\rm{y}}$. In principle, we can specify $\phi_{\rm{y}}$ as a spatially-varying parameter. However, we treat it as an integrated quantity for this work since variations in $\phi_{\rm{y}}$ with $x$ remain unexplored. As we will discuss below, $\mathcal{P}, \mathcal{S}$ and $\mathcal{A}$ form the cornerstone of our metallicity analysis and act as useful diagnostics to understand what sets metallicity gradients in galaxies.

Note that we only consider elements produced by core collapse supernovae in this work, since the observable we study is the oxygen abundance gradient in the ISM. Therefore, we cannot comment on the evolution of $\phi_{\rm{y}}$ for elements produced by other nucleosynthetic sources (e.g., AGB stars or Type Ia supernovae). In fact, simulations of low mass galaxies that explicitly resolve feedback find metal enrichment of galactic winds to be dependent on the nucleosynthetic source \citep{2018ApJ...869...94E}. Similarly, metal mixing in the ISM is also sensitive to the nucleosynthetic source \citep{2018MNRAS.475.2236K}. 

\subsubsection{Metal equilibration timescale}
\label{s:model_gasphasemetal_teqbm}
Because we are interested in searching for steady-state solutions to \autoref{eq:ode_Z}, we first estimate the time it takes for a given metal distribution to reach equilibrium, $t_{\rm{eqbm}}$. Following \citet[equation 19]{2021MNRAS.502.5935S}, $t_{\rm{eqbm}}$ is given by
\begin{equation}
\frac{1}{t_{\mathrm{eqbm}}} = \Omega_0 \frac{\big\lvert \frac{\mathcal{P}}{x}\frac{\partial \mathcal{Z}}{\partial x}\big\rvert + \big\lvert\frac{1}{x}\frac{\partial}{\partial x}\left(x k s_g\frac{\partial \mathcal{Z}}{\partial x}\right)\big\rvert + \big\lvert\mathcal{S}\dot s_{\star}\big\rvert + \big\lvert\mathcal{Z}\mathcal{A} \dot c_{\star}\big\rvert}{\mathcal{Z}s_g\mathcal{T}}\,.
\label{eq:teqbm}    
\end{equation}
The physical interpretation of the above equation is that we compare the time taken by gas accretion, star formation, outflows, metal advection and metal diffusion to induce changes in metallicity with the rate at which metallicity evolves with time as given by the leading term in \autoref{eq:ode_Z}. Under equilibrium, $\partial \mathcal{Z} / \partial \tau \to 0$. This is a common practice in chemical evolution models as it allows us to estimate if metallicity reaches equilibrium simultaneously with other galaxy properties \citep{2012MNRAS.421...98D,2013ApJ...772..119L,2015MNRAS.449.3274F,2018MNRAS.475.2236K}.

If the metal distribution does not reach equilibrium within a reasonable time, we consider the resulting gradients to be in non-equilibrium, in which case we cannot apply our model to study gradients. A necessary condition is that the metal equilibration timescale be less than the Hubble time. However, this is not sufficient since physical processes relevant for the model may change on timescales much shorter than the Hubble time. Out of all the physical processes we include, the timescale for star formation, quantified by the molecular gas depletion timescale, $t_{\rm{dep}}$, is usually observed to be the shortest \citep{2015ApJ...800...20G,2017ApJ...837..150S,2017ApJ...849...26U,2018ApJ...853..179T,2019ApJ...887..235L}. In nearby galaxies, observations find $t_{\rm{dep}} \approx 2\,\rm{Gyr}$ \citep{2008AJ....136.2782L,2011ApJ...730L..13B,2011MNRAS.415...61S,2023arXiv230212267S}. Keeping this in mind, we consider the metallicity to be in non-equilibrium if $t_{\rm{eqbm}} \gg t_{\rm{dep}}$. To ensure that local fluctuations in metallicity are not responsible for setting the metallicity gradient (that would otherwise generate a random metallicity gradient set by a stochastic metal field), we investigate if $t_{\rm{eqbm}}$ is longer than the timescale for such fluctuations to become steady. Typically, the latter is of the order of $t_{\rm{fluc}} \approx 300\,\rm{Myr}$ in local galaxies \citep[fig. 7]{2018MNRAS.475.2236K}, and we check if $t_{\rm{eqbm}} \gtrsim t_{\rm{fluc}}$. Given the self-consistent evolution of gas parameters with metals, the model we present here allows for inverted gradients to form in equilibrium, in contrast to the \cite{2021MNRAS.502.5935S} model.

\begin{table*}
\centering
\caption{Summary of differences in the models of \protect\citetalias{2018MNRAS.477.2716K},  \protect\cite{2021MNRAS.502.5935S}, \protect\cite{2022MNRAS.TMP.1282G}, and this work. Parameter definitions are available in \autoref{tab:tab1} and \autoref{tab:tab2}.}
\begin{tabular}{|l|c|c|c|c}
\hline
Parameter & \citetalias{2018MNRAS.477.2716K} & \cite{2021MNRAS.502.5935S} & \cite{2022MNRAS.TMP.1282G} & This work\\
\hline
Evolution of $\dot M_{\rm{g}}$ & \XSolid & \XSolid & \Checkmark & \Checkmark \\
Self-consistent $\dot M_{\rm{g}}$ and $\sigma_{\rm{g}}$ & \XSolid & \XSolid & \Checkmark & \Checkmark \\
Radial resolution & \XSolid & \Checkmark & \XSolid & \Checkmark \\
Include GMC regime & \Checkmark & \XSolid & \XSolid & \Checkmark \\
Include $\xi_{\rm{a}}$ & \XSolid & \XSolid & \Checkmark & \Checkmark \\
Include $\sigma_{\rm{acc}}$ & \XSolid & \XSolid & \Checkmark & \Checkmark \\
Include $\eta_{\rm{w}}$ & \XSolid & \Checkmark & \XSolid & \Checkmark \\
Include $\mathcal{Z}$ & \XSolid & \Checkmark & \XSolid & \Checkmark \\
Include $\phi_{\rm{y}}$ & \XSolid & \Checkmark & \XSolid & \Checkmark \\
Include $\phi_{\rm{a}}$ & \Checkmark &  \XSolid & \XSolid & \Checkmark \\
Allow $Q \geq Q_{\rm{min}}$ & \Checkmark & \XSolid & \XSolid & \Checkmark \\
Allow $\beta \neq 0$ & \Checkmark & \Checkmark & \XSolid & \Checkmark \\
Allow radial variations in $v_{\phi}$ & \Checkmark & \XSolid & \XSolid & \Checkmark \\
Allow $\phi_{\rm{nt}} \leq 1 $ & \Checkmark & \XSolid & \XSolid & \Checkmark \\
Allow $f_{\rm{sf}} \leq 1 $ & \Checkmark & \Checkmark & \XSolid & \Checkmark \\
\hline
\end{tabular}
\label{tab:diffgal}
\end{table*}

\begin{table}
\centering
\caption{List of parameter values for metallicity gradients in two representative galaxies as shown in \autoref{fig:newgrad}.}
\begin{tabular}{|l|c|c}
\hline
Parameter & Massive galaxy & Low-mass galaxy \\
\hline
$M_{\star}$ & $10^{10.8}\,\rm{M_{\odot}}$ & $10^{9}\,\rm{M_{\odot}}$\\
$x_{\rm{max}}$ & $15$ & $5.5$\\
$f_{\rm{g,Q}}$ & 0.5 & 0.9\\
$\epsilon_{\rm{in}}$ & $0.24$ & $0.46$\\
$c$ & 10 & 15 \\
$\beta$ & 0 & 0.5\\
$f_{\mathrm{sf}}$ & 0.6 & 0.4\\
$f_{\rm{g,P}}$ & 0.5 & 0.9\\
\hline
\end{tabular}
\label{tab:tab3}
\end{table}

\begin{figure}
\includegraphics[width=1.0\columnwidth]{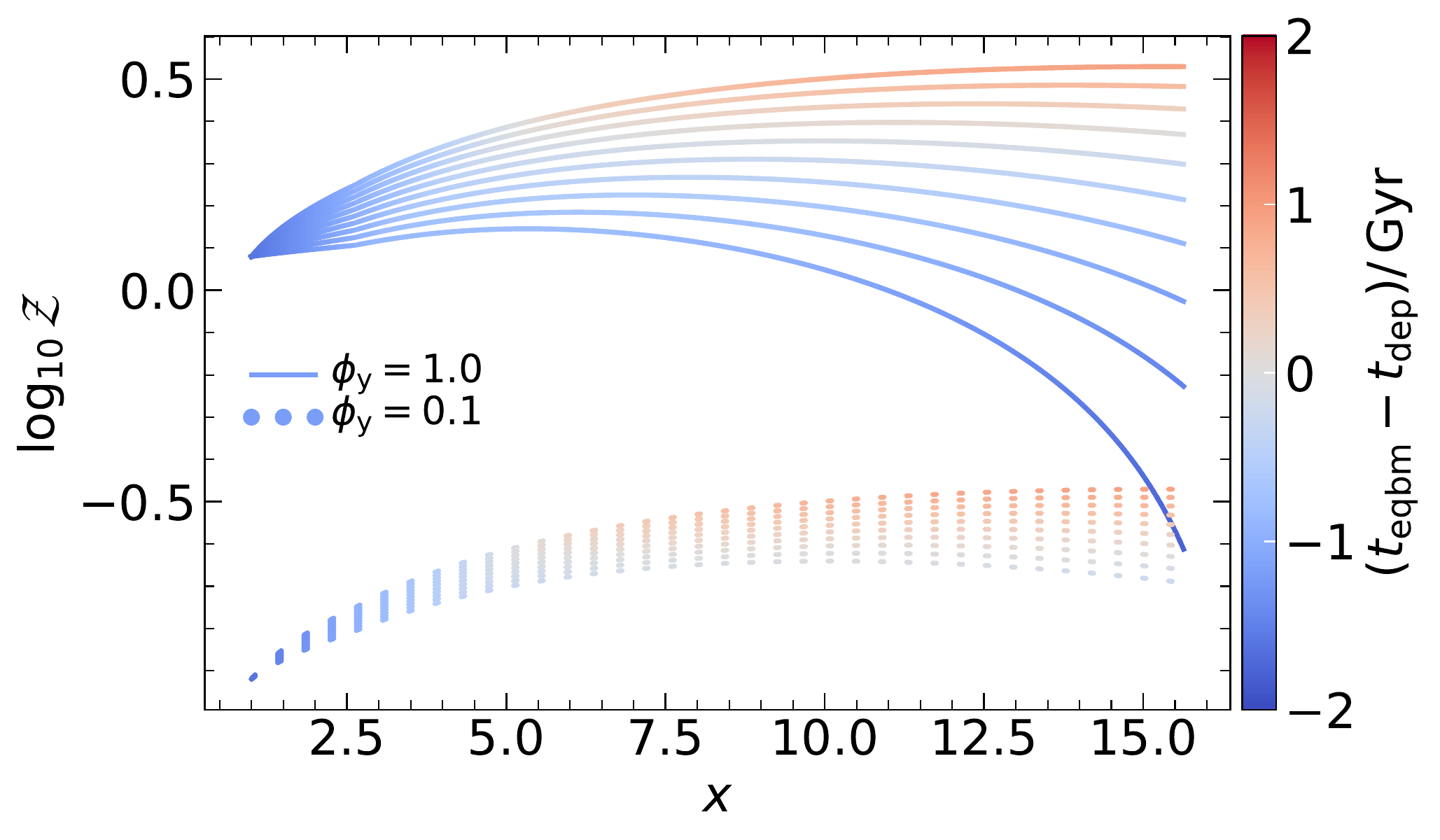}
\includegraphics[width=1.0\columnwidth]{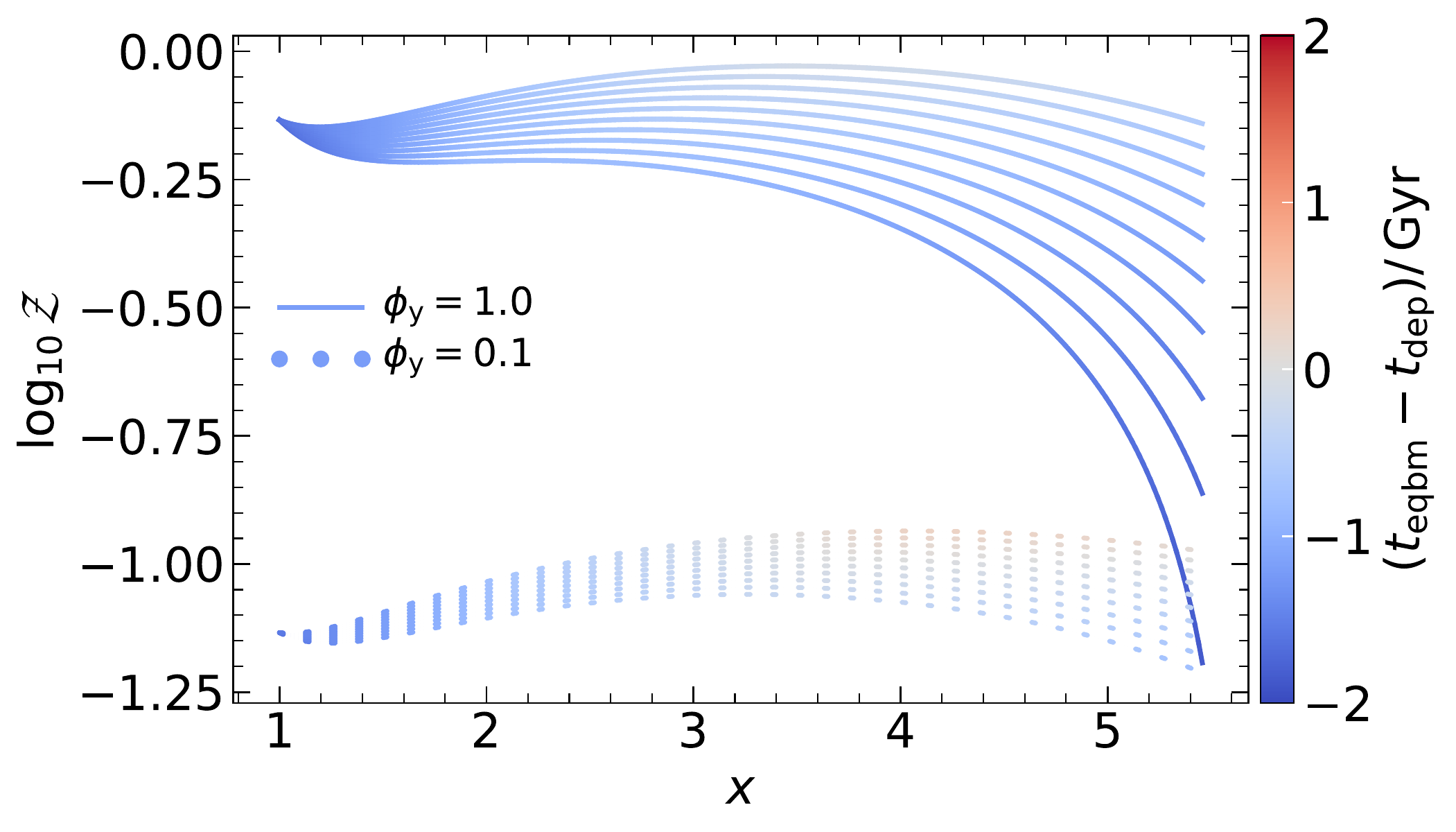}
\caption{\textit{Top panel:} ISM metallicity profiles ($\mathcal{Z} = Z/Z_{\odot}$) as a function of the dimensionless radius ($x = r/r_0$ with $r_0 = 1\,\rm{kpc}$) for the fiducial case of a massive galaxy ($M_{\star} = 10^{10.8}\,\rm{M_{\odot}}$) with $\phi_{\rm{y}} = 1.0$ (solid) and $0.1$(dotted), respectively. $\phi_{\rm{y}}$ describes the preferential metal enrichment of galactic winds. The family of curves for each $\phi_{\rm{y}}$ arise from constraints on the constant of integration $c_1$ in the solution for metallicity. Colorbar represents the difference between the metal equilibration timescale, $t_{\rm{eqbm}}$, and the gas depletion timescale, $t_{\rm{dep}}$, with increasingly red color representing larger deviations from equilibrium. \textit{Bottom panel:} Same as the top panel but for a low-mass galaxy with $M_{\star} = 10^{9}\,\rm{M_{\odot}}$.}
\label{fig:newgrad}
\end{figure}

\subsubsection{Equilibrium solutions}
\label{s:model_gasphasemetal_eqbmsol}
\citet{2021MNRAS.502.5935S} show that most galaxies tend to achieve a steady-state metallicity gradient within a timescale much shorter than the Hubble time, and comparable or shorter than $t_{\rm{dep}}$. When this condition is satisfied, we can search for equilibrium solutions for \autoref{eq:ode_Z} by setting $\partial \mathcal{Z} / \partial \tau \to 0$. 

Since the functional form of the star formation term, $\dot s_{\star}(x)$, is different in the Toomre and the GMC regimes of star formation, the resulting solution for $\mathcal{Z}(x)$ is also different. $\mathcal{Z}(x)$ in the Toomre regime is given by
\begin{eqnarray}
    \lefteqn{\mathcal{Z}(x) = \frac{\mathcal{S}x^{2\beta}}{\mathcal{A}-2\beta\left(\mathcal{P}+2\beta\right)} + c_1 x^{\frac{1}{2}\left[\sqrt{\mathcal{P}^2+\,4\mathcal{A}}-\mathcal{P}\right]}
    }
    \nonumber \\
    & {}  + \left(\mathcal{Z}_{r_0} - \frac{\mathcal{S}}{\mathcal{A}-2\beta\left(\mathcal{P}+2\beta\right)} - c_1\right) x^{\frac{1}{2}\left[-\sqrt{\mathcal{P}^2+\,4\mathcal{A}}-\mathcal{P}\right]}\,,
    \label{eq:Z_Toomre}
\end{eqnarray}
and in the GMC regime is given by
\begin{eqnarray}
    \lefteqn{\mathcal{Z}(x) = \frac{\mathcal{S}x^{1+\beta}}{\mathcal{A}-\left(1+\beta\right)\left(1+\mathcal{P}+\beta\right)} + c_1 x^{\frac{1}{2}\left[\sqrt{\mathcal{P}^2+\,4\mathcal{A}}-\mathcal{P}\right]}
    }
    \nonumber \\
    & {}  + \left(\mathcal{Z}_{r_0} - \frac{\mathcal{S}}{\mathcal{A}-\left(1+\beta\right)\left(1+\mathcal{P}+\beta\right)} - c_1\right) x^{\frac{1}{2}\left[-\sqrt{\mathcal{P}^2+\,4\mathcal{A}}-\mathcal{P}\right]}\,.
    \label{eq:Z_GMC}
\end{eqnarray}
Here, $\mathcal{Z}_{r_0}$ is the metallicity at $r_0$, and $c_1$ is a constant of integration. Physically, $c_1$ reflects the strength of the metallicity gradient at $r_0$.  Galaxies tend to achieve a particular value of $\mathcal{Z}_{r_0}$ that is governed by the competition between terms that dominate at $r_0$. For massive local galaxies, $\mathcal{Z}_{r_0}$ is set by the competition between source (star formation and outflows) and accretion. For low-mass galaxies, $\mathcal{Z}_{r_0}$ is set by the interplay between advection and diffusion if $\dot M_{\rm{trans}} > 0$. In the case where there is no mass transport ($\dot M_{\rm{trans}} = 0$), $\mathcal{Z}_{r_0}$ is set by source and diffusion. We provide the equations for $\mathcal{Z}_{r_0}$ in terms of $c_1$ in \aref{s:app_Zr0}. Thus, $c_1$ is the only unknown parameter in practice. We also see that \autoref{eq:Z_Toomre} reduces to equation 41 of \citet{2021MNRAS.502.5935S} when $\beta \to 0$.

With these solutions, we can develop some intuition for the steepness of the resulting metallicity profiles. If $\mathcal{S} > \mathcal{P} + \mathcal{A}$, the profiles are steeper, thus giving strong negative metallicity gradients. On the contrary, if $\mathcal{P} + \mathcal{A} > \mathcal{S}$, the metal distribution is quite homogeneous, giving flat metallicity gradients. In the extreme case where $\mathcal{P} + \mathcal{A} \gg \mathcal{S}$, this can even lead to an inversion in the metallicity profile where the inner regions of the galaxy are more metal-poor as compared to the outskirts, thereby driving an inverted metallicity gradient. The inversion is more likely to occur for low-mass galaxies because of the leading term that scales as $x^{\beta}$ in both the Toomre and GMC regimes. The strength of the gradient is also modulated by $c_1$, which we constrain using physically meaningful boundary conditions in \autoref{s:model_gasphasemetal_BCs} below.

\subsubsection{Boundary conditions for equilibrium solutions}
\label{s:model_gasphasemetal_BCs}
We follow \cite{2021MNRAS.502.5935S} to define boundary conditions for the equilibrium solutions we obtain in \autoref{s:model_gasphasemetal_eqbmsol}. These boundary conditions constrain $c_1$ to a finite range of values that in turn give rise to a family of metallicity profiles (hence, gradients) for each solution. Note that if we excluded metal diffusion, \autoref{eq:ode_Z} would reduce to first order, which will only have one constant of integration that can be fully specified using $\mathcal{Z}_{r_0}$. So, we would only obtain one profile per solution instead of a family of profiles.

We first require that $\mathcal{Z}(x)$ is greater than some minimum value, $\mathcal{Z}_{\rm{min}}$ at all $x$. This condition provides a lower bound on $c_1$. We also demand that the metal flux crossing into the disc is at most equal to advection of metals from the CGM (with metallicity $\mathcal{Z}_{\rm{CGM}}$). Following \citet[equation 43]{2021MNRAS.502.5935S}, we express this condition as
\begin{equation}
    -\mathcal{P}\mathcal{Z}(x) - \frac{\partial}{\partial x} \mathcal{Z}(x) \geq -\mathcal{P}\mathcal{Z}_{\rm{CGM}}\,.
\label{eq:outerbc}
\end{equation}
This condition ensures that most metals present in the disc belong to the \textit{in-situ} population of metals produced by star formation (which is generally true unless galactic fountains recycle a significant amount of metals), and provides an upper bound on $c_1$. We do not show the resulting equations for $c_1$ here as they consist of several non-linear functions that are not illuminating, but they can be directly obtained by the applying the above constraints.

We set $\mathcal{Z}_{\rm{CGM}} = 0.05$ for low-mass galaxies, $0.2$ for massive galaxies, and create a linear ramp in $\log_{10} M_{\star}$ between these two for intermediate mass galaxies. Such a setup roughly reproduces the metallicity of metal-poor inflows onto the disc seen in simulations \citep{2017MNRAS.468.4170M}. As more measurements of $\mathcal{Z}_{\rm{CGM}}$ become available \citep[e.g.,][]{2017ApJ...837..169P,2019ApJ...886...91K}, it will be possible to refine the prescription we use. 

This completes the formulation of the model. In \autoref{tab:diffgal}, we provide a summary of key differences between our approach and earlier works we make use of.

\section{Equilibrium metallicity gradients}
\label{s:results}
We now present resulting ISM metallicity profiles and gradients from the model for two classes of galaxies. \autoref{tab:tab3} lists the values of model parameters for the two representative galaxies. The top panel of \autoref{fig:newgrad} presents the resulting family of metallicity profiles from the model for a fiducial local massive galaxy with $M_{\star} = 10^{10.8}\,\rm{M_{\odot}}$ at $z = 0$ with $\phi_{\rm{y}} = 1.0$ (solid curves) and $0.1$ (dotted curves). We include all three sources of turbulence (feedback, accretion, and transport) to create these profiles. The family of curves results from the permissible range of values of the constant $c_1$ in the metallicity equation. However, not all the gradients are in equilibrium. The colorbar in \autoref{fig:newgrad} informs on the time taken by the metal distribution to reach equilibrium compared to $t_{\rm{dep}}$; the redder the color, the larger are the deviations from equilibrium. As expected, we see that the metallicity at the inner edge of the disc naturally approaches the value $\mathcal{Z}_{r_0}$ set by the terms that dominate at small radii. The inner regions ($x < 5$) of all the profiles are in the Toomre regime of star formation whereas the outer regions are in the GMC regime, as expected from the \citetalias{2018MNRAS.477.2716K} model. The bottom panel of \autoref{fig:newgrad} presents the family of metallicity profiles for a fiducial local low-mass galaxy with $M_{\star} = 10^{9}\,\rm{M_{\odot}}$. In this case, the Toomre regime exists for $x < 3.5$, beyond which star formation occurs in GMCs. For the lowest mass galaxies we study, the entire star-forming disc is in the GMC regime at $z = 0$. For the massive galaxy, the radial metallicity profiles for high values of $\phi_{\rm{y}}$ are consistent with the average metallicity profile observed in nearby galaxies of similar mass \citep{2017MNRAS.469..151B}. On the other hand, model profiles with low values of $\phi_{\rm{y}}$ in the low-mass galaxy better match the observed profiles for similar mass \citep[see also, ][figures 2 and 5]{2021MNRAS.502.5935S}.

To find a 1D gradient, we fit the logarithmic metallicity profiles with a linear function in $x$ from $1$ to $x_{\rm{max}}$ (see \autoref{tab:tab3}), the slope of which is the metallicity gradient. This is an oversimplification because the profiles are typically more complex given the non-linearity of $\mathcal{Z}(x)$, so a gradient returned by the fit is often a poor representation of the underlying profile. However, we use this approach to mimic observational measurements where the 2D maps are reduced to 1D by stacking metallicity profiles as a function of galactocentric distance, and then fit with a linear function in $\log_{10} \mathcal{Z}$. For the massive galaxy, we find gradients to be in the range $-0.18\,\mathrm{dex}\,r^{-1}_{\rm{e}}$ to $0.08\,\mathrm{dex}\,r^{-1}_{\rm{e}}$, where we take $r_{\rm{e}} \approx R/2$ in the model; adopting $r_{\rm{e}}$ from \cite{2014ApJ...788...28V} (as in \citealt{2021MNRAS.502.5935S}) gives similar results. Similarly, we obtain gradients in the range $-0.23\,\mathrm{dex}\,r^{-1}_{\rm{e}}$ to $0.13\,\mathrm{dex}\,r^{-1}_{\rm{e}}$ for the low-mass galaxy. Typically, profiles given by the maximum possible value of $c_1$ (especially for massive galaxies) are out of equilibrium, and these profiles lead to inverted gradients. While most inverted metallicity gradients are out of equilibrium in the model, we do find some steady-state profiles where the metallicity in the outer regions is larger than that in the inner regions. This differs from our earlier models in \cite{2021MNRAS.502.5935S}, which did not include the GMC regime of star formation and did not find any inverted gradient equilibria.

Given this new finding, it is worth investigating how these inverted gradients form and exist in equilibrium. Let us consider a simpler case of a massive galaxy with no advection, so $\beta = \mathcal{P} = 0$. In such a case, at large radii, accretion and diffusion compete with star formation to set the metallicity. However, star formation in the GMC regime only declines with radius as $1/x$, while accretion and diffusion go as $1/x^2$. So, even though fewer metals are being produced at large radii, the rate at which they are diluted or transported is even smaller. This is why metallicity in the outskirts can be higher, leading to inverted gradients in the model in equilibrium. However, there are two important corollaries. First, this result is sensitive to our assumed radial profile of accretion, $\dot c_{\star} = 1/x^2$ \citep{2008A&A...483..401C}. If we assume that accretion goes as $1/x$ instead, for example, the resulting gradient will likely be close to zero. Second, if $f_{\rm{sf}}$ declines strongly with radius \citep[e.g.,][]{2009ApJ...693..216K,2013MNRAS.436.2747K,2015A&A...580A.127K}, the source term $\mathcal{S}$ would decrease more rapidly than $1/x$, and we would again have no or less inversion. We also do not invoke radial variations in $\phi_{\rm{y}}$ or $\eta_{\rm{w}}$ below \citep[e.g.,][]{2021MNRAS.508.4484J}. Exploring radial variations in these parameters requires direct measurements of both the molecular and atomic gas surface densities, as well as metal enrichment of galactic winds at different radii. This is an area where a multi-wavelength observations, combining JWST IFU instruments, VLT/MUSE and ALMA \citep[e.g.,][]{2021ApJS..257...43L,2022A&A...659A.191E} have the potential to make a big impact on our understanding of inverted metallicity gradients. 

Note that our model is based on a disc flow model, but (major) mergers can significantly impact the disc, and likely reset the metallicity gradients. We showed in \citet{2021MNRAS.502.5935S} that the metallicity distribution falls out of equilibrium during a major merger (a result corroborated by observations -- \citealt{2023arXiv230614843P}), and thus our equilibrium model cannot be applied in such cases. It is for this reason that we do not study merging galaxies (e.g., ULIRGs) in this work. The impact of minor mergers is small at resetting the gas-phase metallicity gradient, especially at $z = 0$, which is the focus of this work. There is an associated question of the impact of \textit{past} major mergers on the \textit{present day} metallicity gradient in the disc, answering which requires using a combination of hydrodynamic simulations and SAMs like ours (Sharda et al., in prep.), and is beyond the scope of this work.

\begin{figure*}
\includegraphics[width=0.85\textwidth]{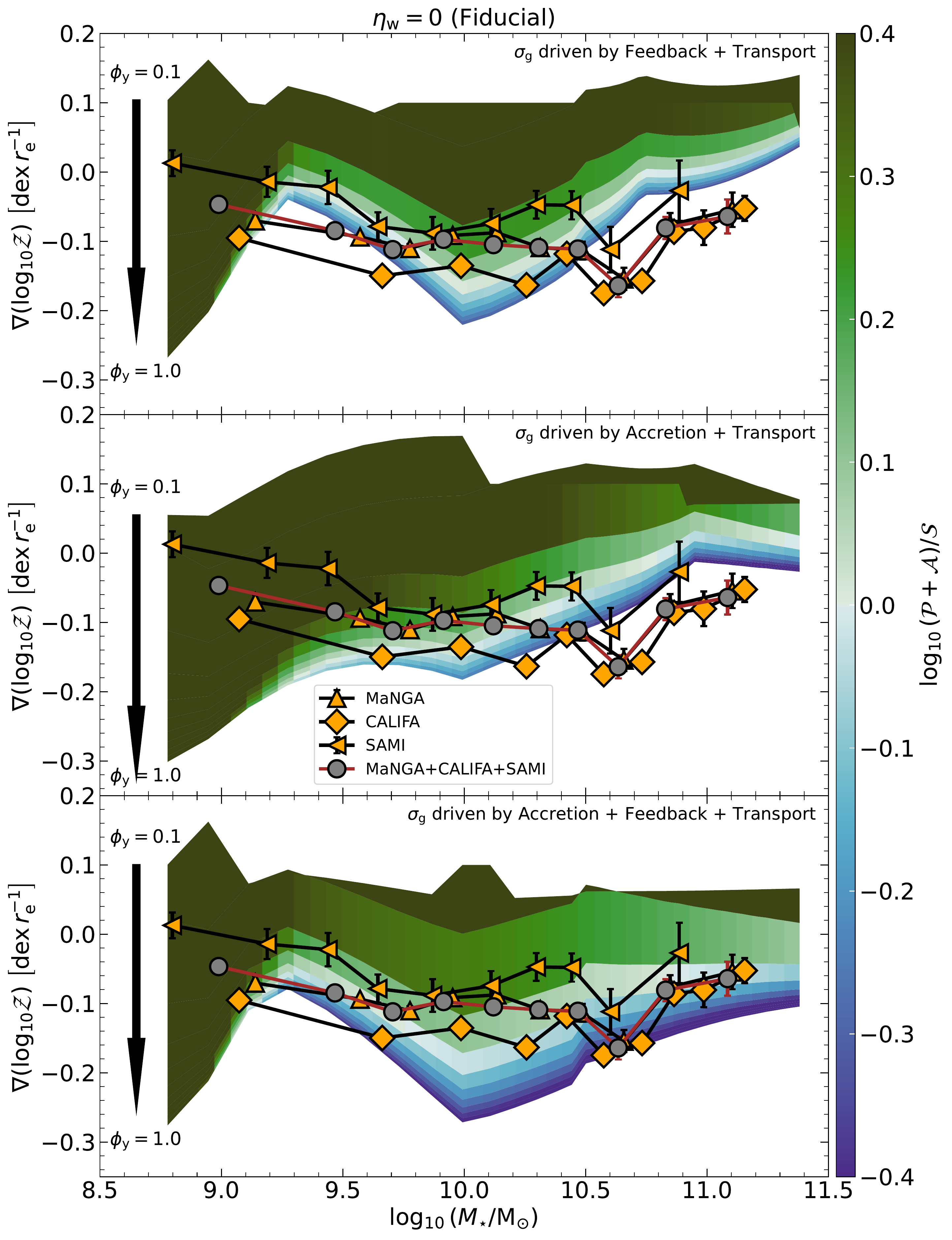}
\caption{The local mass-metallicity gradient relation (MZGR) as predicted from the fiducial model. It plots metallicity gradients ($\nabla \log_{10}\,\mathcal{Z}$) as a function of the stellar mass ($M_{\star}$). The colorbar indicates the ratio of the strength of processes that create flat metal distributions ($\mathcal{P}$ and $\mathcal{A}$) to those that create steep metallicity gradients ($\mathcal{S}$) in equilibrium, as defined in \autoref{s:model}. The top panel plots the model MZGR for the case where only gas accretion and transport drive turbulence in the galaxy. The middle panel corresponds to the case where only star formation feedback and gas transport drive turbulence. The bottom panel corresponds to the case where all three factors are included as drivers of gas turbulence. The scatter in the model at fixed $M_{\star}$ corresponds to the results of varying $\phi_y$, the parameter that describes the preferential ejection of metals via galactic winds, from $0.1 - 1$, as indicated. Overplotted (orange markers) are average metallicity gradients in bins of $M_{\star}$ obtained from IFU surveys -- MaNGA \protect\citep{2017MNRAS.469..151B,2020A&A...636A..42M}, CALIFA \protect\citep{2014A&A...563A..49S,2016A&A...587A..70S}, and SAMI \protect\citep{2018MNRAS.479.5235P}, as well as mean of the three (grey markers).}
\label{fig:mzgr_basic}
\end{figure*}

\begin{figure*}
\includegraphics[width=1.0\textwidth]{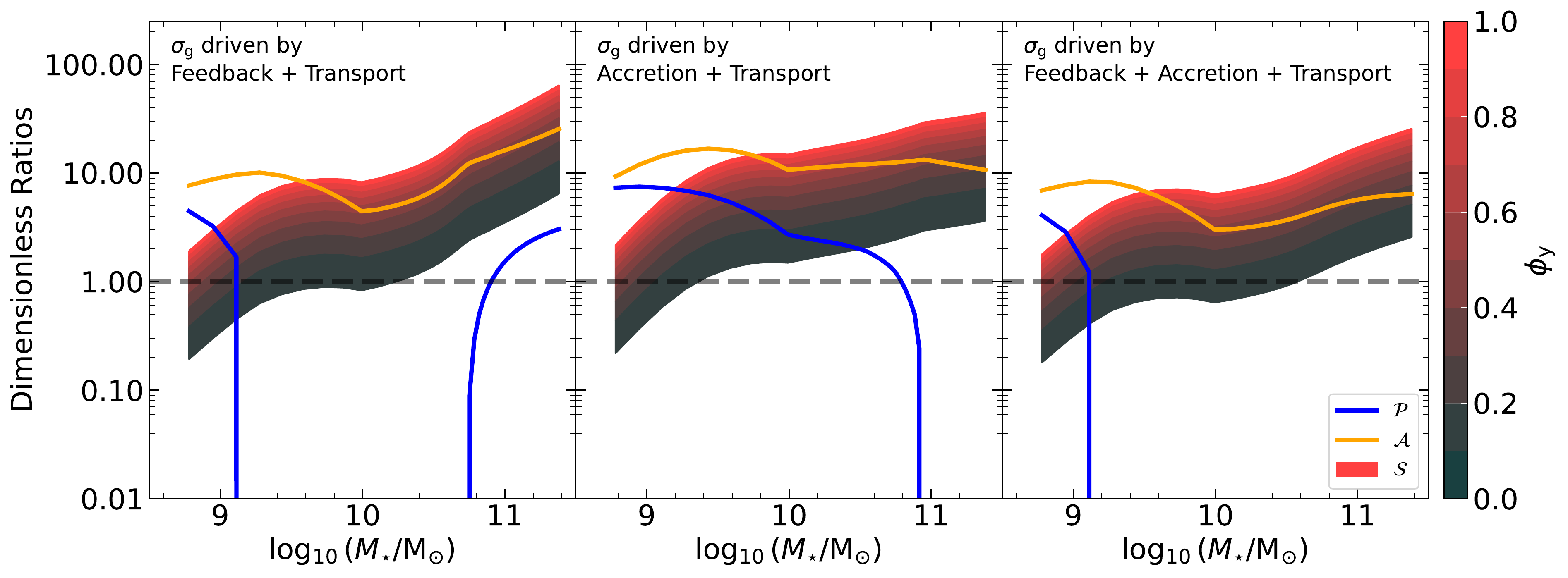}
\caption{\textit{Left panel:} The three dimensionless ratios ($\mathcal{P}, \mathcal{A}$, and $\mathcal{S}$) that set the metal distribution at $z = 0$ in the fiducial model presented in \autoref{s:mzgr}, as a function of stellar mass, for the case where turbulence is driven by star formation feedback and gas transport. The value of $\mathcal{S}$ is directly proportional to the parameter $\phi_{\rm{y}}$ that describes the preferential enrichment of galactic winds (see \autoref{eq:S}), so for this parameter we plot a vertical range corresponding to varying $\phi_\mathrm{y}$ from $0.1-1$, as indicated by the colour bar. Metal transport due to radial gas flows ($\mathcal{P}$) is only active for the lowest and most massive galaxies in this case. The dashed line denotes a value of unity, with values of a parameter less than unity indicating that turbulent diffusion is a more important process than the one described by that parameter. \textit{Middle panel:} Same as the left panel but now turbulence is driven by gas accretion and transport. Metal advection is active for most galaxies in this case. \textit{Right panel:} Same as the other two panels but turbulence is now driven by all three mechanisms -- feedback, accretion, and transport. Metal advection is active only in the lowest mass galaxies. These diagnostic plots aid in understanding the nature of metal distributions and metallicity gradients in each of the corresponding models presented in \autoref{fig:mzgr_basic}.}
\label{fig:ratios_PSA}
\end{figure*}

\section{Mass-metallicity gradient relation (MZGR) with the fiducial model}
\label{s:mzgr}
We show the observed MZGR (corrected for spatial resolution) from three different surveys -- MaNGA \citep[Mapping Nearby Galaxies at Apache Point Observatory,][]{2015ApJ...798....7B}, CALIFA \citep[Calar Alto Legacy Integral Field Area,][]{2012A&A...538A...8S}, and SAMI \citep[Sydney-AAO Multi-object Integral-field spectrograph,][]{2015MNRAS.447.2857B} in \autoref{fig:mzgr_basic}. Each marker denotes the average metallicity gradient (normalized by $r_{\rm{e}}$) in different bins of stellar mass. It is immediately clear that even between the three surveys, average gradients differ by as much as $0.1\,\mathrm{dex\,}r^{-1}_{\rm{e}}$. In fact, the galaxy-to-galaxy scatter at fixed $M_{\star}$ is as high as $0.3\,\mathrm{dex\,}r^{-1}_{\rm{e}}$. The scatter is larger at the low and high mass ends than at intermediate masses \citep[fig. 10]{2021MNRAS.502.3357P}. While some of this scatter can be physical (especially at the low-mass end), it is also caused by systematic differences between gas-phase metallicity calibrations \citep{2008ApJ...681.1183K}, limited spatial resolution and S/N ratio \citep{2014A&A...561A.129M,2020MNRAS.495.3819A}, the sensitivity of metallicity diagnostics to physical properties of \ion{H}{ii} regions such as ionisation parameter \citep[e.g.,][]{2004MNRAS.348L..59P,2004ApJ...617..240K,2016MNRAS.457.3678P,2017MNRAS.465.1384C,2020A&A...636A..42M}, and to poorly-understood contributions from diffuse ionized gas \citep[DIG,][]{2017ApJ...850..136S,2017MNRAS.466.3217Z,2019MNRAS.487...79P,2019MNRAS.489.4721V}. An important effect of this discrepancy is that it is not clear if the MZGR shows an inflection at $9.5 < \log_{10} M_{\star}/\rm{M_{\odot}} < 10.5$ \citep{2017MNRAS.469..151B,2020A&A...636A..42M}; some studies favor such a break whereas others find a constant, characteristic gradient for all galaxies \citep{2014A&A...563A..49S,2016A&A...587A..70S,2018A&A...609A.119S,2018MNRAS.479.5235P}.

Using the second data release of SAMI data, \citet{2021MNRAS.502.3357P} reduce the discrepancy due to different emission line ratios and metallicity calibrations to $0.02\,\mathrm{dex\,}r^{-1}_{\rm{e}}$ by using a carefully-chosen combination of various diagnostics that include more than two emission line ratios \citep[e.g.,][]{2016MNRAS.457.3678P}. These authors find that there is indeed evidence for a break in the MZGR between $9.5 < \log_{10} M_{\star}/\rm{M_{\odot}} < 10.5$, however, the feature is rather weak as compared to the curvature observed in the MZR. \cite{2021ApJ...923...28F} reach the same conclusion based on their analysis of the MZGR in the MaNGA and GASP \citep[GAs Stripping Phenomena in galaxies with MUSE,][]{2017ApJ...844...48P} surveys. Deciphering the origin of the MZGR and determining whether there is an inflection at intermediate masses is important to understanding how various physical processes we describe in \autoref{s:model} impact metal distribution in galaxies.

To construct an MZGR from our model, we specify a range of halo masses at high-$z$ such that the resulting halo mass at $z = 0$ span $10.9 < \log_{10} M_{\rm{h}}/\rm{M_{\odot}} < 14$.\footnote{Note that most local galaxies with $M_{\rm{h}} \sim 10^{14}\,\rm{M_{\odot}}$ are usually ellipticals. We deliberately cover such high $M_{\rm{h}}$ to compare against the few massive star-forming galaxies with measured gas-phase metallicity gradients in the CALIFA survey \citep{2014A&A...563A..49S,2016A&A...587A..70S}}. As we saw in \autoref{s:results}, the model produces a family of metallicity profiles. Below, we find the MZGR for different cases corresponding to the different drivers of turbulence (feedback, accretion, and transport).

\subsection{Gas turbulence driven by feedback and transport}
\label{s:mzgr_ft}
In this case, the Péclet number $\mathcal{P}$ that describes the ratio of metal advection to diffusion is set only by star formation feedback and gas transport ($\sigma_{\rm{g}}$ corresponding to the blue curve in \autoref{fig:sigmag}). This is the case studied in \cite{2021MNRAS.502.5935S} since the \citetalias{2018MNRAS.477.2716K} model we used only considers feedback and transport as sources of turbulence. However, there are several differences between this and earlier work, as we point out in \autoref{s:model}.

The top panel of \autoref{fig:mzgr_basic} plots the resulting MZGR from the model in this case. We color-code the model MZGR by the ratio $(\mathcal{P} + \mathcal{A}) /\mathcal{S}$; the reason for this coding is that the processes parameterised by the numerator (advection and accretion) tend to flatten the gradient, while the one parameterised by the denominator (star formation) tends to steepen it. Thus this ratio is effectively the ratio of flattening to steepening processes. We show the resulting values of $\mathcal{P}$, $\mathcal{A}$, and $\mathcal{S}$ for this case in the left panel of \autoref{fig:ratios_PSA}. The scatter in $\mathcal{S}$ at fixed $M_{\star}$ is caused by $\phi_{\rm{y}}$ (see \autoref{eq:S}). In line with our expectations, we see that steep negative gradients develop when $\mathcal{S} > \mathcal{P} + \mathcal{A}$; this process is largely controlled by the parameter $\phi_{\rm y}$, since at fixed $M_{\star}$ increasing $\phi_{\rm{y}}$ leads to larger $\mathcal{S}$ (as indicated by colour in \autoref{fig:mzgr_basic}) and to steeper, negative gradients (corresponding to moving downward in \autoref{fig:mzgr_basic}). The exception to this trend is the least massive galaxies with $M_{\star} < 10^{9.1}\,\rm{M_{\odot}}$ -- these galaxies lie in the GMC regime of star formation, as against galaxies with $M_{\star} \geq 10^{9.1}\,\rm{M_{\odot}}$ where the inner regions are in the Toomre regime and the outer regions are in the GMC regime. Since the star formation timescale is capped by $t_{\rm{SF,max}}$ in the GMC regime, the resulting $\mathcal{S} < \mathcal{P} + \mathcal{A}$ even with $\phi_{\rm{y}} = 1$, as we can also read off from the left panel of \autoref{fig:ratios_PSA}. We also find that $\mathcal{A} > 1$ for all $M_{\star}$ in this case, implying that metal diffusion is weak as compared to gas accretion. However, metal diffusion dominates over metal advection and star formation in some low-mass galaxies that have $\mathcal{P} < 1$ and $\mathcal{S} < 1$. Since diffusion tends to homogenize the metal distribution by moving metals from high to low concentrations, strong diffusion leads to flatter gradients. The inflection we see in the MZGR at $M_{\star} \approx 10^{9.2}\,\rm{M_{\odot}}$ occurs due to mass transport shutting off.

We further see that there is a large scatter in the gradient at fixed stellar mass (especially in low-mass galaxies) that narrows at the high-mass end. This scatter occurs because the gradients are quite sensitive to the choice of $\phi_{\rm{y}}$ -- low $\phi_{\rm{y}}$ lowers $\mathcal{S}$, driving flat or even inverted gradients, whereas high $\phi_{\rm{y}}$ drives steep negative gradients. At the low-mass end, the data prefers low values of $\phi_{\rm{y}}$, implying a significant boost in preferential metal ejection in winds of low-mass galaxies. In the absence of accretion as a source of turbulence, $\sigma_{\rm{g}}$ equilibrates to a lower value that is closer or equal to $\sigma_{\rm{SF}}$. This in turn causes $\mathcal{P}$ to go to zero, thus shutting off mass transport and driving flatter (and even inverted) metallicity gradients at the high-mass end. This is complimented by the dramatic increase in $\mathcal{S}$ and an even larger increase in $\mathcal{A}$ because $\mathcal{S} \propto \sigma^2_{\rm{g}}$ and $\mathcal{A} \propto \sigma^3_{\rm{g}}$. We find that this model fails to reproduce the observed MZGR, especially at the high-mass end.

At first look, this result seems contradictory to the MZGR presented in \citet{2020aMNRAS.xxx..xxxS}, because these authors could also reproduce the high-mass end of the observed MZGR with just feedback + transport. However, their \textit{ad-hoc} choice of $\sigma_{\rm{g}}$ was larger than that we derive in this work (blue curve in \autoref{fig:sigmag}). Larger $\sigma_{\rm{g}}$ led to $\mathcal{P} > 0$ and smaller $\mathcal{A}$ in massive galaxies in \cite{2020aMNRAS.xxx..xxxS}, yielding flatter gradients in good agreement with those observed. As we will discuss below, our self-consistent solution for $\sigma_{\rm{g}}$ which includes accretion as a source of turbulence does reproduce the MZGR at the high-mass end because of larger $\sigma_{\rm{g}}$. This highlights the importance of accretion and turbulence in setting ISM metallicity distributions and gradients in massive galaxies.

\subsection{Gas turbulence driven by accretion and transport}
\label{s:mzgr_at}
Next, we discuss the case where gas turbulence is driven by a combination of accretion and transport (orange curve in \autoref{fig:sigmag}). We show the predictions from the model in the middle panel of \autoref{fig:mzgr_basic}, and the corresponding dimensionless ratios in the middle panel of \autoref{fig:ratios_PSA}. Overall, this model produces flatter metallicity gradients as compared to the feedback + transport model in \autoref{s:mzgr_at} above. It can also reproduce the diversity of metallicity gradients in low-mass galaxies, although the range of $\phi_{\rm{y}}$ that best matches the data is narrower as compared in the feedback + transport case. Including accretion as a source of turbulence brings the model gradients closer to the data, but agreement is still quite poor for massive galaxies. This is because $\mathcal{P} > 1$ in most galaxies where transport is active, so a combination of strong metal advection and relatively higher gas accretion drives the flat and inverted gradients in this model. Thus, we learn that removing feedback and accretion as sources of turbulence results in model gradients that are too flat or inverted at the high-mass end as compared to the observed MZGR.

\subsection{Gas turbulence driven by accretion, feedback and transport}
\label{s:mzgr_aft}
We have seen that if feedback and accretion are not included as sources of turbulence in the disc, the resulting model MZGRs do not reproduce the observed MZGR at the high-mass end, even with the uncertainty in $\phi_{\rm{y}}$. Now, we consider the more general case where $\sigma_{\rm{g}}$ is driven by feedback, accretion, and transport acting together (green curve in \autoref{fig:sigmag}). The bottom panel of \autoref{fig:mzgr_basic} plots the model MZGR with the data. We find that the agreement between the model and the data is much better at all $M_{\star}$ as compared to the cases above where we exclude either feedback or accretion as sources of turbulence. Higher $\sigma_{\rm{g}}$ caused by turbulence due to all three sources (under energy balance) leads to smaller $\mathcal{S}$ and even smaller $\mathcal{A}$ that results in steeper gradients, particularly at the high mass end. We can also discern the same from the right panel of \autoref{fig:ratios_PSA}. As compared to the other two models, we see that $\mathcal{A}$ is smaller due to larger $\sigma_{\rm{g}}$. The inflection in the MZGR at $M_{\star} \approx 10^{9.2}\,\rm{M_{\odot}}$ that we noticed in the feedback+transport model is also present in this case, due to $\mathcal{P} \to 0$ as mass transport shuts off. In equilibrium, the strength of mass transport is very sensitive to $\dot M_{\rm{sf}}$ at $z=0$, and thus to variations in the star-forming gas fraction $f_{\rm{sf}}$ to which $\dot M_{\rm{sf}}$ is proportional. We explore alternate models of $f_{\rm{sf}}$ in \aref{s:app_fsf} and show that the model MZGR at the high mass end (where transport may or may not be shut off) remains qualitatively similar.

We also see that low-mass galaxies strongly prefer a low $\phi_{\rm{y}}$, implying significant metal enrichment of galactic winds. Below, we will see that this conclusion holds irrespective of the mass-loading of galactic winds in low-mass galaxies. We remind the reader that we adopt $\xi_{\rm{a}} = 0.2$ for local galaxies, and show in \aref{s:app_xia} how the results change in the unlikely case where $\xi_{\rm{a}}$ is higher. As we show in \aref{s:app_mingozzi} using MaNGA data, these interpretations are robust to scatter caused by different metallicity diagnostics used to measure metallicity gradients in the data \citep[see also,][]{2021MNRAS.502.3357P,2021MNRAS.502.5935S}.

\section{Role of galactic winds}
\label{s:galactic_winds}
As we mention in \autoref{s:model_gasphase_winds}, the fiducial model we have presented so far is limited, and to some extent, inconsistent, because we include mass-loading of galactic winds ($\eta_{\rm{w}}$) in the evolution of metallicity (cf. \autoref{eq:Zw}) but not in the evolution of the gas mass. There are several different scalings of $\eta_{\rm{w}}$ available in the literature, which can broadly be grouped into two categories: one where $\eta_{\rm{w}}$ is related to the large-scale properties of the galaxy (e.g., $M_{\rm{h}}$ and virial velocity,\, \citealt{2005ApJ...618..569M,2015MNRAS.454.2691M,2018MNRAS.473.4077P,2020MNRAS.494.3971M,2021MNRAS.508.2979P}), and another where it is described by properties internal to the disc (e.g., $\Sigma_{\rm{g}}$, $\Sigma_{\rm{SFR}}$ and $f_{\rm{g}}$, \citealt{2013MNRAS.432..455D,2017MNRAS.465.1682H,2020ApJ...900...61K,2021MNRAS.508.2979P,2022arXiv221108423B}). To rectify the inconsistency between metals and the gas, we discuss three cases below where we adopt $\eta_{\rm{w}}$ from the two groups of works as above. Our choice of adopted models is such that we can cover a broad range in $\eta_{\rm{w}}$ at fixed $M_{\star}$. While this approach of adopting $\eta_{\rm{w}}$ from other works still renders the treatment of outflows decoupled from the rest of the model, the advantage of using these recipes is that they can easily be incorporated into semi-analytical models like ours.

\begin{figure}
\includegraphics[width=\columnwidth]{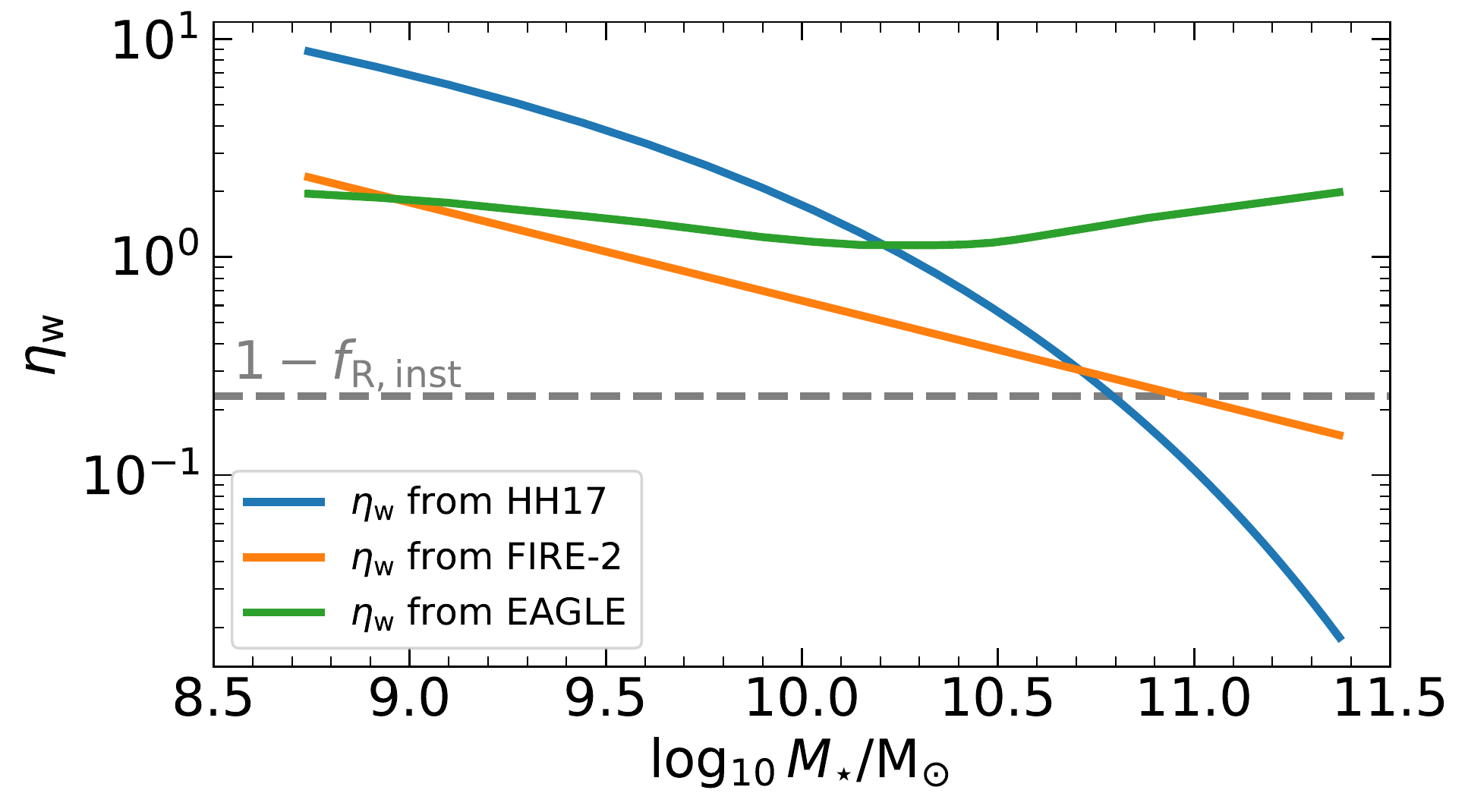}
\caption{Mass loading factor, $\eta_{\rm{w}}$, as a function of stellar mass at $z = 0$ from the three different cases we consider: (1.) analytical model for star formation-driven winds \citep[blue,][]{2017MNRAS.465.1682H}, (2.) FIRE-2 cosmological zoom-in simulations without AGN feedback \citep[orange,][]{2021MNRAS.508.2979P}, and (3.) EAGLE cosmological simulations including AGN feedback \citep[green,][]{2020MNRAS.494.3971M}. The dashed grey line demarcates $1-f_{\rm{R,inst}}$, the fraction of newly formed metals not locked in low mass stars \citep{1980FCPh....5..287T}.}
\label{fig:etaw}
\end{figure}

\begin{figure*}
\includegraphics[width=0.9\textwidth]{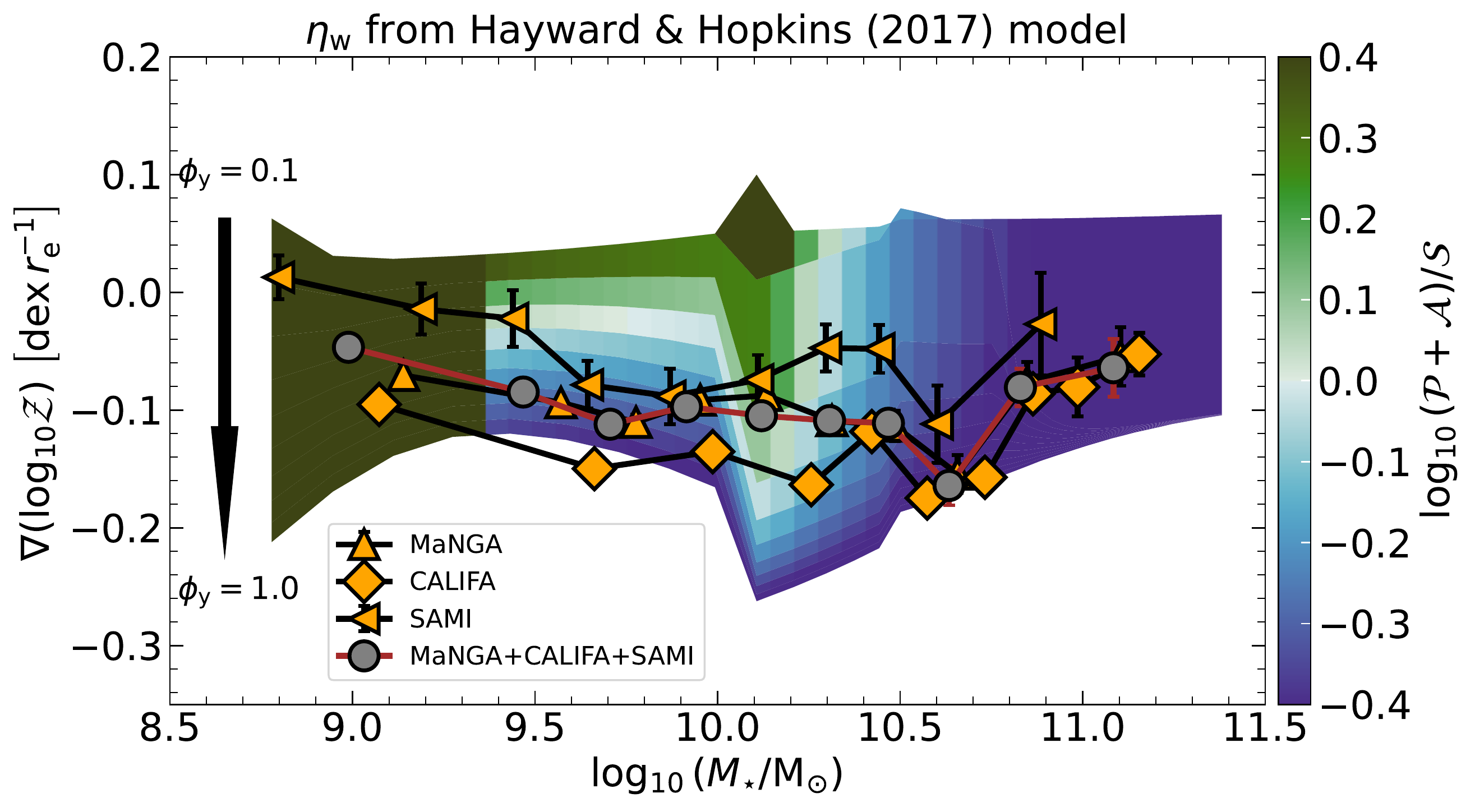}
\caption{Same as the bottom panel of  \autoref{fig:mzgr_basic} but with the mass-loading factor ($\eta_{\rm{w}}$) estimated from the analytical model of \citep{2017MNRAS.465.1682H}.}
\label{fig:mzgr_HH17}
\end{figure*}

\begin{figure*}
\includegraphics[width=0.9\textwidth]{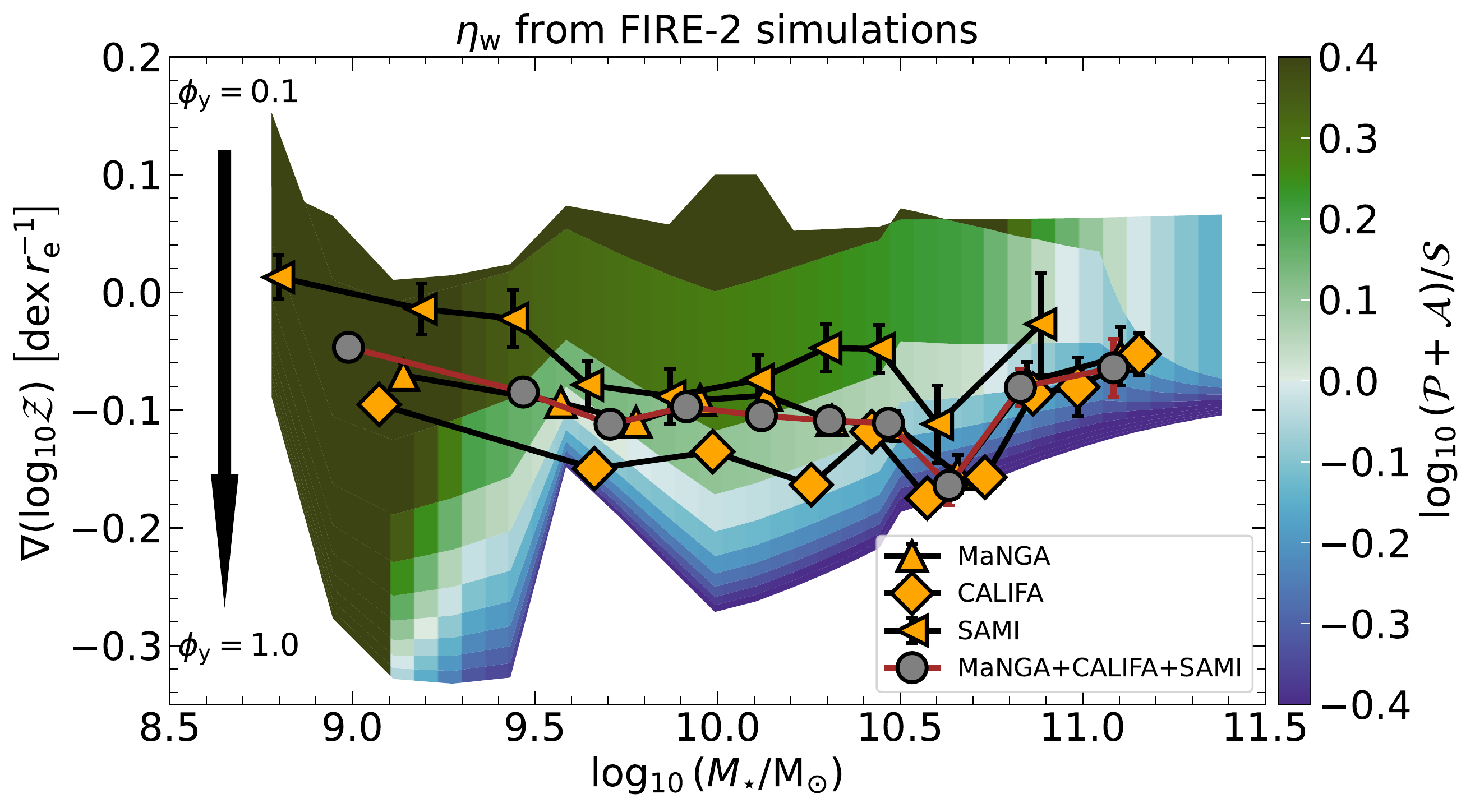}
\caption{Same as the bottom panel of \autoref{fig:mzgr_basic} but with the mass-loading factor ($\eta_{\rm{w}}$) estimated from the FIRE-2 cosmological zoom-in simulations without AGN feedback \citep{2021MNRAS.508.2979P}.}
\label{fig:mzgr_FIRE2}
\end{figure*}

\begin{figure*}
\includegraphics[width=0.9\textwidth]{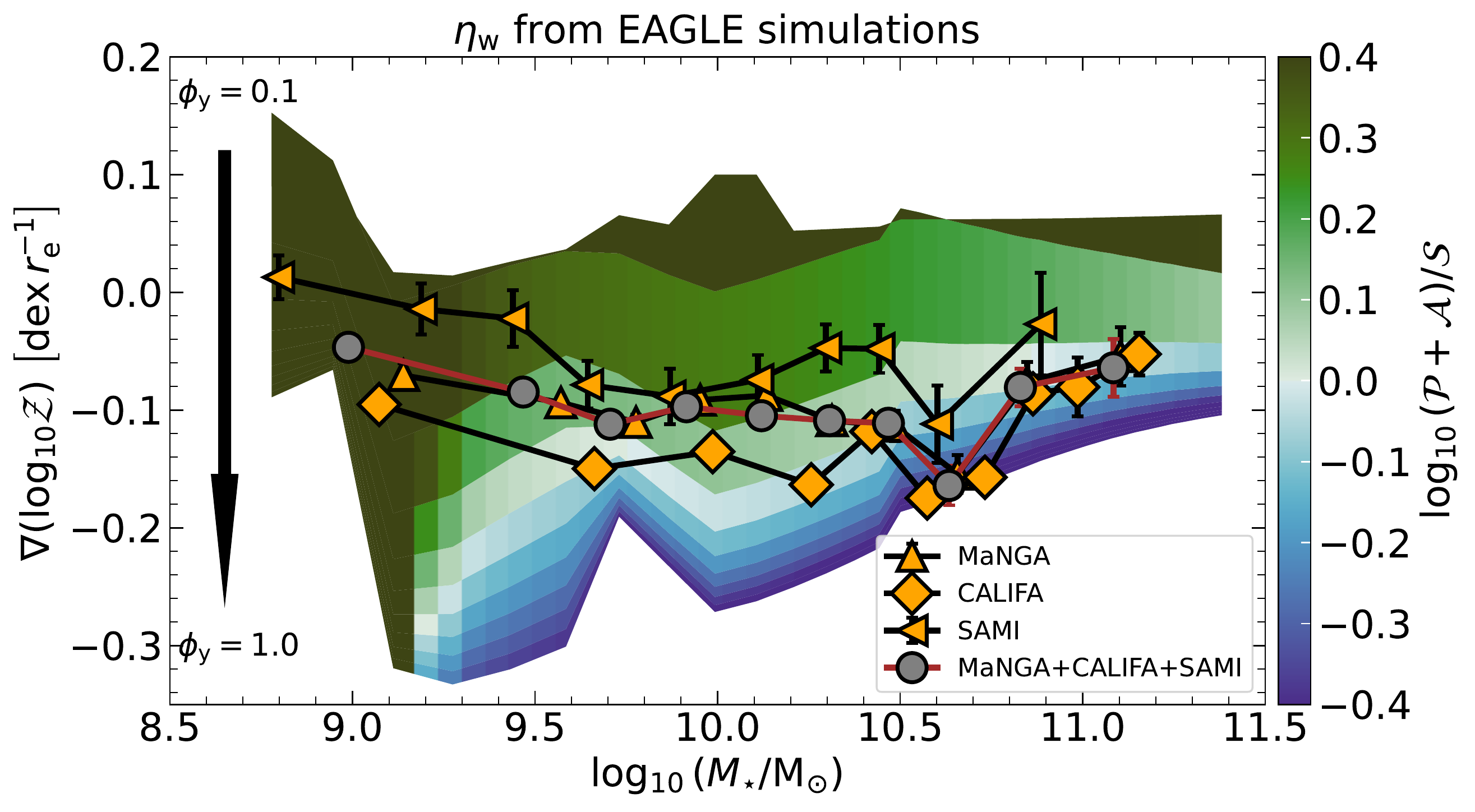}
\caption{Same as the bottom panel of \autoref{fig:mzgr_basic} but with the mass-loading factor ($\eta_{\rm{w}}$) estimated from the EAGLE cosmological simulations including AGN feedback \citep{2020MNRAS.494.3971M}.}
\label{fig:mzgr_EAGLE}
\end{figure*}

\begin{figure*}
\includegraphics[width=1.0\textwidth]{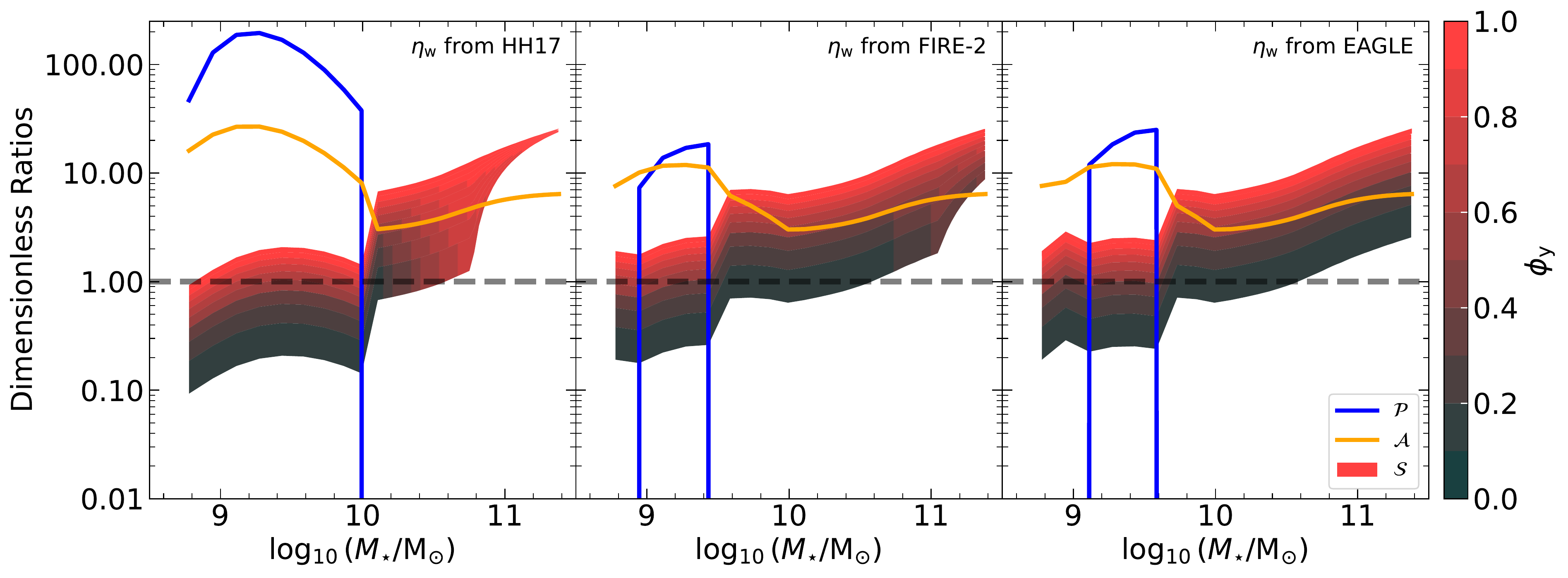}
\caption{\textit{Left panel:} Strength of the dimensionless ratios in the metallicity model ($\mathcal{P}, \mathcal{A}$, and $\mathcal{S}$) as a function of $M_{\star}$ for the case where turbulence is driven by feedback, accretion, and transport, and the mass-loading factor, $\eta_{\rm{w}}$, scales with the gas fraction and $M_{\star}$, following an analytical model of star formation-driven winds \citep{2017MNRAS.465.1682H}. Colorbar denotes the permitted range of $\phi_{\rm{y}}$ that describes the preferential enrichment of galactic winds (see \autoref{eq:phiy}). \textit{Middle panel:} Same as the left panel but $\eta_{\rm{w}}$ is scaled with $M_{\star}$ following FIRE-2 cosmological zoom-in simulations without AGN feedback \citep{2021MNRAS.508.2979P}. \textit{Right panel:} Same as the other two panels but $\eta_{\rm{w}}$ is scaled with $M_{\star}$ following EAGLE cosmological simulations that include AGN feedback \citep{2020MNRAS.494.3971M}.}
\label{fig:ratios_etaw}
\end{figure*}

\subsection{Effects of different mass loading parameterisations}
\label{s:galactic_winds_scalings}
\begin{enumerate}

    \item {\textit{Mass-loading factor from \cite{2017MNRAS.465.1682H}.} Using an analytical model that describes the dual role of star formation feedback in regulating star formation as well as launching winds from the galactic disc, \cite{2017MNRAS.465.1682H} show that the gas fraction is a critical factor that determines $\eta_{\rm{w}}$ in star-forming galaxies. These authors find that gas-rich galaxies are typically more turbulent, and turbulence causes sharper density contrasts in the ISM \citep[e.g.,][]{2010A&A...512A..81F,2012ApJ...761..156F,2018ApJ...863..118B}. As a result, low-density material is more efficiently accelerated out of the disc by star formation feedback. We adopt the approximate relation between the gas fraction, $f_{\rm{g,HH17}}$, and $\eta_{\rm{w}}$ provided by \citeauthor{2017MNRAS.465.1682H} that effectively captures their detailed calculations
    \begin{equation}
        \eta_{\rm{w}} = 14 \left(f_{\rm{g,HH17}} \frac{M_{\star}}{10^{10} \rm{M_{\odot}}}\right)^{-0.23} e^{-0.75/f_{\rm{g,HH17}}}\,,
    \label{eq:etaw_HH17}
    \end{equation}
    where $f_{\rm{g,HH17}} = M_{\rm{g}} / (M_{\rm{g}} + M_{\star})$ is given by \citep{2009ApJ...691.1424H,2010MNRAS.402.1693H}
    \begin{eqnarray}
        \lefteqn{f_{\rm{g,HH17}} = f_{\rm{0,HH17}} \left[1 - \tau(z)\left(1 - f^{3/2}_{\rm{0,HH17}}\right)\right]^{-2/3}
        }
        \nonumber \\
        &&    f_{\rm{0,HH17}} =  \left[1 + \left(\frac{M_{\star}}{10^{9.15}\,\rm{M_{\odot}}}\right)^{0.4}\right]^{-1}\,,
    \label{eq:fgHH17}
    \end{eqnarray}
    where $\tau(z)$ is the fractional redshift given by the ratio of lookback time at given $z$ to that at $z \to \infty$. However, this model uses a scaling $\Sigma_{\rm{SFR}} \propto \Sigma^{1.7-2}_{\rm{g}}$ \citep{2013MNRAS.433.1970F,2016MNRAS.455..334T} that is too steep as compared to the well-known Kennicutt-Schmidt relation \citep{1998ARA&A..36..189K}.}

    \item{ \textit{Mass-loading factor from FIRE-2 simulations.} We use the best-fit scaling of the multi-phase ISM $\eta_{\rm{w}}$ (measured in a radial shell between $0.1-0.2\,R_{\rm{vir}}$) as a function of $M_{\star}$ from FIRE-2 zoom-in simulations \citep{2018MNRAS.480..800H}
    \begin{equation}
        \eta_{\rm{w}} = 10^{4.3} \left(\frac{M_{\star}}{\rm{M_{\odot}}}\right)^{-0.45}\,.
    \label{eq:etaw_FIRE2}
    \end{equation}
    This scaling is based on an analysis of multi-phase winds in central galaxies of 13 haloes spanning a wide range of $M_{\rm{h}}$ \citep[equation 15]{2021MNRAS.508.2979P}. A key improvement over similar studies is that \citeauthor{2021MNRAS.508.2979P} define a more physically-meaningful criteria to define outflowing gas particles that also encaptures the slow-moving hot wind. However, the set of simulations used for this analysis does not include AGN feedback and turbulent metal diffusion.
    
    \item \textit{Mass-loading factor from EAGLE simulations:} Lastly, we adopt the relation between $\eta_{\rm{w}}$ measured in EAGLE galaxies \citep{2015MNRAS.446..521S} as a function of $M_{\star}$ from \citet[fig. 15]{2020MNRAS.494.3971M}. Similar to \cite{2021MNRAS.508.2979P}, this $\eta_{\rm{w}}$ corresponds to mass leaving the ISM of the galaxy, and not the halo, and is measured for central galaxies. Crucially, the simulations used for this work by \citeauthor{2020MNRAS.494.3971M} include AGN feedback. However, the criteria used by EAGLE to select outflowing gas particles is different than FIRE-2 (see section 2.6 of \citealt{2020MNRAS.494.3971M}).}

\end{enumerate}

These scalings of $\eta_{\rm{w}}$ also constrain the range over which $\phi_{\rm{y}}$ can vary, as against the fiducial model where we swept across $\phi_{\rm{y}} = 0 - 1$. This is simply because a fixed value of $\eta_{\rm{w}}$ limits the range of $\xi_{\rm{w}}$ in \autoref{eq:Zw} that in turn limits $\phi_{\rm{y}}$. However, as we can expect from \autoref{eq:Zw}, this constraint is only used in practice in galaxies where $\eta_{\rm{w}} < 1 - f_{\rm{R,inst}}$.

\autoref{fig:etaw} unifies these scaling relations to represent them as a function of $M_{\star}$, for illustration. We find that, at the low mass end, $\eta_{\rm{w}}$ from all the three works are qualitatively consistent with each other. While different simulations qualitatively agree on the behaviour of $\eta_{\rm{w}}$ with $M_{\star}$ for low-mass galaxies, they do not converge on the partition of mass and metal loading in the different phases of galactic winds as this is sensitive to the distance from the midplane where $\eta_{\rm{w}}$ is measured. There also exists tension between $\eta_{\rm{w}}$ measured in simulations and observations \citep[e.g.,][]{2022MNRAS.513.2535C,2023A&A...670A..92M}. Simulations of isolated galaxies that can afford very high resolution tend to measure mass fluxes few $\rm{kpc}$ from the midplane \citep[e.g.,][]{2020ApJ...900...61K,2020ApJ...894...12V,2022MNRAS.tmp.2441W}, whereas cosmological simulations tend to measure much farther away from the disc \citep[see the discussion in][]{2021MNRAS.508.2979P}. However, there is a large scatter in $\eta_{\rm{w}}$ for simulated massive galaxies. The key reason for this difference is that EAGLE includes AGN feedback that boosts mass-loading in massive galaxies at $z = 0$. In the absence of AGN feedback, it is difficult to drive strong winds in massive galaxies due to their deep potential wells. While our model does not include AGN feedback and is closer in spirit to \citet{2017MNRAS.465.1682H}, it is insightful to test how AGN feedback-dominated $\eta_{\rm{w}}$ impacts the MZGR.

\subsection{Results}
\label{s:galactic_winds_results}
We know from \autoref{eq:dotMg} that introducing $\eta_{\rm{w}} \neq 0$ activates the outflow sink term that takes mass out of the disc and thus impacts $M_{\rm{g}}$. Since $\sigma_{\rm{g}} \propto M_{\rm{g}}$, and the dimensionless ratios that govern the metallicity profile are all sensitive to $\sigma_{\rm{g}}$, we expect wind mass loading to non-linearly modulate the metallicity profiles and resulting metallicity gradients. \autoref{fig:mzgr_HH17}, \autoref{fig:mzgr_FIRE2}, and \autoref{fig:mzgr_EAGLE} plot the resulting model MZGR where we scale $\eta_{\rm{w}}$ following \citet{2017MNRAS.465.1682H}, \citet{2021MNRAS.508.2979P}, and \citet{2020MNRAS.494.3971M}, respectively. We only plot the model for the case where turbulence is driven by a combination of feedback, accretion, and transport, as we have seen in \autoref{s:mzgr} that neglecting either feedback- or accretion-driven turbulence does not reproduce the observed MZGR even in the fiducial case with $\eta_{\rm{w}}$ undefined.

We first discuss results for the scaling of $\eta_{\rm{w}}$ from \cite{2017MNRAS.465.1682H}. We notice from \autoref{fig:mzgr_HH17} that, as compared to the fiducial model (bottom panel of \autoref{fig:mzgr_basic}), the range of metallicity gradients at the high mass end of the MZGR remains unaffected by this scaling. This is because even at fixed $\phi_{\rm{y}}$, the model produces a family of gradients due to constraints on $c_1$ (see \autoref{s:model_gasphasemetal_BCs}). However, the range of gradients at the low mass end is reduced by more than $0.1\,\rm{dex}\,r^{-1}_{\rm{e}}$, due to a high $\eta_{\rm{w}} > 1$ for low-mass galaxies. While this scaling of $\eta_{\rm{w}}$ can capture the observed MZGR, the mechanism giving rise to the agreement between the observations and the model is significantly different as compared to the fiducial model, as we see from the differences in the ratio $(\mathcal{P} + \mathcal{A})/\mathcal{S}$ denoted by the colorbar. The transition in color from green to blue as a function of increasing $M_{\star}$ implies a transition in the dominant mechanism that sets metallicity gradients in low-mass versus massive galaxies, and is responsible for driving the model MZGR. Additionally, the best match between the model and the data reveals a large scatter in the preferred value of $\phi_{\rm{y}}$ for galaxies with $M_{\star} < 10^{9.4}\,\rm{M_{\odot}}$: low-mass galaxies in the SAMI survey prefer $\phi_{\rm{y}} \sim 0.1$, whereas MaNGA and CALIFA dwarfs prefer $\phi_{\rm{y}} \sim 0.4$. The average measurement for the three surveys prefers $\phi_{\rm{y}} \sim 0.25$. Massive galaxies ($M_{\star} \gtrsim 10^{10}\,\rm{M_{\odot}}$) prefer $\phi_{\rm{y}} \sim 0.8-1.0$.

To further diagnose the cause for these differences, we plot the dimensionless ratios as a function of $M_{\star}$ for this scaling in the left panel of \autoref{fig:ratios_etaw}. At the high mass end, $\phi_{\rm{y}}$ is restricted to a very narrow range close to unity because $\eta_{\rm{w}}$ in the \citeauthor{2017MNRAS.465.1682H} model is less than $1 - f_{\rm{R,inst}}$, which limits the allowed range of $\mathcal{Z}_{\rm{w}}$ (cf. \autoref{eq:Zw}). Physically, we can understand this result very simply: for these galaxies the mass flux carried by the winds is so small that, even if this mass were to consist of pure SN ejecta, winds would carry away at most a small fraction of the metals being produced. Thus most metals must remain in the galaxy, corresponding to $\phi_\mathrm{y}$ close to unity. Since the dimensionless ratio $\mathcal{S} \propto \phi_{\rm{y}}$, this leads to a decrease in the quantity $(\mathcal{P} + \mathcal{A})/\mathcal{S}$, which in turn drives the colorbar to blue at the high mass end in \autoref{fig:mzgr_HH17}. We also see from the left panel of \autoref{fig:ratios_etaw} that $\mathcal{P} > 0$ only for galaxies with $M_{\star} < 10^{10}\,\rm{M_{\odot}}$, implying that mass transport shuts off in massive galaxies because accretion and feedback can drive sufficient levels of turbulence in these galaxies. Lastly, metal diffusion dominates over star formation for a wider range of $M_{\star}$ in this case as compared to the fiducial model.

Next, we examine the MZGR using the scaling of $\eta_{\rm{w}}$ from FIRE-2 zoom-in simulations. \autoref{fig:mzgr_FIRE2} shows the results. The scatter in metallicity gradients at the low mass end is larger than both the fiducial model and the model above where $\eta_{\rm{w}}$ scales according to the \citeauthor{2017MNRAS.465.1682H} model. This correlates with the non-zero $\mathcal{P}$ as we show in the middle panel of \autoref{fig:ratios_etaw}, and occurs because $\eta_{\rm{w}}$ measured in FIRE-2 dwarf galaxies is smaller by almost an order of magnitude as compared to \citeauthor{2017MNRAS.465.1682H}. On the contrary, the scatter at the high mass end remains identical to the other models, although the relative contribution of the dimensionless ratio $\mathcal{S}$ is larger than the fiducial model but smaller than the model with $\eta_{\rm{w}}$ from \citeauthor{2017MNRAS.465.1682H}. This occurs because the larger $\eta_{\rm{w}}$ in FIRE-2, as compared to \citeauthor{2017MNRAS.465.1682H}, leads to a wider range of $\phi_{\rm{y}}$, as we read off from the middle panel of \autoref{fig:ratios_etaw}. Nevertheless, $\eta_{\rm{w}}$ from FIRE-2 simulations also results in an MZGR that can reproduce the observed data, but the preferred value of $\phi_{\rm{y}}$ for low-mass galaxies is lower as compared to the case above with $\eta_{\rm{w}}$ from \citeauthor{2017MNRAS.465.1682H}. Specifically, models with $\phi_{\rm{y}} \sim 0.1-0.2$ best reproduce metallicity gradients in low-mass galaxies in all the three IFU surveys in this case. The preferred values of $\phi_{\rm{y}}$ for massive galaxies remain similar to the case above.

Finally, we examine the case where we scale $\eta_{\rm{w}}$ following the EAGLE simulations \citep{2020MNRAS.494.3971M}. We plot the resulting model MZGR in \autoref{fig:mzgr_EAGLE}, and the corresponding diagnostic plot for the dimensionless ratios in the right panel of \autoref{fig:ratios_etaw}. For $10^{9}\,\mathrm{M_{\odot}} < M_{\star} < 10^{10.7}\,\mathrm{M_{\odot}}$, the MZGR from EAGLE is similar to that we obtain from FIRE-2 results above. The difference in the range of gradients at $M_{\star} < 10^{9}\,\mathrm{M_{\odot}}$ between the two is due to a non-zero $\mathcal{P}$ in FIRE-2. Further, the best match between the observed and the EAGLE MZGR also leads to $\phi_{\rm{y}} \sim 0.1-0.2$ for low-mass galaxies. This is not surprising given $\eta_{\rm{w}}$ from both these simulations for low- and intermediate-mass galaxies is within a factor of few (\autoref{fig:etaw}). While the range of metallicity gradients predicted for massive galaxies is also the same, the underlying mechanism is qualitatively different. As compared to \autoref{fig:mzgr_FIRE2}, we see that $(\mathcal{P}+\mathcal{A}) > \mathcal{S}$ for some gradients at large $M_{\star}$. The reason behind this is the fact that EAGLE includes AGN feedback that boosts $\eta_{\rm{w}}$ above $1 - f_{\rm{R,inst}}$ for massive galaxies. As a consequence, EAGLE allows $\phi_{\rm{y}}$ to be as low as possible, which in turn decreases $\mathcal{S}$. We confirm this from the range of $\phi_{\rm{y}}$ and $\mathcal{S}$ we observe from the right panel of \autoref{fig:ratios_etaw}. 

We can summarise the overall conclusions from using different scalings of $\eta_{\rm{w}}$ as follows: (1.) there is a transition in the dominant process setting metallicity gradients in local galaxies as a function of $M_{\star}$; gradients are set by $\mathcal{P}$ (advection) and $\mathcal{A}$ (accretion) in low-mass galaxies, and $\mathcal{S}$ (production) begins to play an important role in setting the gradients in intermediate-mass and massive galaxies, (2.) discrepancies in $\eta_{\rm{w}}$ that exist for massive galaxies due to AGN feedback lead to different mechanisms setting the high-mass end of the MZGR, (3.) mass-loading is less critical for the MZGR as compared to metal-loading, and (4.) low-mass galaxies seem to prefer a low $\phi_{\rm{y}}$ irrespective of our adopted scaling of $\eta_{\rm{w}}$, implying winds in low-mass galaxies are metal-enriched compared to their ISMs. 

\subsection{Predicted scaling of metal enrichment of galactic winds}
\label{s:galactic_winds_phiy}
Based on this analysis, we can now predict a scaling of $\phi_{\rm{y}}$ that describes the preferential metal enrichment of galactic winds using $\eta_{\rm{w}}$ from the three studies we discuss above in \autoref{s:galactic_winds_results}. Such an exercise is similar in spirit to \cite{2011MNRAS.417.2962P} who predict the scaling of the ratio $\mathcal{Z}_{\rm{w}}/\mathcal{Z}$ based on the best match between their chemical evolution model and the MZR from Sloan Digital Sky Survey \citep[SDSS,][]{2004ApJ...613..898T}. For this purpose, we use the observed MZGR from the three IFU surveys as well as their overall mean, and calculate the range of $\phi_{\rm{y}}$ needed to reproduce the average metallicity gradient observed per $M_{\star}$ in the model with $\eta_{\rm{w}}$ from \cite{2017MNRAS.465.1682H}, \cite{2021MNRAS.508.2979P}, and \cite{2020MNRAS.494.3971M}. Specifically, we start with a uniform distribution of $\phi_{\rm{y}}$ from $0$ to $1$; for each $\phi_{\rm{y}}$, we check if the model in question can reproduce the gradients observed in a given survey. If so, then it ends up in our final $\phi_{\rm{y}}$ distribution. It is difficult to put strong constraints on the range of $\phi_{\rm{y}}$ (especially for massive galaxies) given the family of gradients produced at fixed $\phi_{\rm{y}}$. Nevertheless, this exercise demonstrates the power of using metallicity gradients as a tracer of metal enrichment of galactic winds, and its subsequent impact on the metallicity of the CGM and the IGM.

\begin{figure}
\includegraphics[width=1.0\columnwidth]{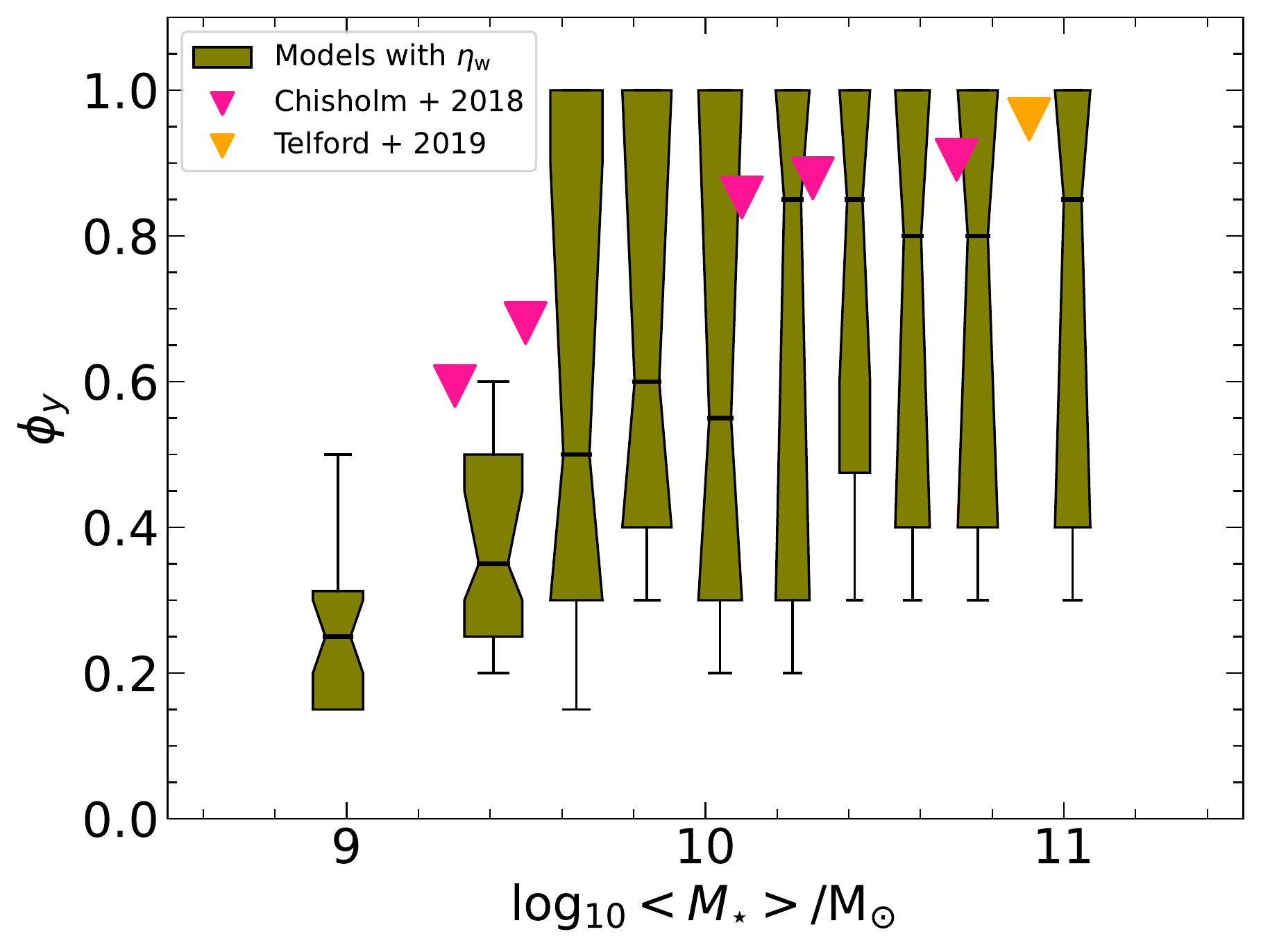}
\caption{Boxplot showing the predicted scaling of $\phi_{\rm{y}}$ that describes the preferential enrichment of galactic winds, as a function of $M_{\star}$ for local galaxies. $\phi_{\rm{y}} \approx 1$ typically corresponds to wind metallicities being similar to ISM metallicities (\autoref{eq:phiy}). $\phi_{\rm{y}} \approx 0$ typically corresponds to winds being very metal enriched as compared to the ISM. The predicted $\phi_{\rm{y}}$ is based on the best match between the models with different mass-loading factors ($\eta_{\rm{w}}$, see \autoref{s:galactic_winds}) and the observed MZGR from the three local IFU surveys (SAMI, MaNGA, CALIFA) as well as their overall mean. Notches denote confidence interval around the median value that is marked in solid black. Errors denote the $5^{\rm{th}}$ and $95^{\rm{th}}$ percentile range. The x-coordinate for each boxplot is the mean $M_{\star}$ across the IFU surveys. The width of the boxplot is the $1\sigma$ deviation from the mean $M_{\star}$. Pink markers denote upper limits on $\phi_{\rm{y}}$ measured by \citet[appendix A]{2020aMNRAS.xxx..xxxS} based on the observations of galactic wind mass-loading and metal-loading in a sample of five local galaxies by \citet{2017MNRAS.469.4831C,2018MNRAS.481.1690C}. Orange marker denotes our estimate of $\phi_{\rm{y}}$ for M31 based on the results of \citet{2019ApJ...877..120T}.}
\label{fig:scaling_phiy}
\end{figure}

\autoref{fig:scaling_phiy} shows the predicted scaling of $\phi_{\rm{y}}$ as a function of $M_{\star}$ from our analysis. The median value of $\phi_{\rm{y}}$ is denoted by the solid black line within the notched boxplots. The errorbars denote the $5^{\rm{th}}$ and the $95^{\rm{th}}$ percentile range. We take the mean and standard deviation of $M_{\star}$ in different bins in the three IFU surveys to specify the $x-$ coordinates and widths for each boxplot, respectively. The key conclusion we draw from \autoref{fig:scaling_phiy} is that $\phi_{\rm{y}}$ is low and well-constrained for galaxies with $M_{\star} < 10^{9.4}\,\rm{M_{\odot}}$ regardless of the IFU survey data, systematics due to metallicity calibrations, adopted scaling of $\eta_{\rm{w}}$, and boundary conditions on the metallicity solution imposed by $c_1$. The median value of $\phi_{\rm{y}}$ increases with $M_{\star}$: it is $\approx 0.5$ for intermediate-mass galaxies and $\approx 0.8$ for massive galaxies. However, the scatter is high, with $\phi_{\rm{y}} = 0.3 - 1.0$ providing a match between the model and the observed MZGRs.

We also overplot $\phi_{\rm{y}}$ for a sample of five local galaxies for which mass- and metal-loading were measured by \cite{2017MNRAS.469.4831C,2018MNRAS.481.1690C} from \ion{Si}{iv} and \ion{Si}{ii} absorption features in the UV using HST/COS spectra. The corresponding $\phi_{\rm{y}}$ values were obtained by \citet[see their appendix A]{2020aMNRAS.xxx..xxxS}.\footnote{The \cite{2018MNRAS.481.1690C} sample has two additional galaxies, however these galaxies have $M_{\star} < 10^8\,\rm{M_{\odot}}$ that is beyond the mass range we consider in our models.} The predictions for $\phi_{\rm{y}}$ are in good agreement with all the galaxies in the \citeauthor{2018MNRAS.481.1690C} sample, however the least massive galaxy (Mrk 1486) prefers a slightly higher $\phi_{\rm{y}}$ as compared to the predictions. \cite{2021ApJ...918L..16C} estimate $\phi_{\rm{y}} \approx 0.8-0.95$ from measurements of ionized gas outflows for Mrk 1486. Since \citeauthor{2018MNRAS.481.1690C} and \citeauthor{2021ApJ...918L..16C} only analyze the ionized phase of winds in these galaxies, it is likely that the $\phi_{\rm{y}}$ estimated for these galaxies is an upper limit, as the hot, X-ray emitting phase can also carry a substantial amount of metals \citep{2020ApJ...904..152L,2022arXiv220909260L}. In addition to these data, we also expect $\phi_{\rm{y}} \sim 1$ based on the results of \citet{2019ApJ...877..120T} for the local massive galaxy M31 with $M_{\star} \approx 10^{11}\,\rm{M_{\odot}}$. Our predicted scaling of $\phi_{\rm{y}}$ is in qualitative agreement with the scaling of $\mathcal{Z}_{\rm{w}}/\mathcal{Z}$ extracted by \cite{2011MNRAS.417.2962P} from the MZR. It is evident that observational measurements of $\phi_{\rm{y}}$ in diverse galaxies will be key to solving the puzzle of the role of galactic winds in driving integrated as well as spatially-resolved gas-phase metal distribution in galaxies.

\section{Conclusions and future outlook}
\label{s:conclusions}
In this work, we create a model of spatially-resolved gas-phase metallicities in galaxies by combining models of turbulent galactic discs in pressure and energy balance (\citetalias{2018MNRAS.477.2716K}, \citealt{2022MNRAS.TMP.1282G}) with a model for ISM metallicities \citep{2021MNRAS.502.5935S}. The evolution of gas-phase metallicity profiles in our model is dictated by both the large-scale properties of galaxies and properties local to the ISM. We include a comprehensive treatment of metal dynamics (advection with radial gas flows and diffusion due to turbulence powered by star formation feedback, gas transport, and cosmic accretion), and allow for preferential enrichment of galactic winds whereby wind metallicities can be higher than the ISM metallicity (see the schematic in \autoref{fig:lilly}). Crucially, the key parameters in our model are constrained by both local and global equilibrium, which ensures that metals are treated self-consistently with the gas and stars in galactic discs. Previous works have shown that our model also reproduces the observed relationship between $\dot M_{\rm{SF}} - M_{\rm{g}}$, $\Sigma_{\rm{SFR}} - \Sigma_{\rm{g}}$, $\sigma_{\rm{g}} - \dot M_{\rm{SF}}$, and $M_{\star} - \mathcal{Z}$. \autoref{tab:diffgal} highlights the key differences in our work as compared to previous models, which are important to self-consistently model spatially-resolved metallicities and metallicity gradients.

We compare the results of our model with observed metallicity gradients in local galaxies. We show that when the gas mass and velocity dispersion are calculated self-consistently, only models where turbulence is driven by a combination of feedback, transport, and accretion can reproduce the local mass-metallicity gradient relation (MZGR; see \autoref{fig:mzgr_basic}). Turbulence driven only by feedback or accretion produces metallicity gradients that are flatter than that observed in massive galaxies. Metal transport, if active (due to radial gas flows), plays an important role in setting metallicity gradients in low-mass galaxies.

The prescriptions of wind mass-loading we adopt from theoretical works \citep{2017MNRAS.465.1682H,2021MNRAS.508.2979P,2020MNRAS.494.3971M} are naturally decoupled from the rest of the model. Regardless of this decoupling, we find strong evidence for the preferential metal enrichment of winds in low-mass galaxies, in line with earlier works based on modeling the MZR \citep{2007ApJ...658..941D,2011MNRAS.417.2962P,2020aMNRAS.xxx..xxxS,2022A&A...657A..19T}. We also find metal loading to be more critical than mass loading in driving the MZGR. The key impact of metal-loaded winds in low-mass galaxies is to reduce the overall level of metal content in the ISM, which is necessary to reproduce the observed flatness in metallicity gradients in these galaxies without violating other gas-phase scaling relations. However, the extent of enrichment of winds in intermediate-mass and massive galaxies remains unclear because metal-loaded winds are less important (compared to feedback and accretion) for driving metallicity gradients in these galaxies (\autoref{fig:scaling_phiy}).

The current sample of direct measurements of wind metallicities and outflow rates is very limited and often only probes a single phase of galactic winds. Given the importance of winds in explaining both the MZR and the MZGR, it is crucial to obtain direct measurements of mass, metal, and energy loading in diverse galaxies. JWST/NIRSpec and VLT/MAVIS are expected to deliver resolved metallicity measurements for a large number of galaxies at high redshifts \citep{2021Msngr.185....7R,2022arXiv220705632R}. On the theoretical front, it is desirable to self-consistently model the CGM and the ISM so that the decoupling that currently exists between the two in chemical evolution models such as ours can be removed \citep{2022arXiv221105115C,2022arXiv221109755P}. Equally important is understanding multiphase metal mixing at the CGM -- ISM interface \citep{2020ApJ...900...61K,2020ApJ...895...43S,2023arXiv230907955V} that occurs at pc scales not currently resolved in cosmological simulations \citep{2019MNRAS.483.3647G}. Such a unification of ISM and CGM models can also provide a self-consistent formalism for galactic fountains that is missing from models like ours. Particularly interesting is the discovery of inverted gradients in galaxies across redshift \citep[e.g.,][]{2010Natur.467..811C,2020MNRAS.492..821C,2019ApJ...882...94W,2022ApJ...938L..16W}. Explaining the origin of inverted gradients within the context of gas regulator models remains a critical task for future theoretical work.

\section*{Acknowledgements}
We thank the referee for their constructive report on this manuscript. We thank Aditi Vijayan and David Weinberg for going through a draft of this work and providing feedback. We also thank Simon Lilly and the AAS Journals for permitting us to adapt and modify the schematic depicting the bathtub model of chemical evolution (\autoref{fig:lilly}). We are grateful to Christopher Hayward, Viraj Pandya, and Joop Schaye for discussions on galactic winds. PS acknowledges support in the form of Oort Fellowship at Leiden Observatory, and the International Astronomical Union -- Gruber Foundation Fellowship. OG was supported by a Milner Fellowship. MRK is supported by the Australian Research Council (ARC) Laureate Fellowship FL220100020 and Future Fellowship FT180100375. JCF is supported by the Flatiron Institute through the Simons Foundation. AD and OG were supported by the Israel Science Foundation Grant ISF 861/20. Parts of this research were supported by the ARC Centre of Excellence for All Sky Astrophysics in 3 Dimensions (ASTRO 3D), through project number CE170100013. Analysis was performed using \texttt{numpy} \citep{oliphant2006guide,2020Natur.585..357H} and \texttt{scipy} \citep{2020NatMe..17..261V}; plots were created using \texttt{Matplotlib} \citep{Hunter:2007}, \texttt{CMasher} \citep{2020JOSS....5.2004V}, and \texttt{astropy} \citep{2013A&A...558A..33A,2018AJ....156..123A}. This research has made extensive use of NASA's Astrophysics Data System Bibliographic Services. 

\section*{Data availability statement}
No data were generated for this work. 

\bibliographystyle{mnras}
\bibliography{references} 

\appendix

\section{Impact of variations in the fraction of star-forming molecular gas}
\label{s:app_fsf}
As we explain in the main text, whether massive galaxies show mass transport or not is sensitive to the star formation rate under equilibrium because transport is not needed if star formation and gas accretion are of similar strength. Thus, the existence of mass transport in the model boils down to parameters that dictate $\dot M_{\rm{sf}}$. 

Variations in the fraction of star-forming molecular gas ($f_{\rm{sf}}$) that sets $\dot M_{\rm{sf}}$ needs further exploration as previous work has shown it varies with galactocentric radius whereas we treat it as a constant in the main text. In principle, we can follow the approach used by \citetalias{2018MNRAS.477.2716K} to define a radially-varying $f_{\rm{sf}}$ that these authors find from the \cite{2013MNRAS.436.2747K} model. However, the \cite{2013MNRAS.436.2747K} model uses metallicity as an input to find $f_{\rm{sf}}$ since the transition from atomic to molecular gas is sensitive to the ISM metallicity \citep{2009ApJ...693..216K,2010ApJ...721..975O}. Since metallicity is the final output of our model, it would require an iterative algorithm over several non-linear terms to model spatial variations in $f_{\rm{sf}}$, which is why we refrain from adopting it as our standard choice.

\begin{figure}
\includegraphics[width=1.0\columnwidth]{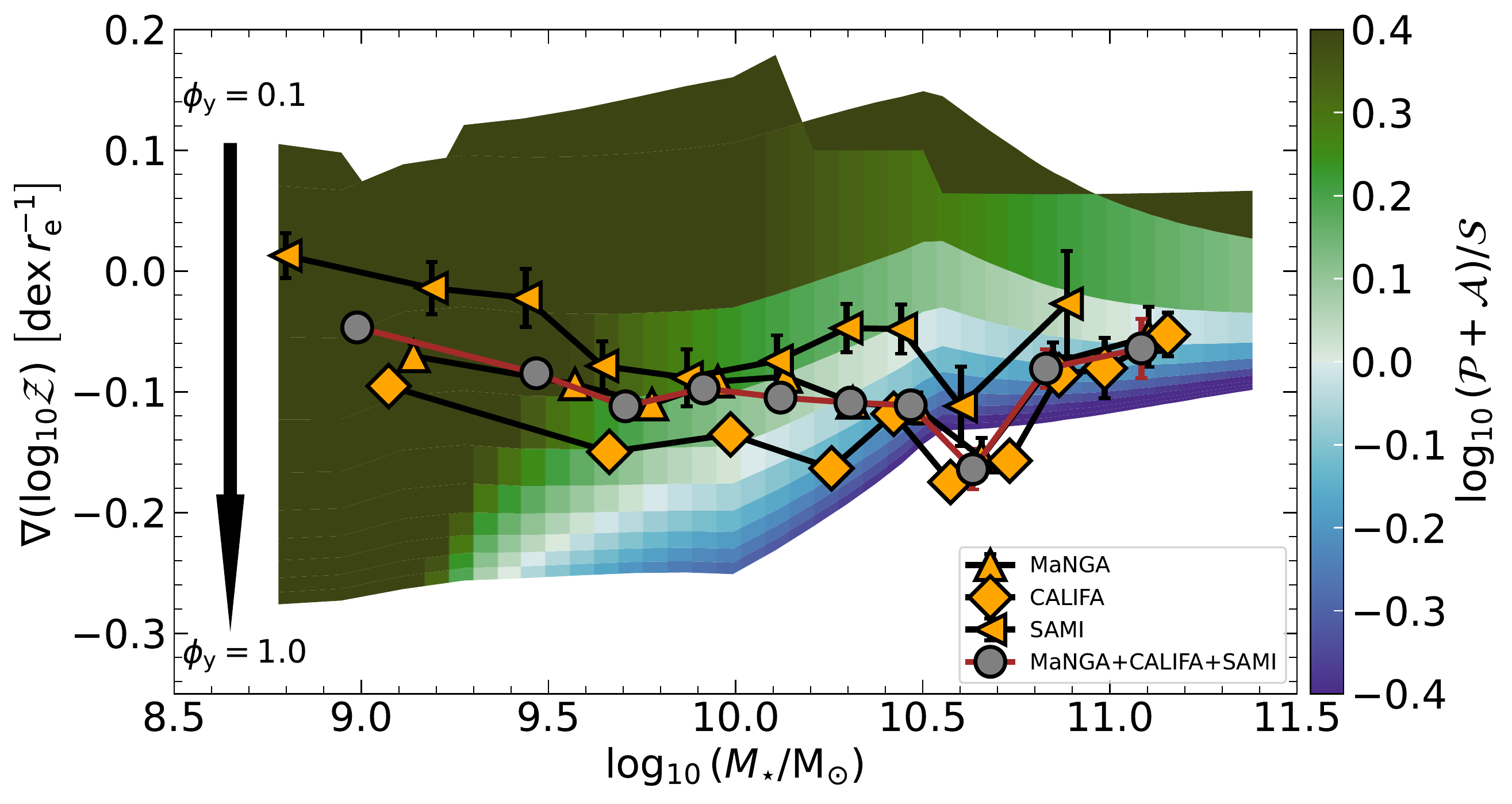}
\caption{Same as the fiducial model in the bottom panel of \autoref{fig:mzgr_basic} but for the fraction of star-forming gas, $f_{\rm{sf}}$, estimated using the approach laid out in \citet{2013MNRAS.436.2747K}.}
\label{fig:app_mzgr_basic_fsf}
\end{figure}

\begin{figure}
\includegraphics[width=1.0\columnwidth]{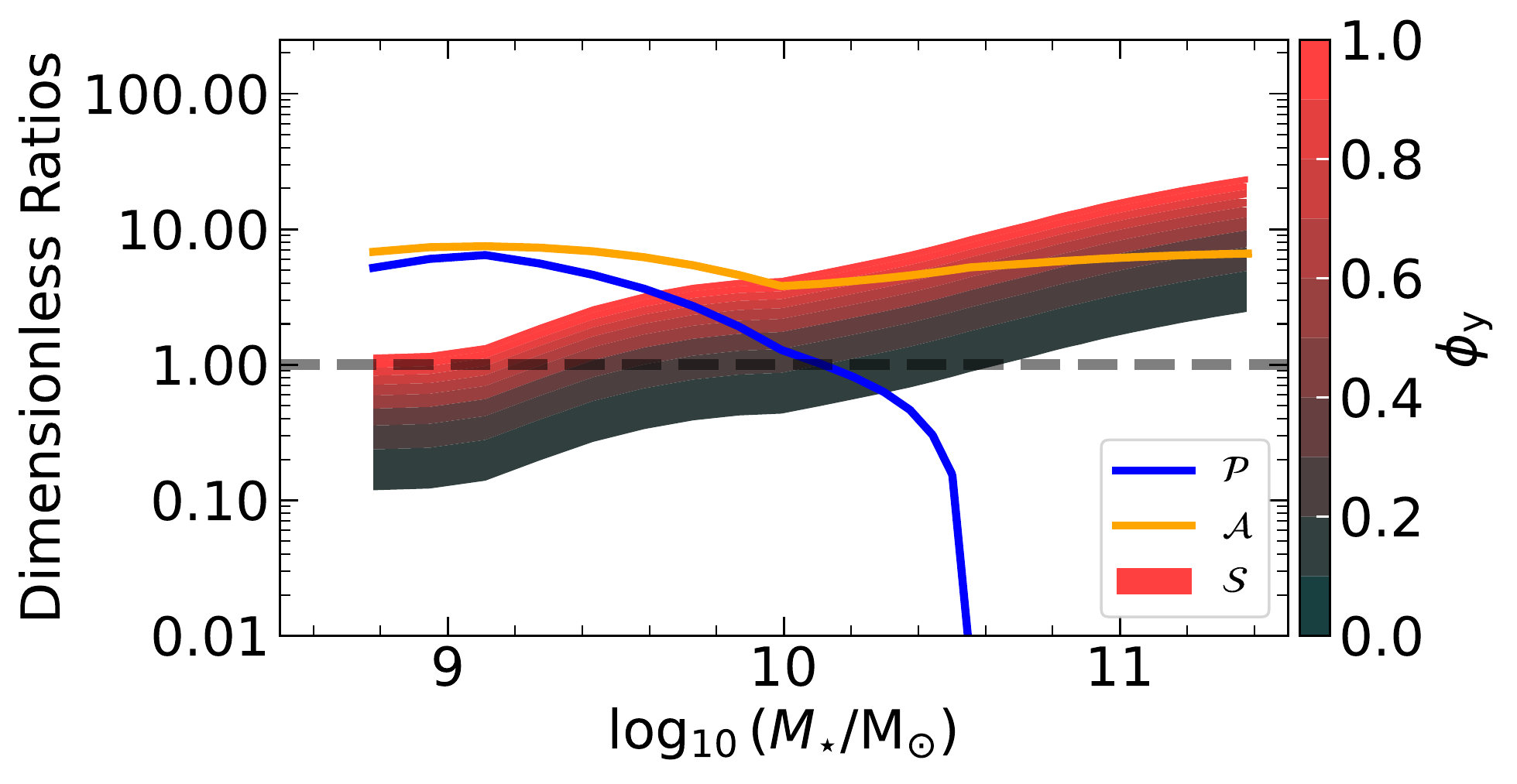}
\caption{Same as \autoref{fig:ratios_PSA} but for the model presented in \autoref{fig:app_mzgr_basic_fsf}.}
\label{fig:app_mzgr_basic_fsf_PSA}
\end{figure}

As a workaround, we now use the \cite{2013MNRAS.436.2747K} model to find $f_{\rm{sf}}$ where we estimate the input metallicity from the MZR. We adopt the value of $f_{\rm{sf}}$ at $r = R/2$ that roughly represents the radially-averaged $f_{\rm{sf}}$ across the disc. \autoref{fig:app_mzgr_basic_fsf} shows the resulting MZGR for the fiducial model we plot in the bottom panel of \autoref{fig:mzgr_basic}. We also plot the corresponding dimensionless ratios that govern the metallicity in \autoref{fig:app_mzgr_basic_fsf_PSA}. We find that transport is now active even in intermediate-mass galaxies because the average $f_{\rm{sf}}$ is lower than the one we use in the main text, which lowers $\dot M_{\rm{sf}}$. However, the gradients produced by the model for $M_{\star} \approx 10^{10.6}\,\rm{M_{\odot}}$ are slightly shallower than that observed, because a lower $f_{\rm{sf}}$ decreases $\mathcal{S}$ (cf. \autoref{eq:S}).

\section{Impact of uncertainties in the efficiency of accretion-induced turbulence}
\label{s:app_xia}
The efficiency with which accreting gas can convert its kinetic into driving turbulence in the disc is described by the parameter $\xi_{\rm{a}}$ in our work. Throughout the main text, we have assumed $\xi_{\rm{a}} = 0.2$ for all galaxies. However, as \citet{2022MNRAS.TMP.1282G} point out, $\xi_{\rm{a}}$ is essentially a free parameter as no constraints or measurements are available for it (except from the analysis of one low-mass galaxy in the IllustrisTNG simulations by \citealt{2022arXiv220405344F} that gives $\xi_{\rm{a}} = 0.2$). It is difficult to constrain $\xi_{\rm{a}}$ because it depends on the angular momentum of the accreting gas as well as its clumpiness \citep{2018ApJ...861..148M,2020MNRAS.498.2415M}. In this section, we explore higher values of $\xi_{\rm{a}} = 0.6$ and $\xi_{\rm{a}} = 1.0$ to understand its impact on the MZGR, noting that $\xi_{\rm{a}} = 1$ in particular is an extreme value since it implies all the kinetic energy of accreting gas goes into driving turbulent motions. We only show results for the fiducial model where $\sigma_{\rm{g}}$ is driven by feedback, accretion, and transport.

\begin{figure}
\includegraphics[width=1.0\columnwidth]{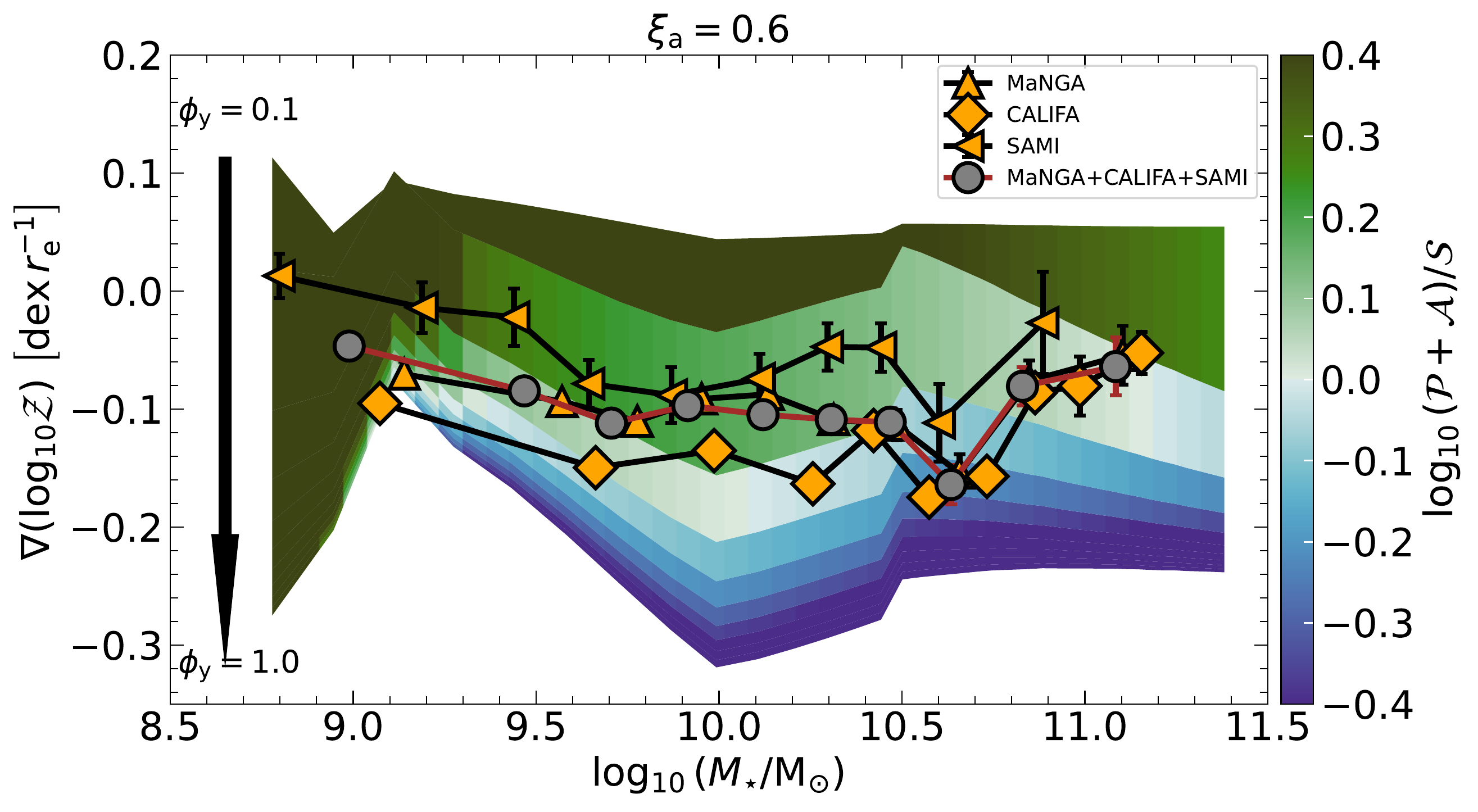}
\includegraphics[width=1.0\columnwidth]{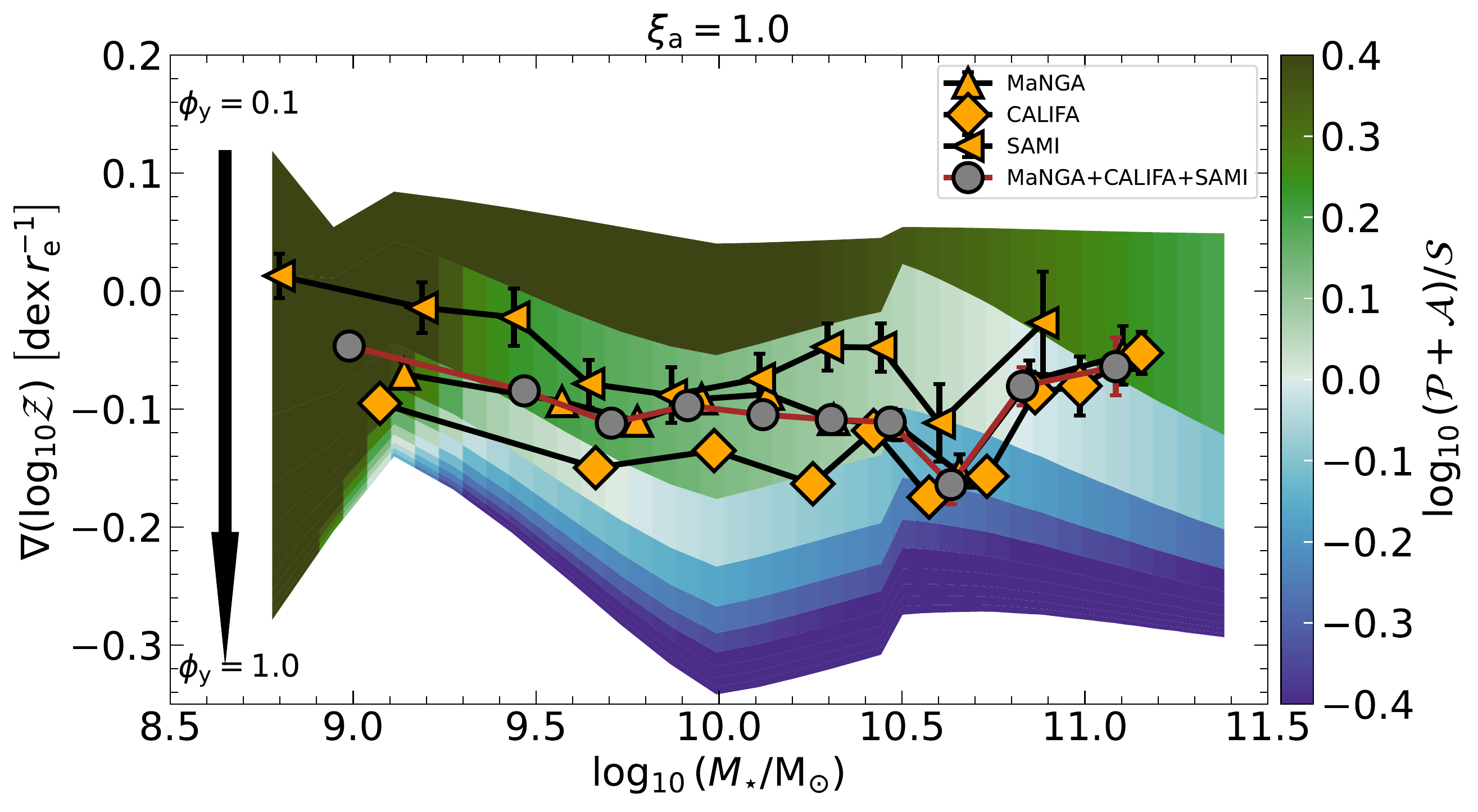}
\caption{Same as the fiducial model in the bottom panel of \autoref{fig:mzgr_basic} but with the $\xi_{\rm{a}} = 0.6$ (top panel) and $\xi_{\rm{a}} = 1.0$ (bottom panel), implying 60 per cent and 100 per cent of the kinetic energy of accreting gas goes into driving turbulent motions in the disc, respectively. The mass-loading factor ($\eta_{\rm{w}}$) is estimated from the EAGLE cosmological simulations \citep{2020MNRAS.494.3971M}.}
\label{fig:app_mzgr_basic_xia}
\end{figure}

The top and bottom panels of \autoref{fig:app_mzgr_basic_xia} plots the MZGRs for the case with $\xi_{\rm{a}} = 0.6$ and $1$, respectively. The key difference between these plots and the ones we present in the main text is that the scatter due to $\phi_{\rm{y}}$ at large $M_{\star}$ is higher. We can understand this trend with the help of $\sigma_{\rm{g}}$ -- higher $\xi_{\rm{a}}$ leads to more turbulence due to gas accretion, and thus, higher $\sigma_{\rm{g}}$. Since $\mathcal{S} \propto \sigma^2_{\rm{g}}$ but $\mathcal{A} \propto \sigma^3_{\rm{g}}$, $\mathcal{A}$ decreases much more as compared to $\mathcal{S}$, and a stronger $\mathcal{S}$ can drive steeper, negative metallicity gradients even at high $M_{\star}$. However, it is unlikely that $\xi_{\rm{a}}$ is high in local massive galaxies, especially if a significant fraction of accreting gas is co-rotating with the disc \citep[e.g.,][]{2015MNRAS.449.2087D,2022MNRAS.509.4149T}.

\section{Equations for central metallicity}
\label{s:app_Zr0}
Following from \autoref{s:model_gasphasemetal_eqbmsol}, we provide here the resulting equations for the central metallicity $\mathcal{Z}_{r_0}$. For the case where $\mathcal{Z}_{r_0}$ is set by the competition between source and accretion (e.g., massive galaxies),
\begin{equation}
\mathcal{Z}_{r_0} = \frac{\mathcal{S}}{\mathcal{A}}\,.
\end{equation}
If $\mathcal{Z}_{r_0}$ is set by the competition between advection and diffusion (e.g., low-mass galaxies), then, in the Toomre regime of star formation,
\begin{align}
\nonumber \mathcal{Z}_{r_0} = \biggl[\mathcal{P}^2 \mathcal{S} + \mathcal{A}\left(2c_1\mathcal{P}\sqrt{4\mathcal{A} + \mathcal{P}^2} + \mathcal{S}\right) - \\
\nonumber 4\mathcal{S}\beta^2 + \mathcal{P}\mathcal{S}\left(\sqrt{4\mathcal{A} + \mathcal{P}^2} + 2\beta\right) -  4c_1\mathcal{P}\sqrt{4\mathcal{A} + \mathcal{P}^2}\left(2\beta^2 + \mathcal{P}\beta\right)\biggr]/\\
\biggl[\left(\mathcal{A} + \mathcal{P}\left(\mathcal{P} + \sqrt{4\mathcal{A} + \mathcal{P}^2}\right)\right)\left(\mathcal{A} - 4\beta^2 - 2\mathcal{P}\beta\right)\biggr]\,,
\end{align}
and, in the GMC regime of star formation,
\begin{align}
\nonumber \mathcal{Z}_{r_0} = \biggl[c_1\mathcal{P}\sqrt{4\mathcal{A} + \mathcal{P}^2}\left(2\mathcal{A} + \mathcal{P}\left(-2 - 2\beta\right) - 2\left(1 + \beta\right)^2\right) \, + \\
\nonumber \mathcal{S}\left(\mathcal{A} + \mathcal{P}^2 - \left(1 + \beta\right)^2 + \mathcal{P}\left(1 + \sqrt{4\mathcal{A} + \mathcal{P}^2} + \beta\right)\right)\biggr]/\\
\biggl[\left(\mathcal{A} + \mathcal{P}\left(\mathcal{P} + \sqrt{4\mathcal{A} + \mathcal{P}^2}\right)\right)\left(\mathcal{A} + \mathcal{P}\left(-1 - \beta\right) - \left(1 + \beta\right)^2\right)\biggr]\,.
\end{align}
Finally, if $\mathcal{Z}_{r_0}$ is set by the competition between source and diffusion (e.g., when transport shuts off),
\begin{align}
\nonumber \mathcal{Z}_{r_0} = \biggl[0.5\biggl(\mathcal{S}\biggl(4\mathcal{A} + \mathcal{P}^2 + \mathcal{P}\left(-2 + \sqrt{4\mathcal{A} + \mathcal{P}^2} - 2\beta\right) - \\
\nonumber 4\left(1 + \beta\right)^2\biggr) + c_1\mathcal{P}\sqrt{4\mathcal{A} + \mathcal{P}^2}\left(2\mathcal{A} + \mathcal{P}\left(-2 - 2\beta\right) - 2\left(1 + \beta\right)^2\right)\biggr)\biggr]/\\
\biggl[\left(\mathcal{A} + 0.5\mathcal{P}\left(\mathcal{P} + \sqrt{4\mathcal{A} + \mathcal{P}^2}\right)\right)\left(\mathcal{A} + \mathcal{P}\left(-1 - \beta\right) - \left(1 + \beta\right)^2\right)\biggr]\,.
\end{align}

\section{Systematic differences in MZGR due to metallicity diagnostics}
\label{s:app_mingozzi}
The gas-phase metallicity is typically measured using a combination of emission lines commonly found in the spectra of \ion{H}{ii} regions. Various diagnostics of ISM metallicity exist that provide a way to convert emission line ratios to metallicity on the absolute scale ($12+\log_{10}\rm{O/H}$; see the reviews by \citealt{2019ARA&A..57..511K} and \citealt{2019A&ARv..27....3M}). These diagnostic have systematic differences that are partially responsible for the observed scatter in metallicity gradients \citep[e.g.,][]{2021MNRAS.502.3357P}. In this section, we study the impacts of these differences on our interpretation of the MZGR. To do so, we plot the MZGR obtained from the MaNGA survey \citep{2020A&A...636A..42M} but for three different calibrators: PP04, based on the \ion{N}{ii} to H$\alpha$ ratio, \citep{2004MNRAS.348L..59P}, M08, based on \ion{O}{III} and \ion{O}{ii} \citep{2008A&A...488..463M}, and IZI \citep{2015ApJ...798...99B}. Briefly, the modified version of IZI that \citeauthor{2020A&A...636A..42M} create uses Bayesian inference to predict the joint posterior probability distribution function (PDF) of the metallicity, ionization parameter, and extinction along the line of sight by comparing the observed emission line ratios with the \cite{2013ApJS..208...10D} photoionization models.

\begin{figure}
\includegraphics[width=1.0\columnwidth]{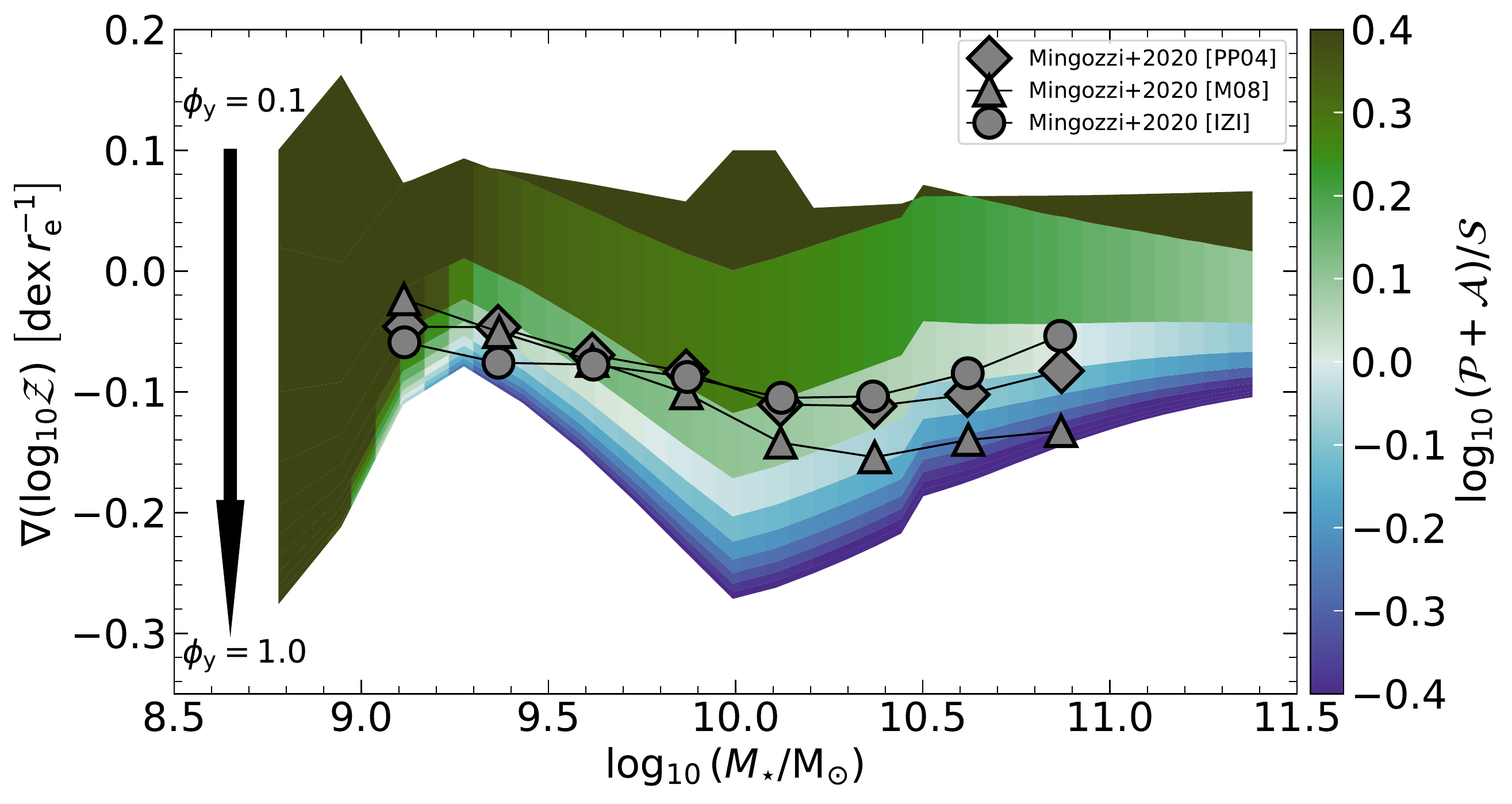}
\caption{Same as the fiducial model in the bottom panel of \autoref{fig:mzgr_basic} but the overplotted data is from the MaNGA survey \citep{2020A&A...636A..42M} using three different metallicity calibrations: PP04 \citep{2004MNRAS.348L..59P}, M08 \citep{2008A&A...488..463M}, and IZI \citep{2015ApJ...798...99B}. The mass-loading factor ($\eta_{\rm{w}}$) is estimated from the EAGLE cosmological simulations \citep{2020MNRAS.494.3971M}.}
\label{fig:app_mzgr_basic_mingozzi}
\end{figure}

\autoref{fig:app_mzgr_basic_mingozzi} shows a comparison of the fiducial model MZGR where $\sigma_{\rm{g}}$ is driven by feedback, accretion, and transport with the MaNGA MZGR derived from these metallicity calibrations. The overall trend in the MZGR remains the same for all the three calibrations, as already noted by \cite[fig. 12]{2020A&A...636A..42M}. In fact, the scatter within different surveys (MaNGA, SAMI, CALIFA) is somewhat larger than that within the same survey but different calibrators. The comparison between the model and the observed MZGR also remains unchanged. Thus, we find that systematic differences arising from different metallicity calibrations has no appreciable impact on the comparison between the model and the data.

\bsp	
\label{lastpage}
\end{document}